\documentclass[a4paper,reqno]{amsart}

\usepackage[english]{babel}
\usepackage[T1]{fontenc}
\usepackage{amsmath,amssymb,amsfonts,amsthm}
\usepackage{hyperref}
\usepackage{slashed}
\usepackage{graphicx,color}
\usepackage[dvipsnames]{xcolor}
\usepackage[color=Apricot]{todonotes}
 \usepackage{stmaryrd} 
 \usepackage{bm}
 \usepackage{pifont}
 \usepackage[bbgreekl]{mathbbol}
\usepackage{mathtools}
\usepackage{tikz-cd}
\usepackage[a4paper]{geometry}
\usepackage{csquotes} %%necessary for biblatex

\theoremstyle{definition} %%upright text, extra space above and below
\newtheorem{definition}{Definition}

\theoremstyle{plain} %% italic text, extra space above and below
\newtheorem{theorem}[definition]{Theorem}
\newtheorem{proposition}[definition]{Proposition}
\newtheorem{lemma}[definition]{Lemma}
\newtheorem{corollary}[definition]{Corollary}

\theoremstyle{remark} %% upright text, no extra space above or below
\newtheorem{remark}[definition]{Remark}

\usepackage[bibstyle=alphabetic,citestyle=alphabetic,useprefix,giveninits=true, sorting=ynt, sortcites, minbibnames=99,maxbibnames=99,backend=biber]{biblatex}  %%other styles: numeric-comp, authoryear 
\renewbibmacro{in:}{} %% removes in: before journals
\bibliography{bibliography}  %% Name of the file with the bibliography
\DeclareSourcemap{
  \maps[datatype=bibtex]{
    \map[overwrite]{ %%If DOI is present, doesn't print arXiv
      \step[fieldsource=doi, final]
      \step[fieldset=url, null]
      \step[fieldset=eprint, null]
      }
     \map[overwrite]{ %% If only arXiv is present, doens't print pages and eid (eliminates duplicates)
      \step[fieldsource=eprint, final]
      \step[fieldset=pages, null]
      \step[fieldset=eid, null]
      \step[fieldset=journal, null]
    }  
  }
}
\bibliography{bibliography}

\newcommand{\qsp}[2]{\,\ensuremath{\raise.5ex\hbox{$#1$}\big\slash\raise-.5ex\hbox{$#2$}}}

\newcommand{\un}[1]{\underline{#1}}
\newcommand{\ol}[1]{\overline{#1}}
\newcommand{\ugam}{\underline{\gamma}}
\newcommand{\igam}{\iota_{\hat{\gamma}}}
\newcommand{\vf}{\iota_\varphi}
\newcommand{\Ker}[1]{\mathrm{Ker}{(#1)}}

\DeclareMathOperator{\Ima}{Im}

    \makeatletter
        \newcommand{\zzlabel}[1]{\ifmeasuring@\else\ltx@label{#1}\fi} %%new label (necessary if amsmath is present)
    \makeatother

    \newcounter{terms}[equation] %%counter for terms in a equation
    \newcommand{\unl}[2]{\underline{#1}_{\refstepcounter{terms} \zzlabel{#2} \theterms}} %%underlines, counts a term and put the corresponding number. Put the term in the first slot and a label in the second

    \newcommand{\reft}[2]{(\ref{#1}.\ref{#2})} %%refers to a term in a equation as (equation.term)

\RequirePackage{tikz-cd}
\RequirePackage{amssymb}
\usetikzlibrary{calc}
\usetikzlibrary{decorations.pathmorphing}

\tikzset{curve/.style={settings={#1},to path={(\tikztostart)
    .. controls ($(\tikztostart)!\pv{pos}!(\tikztotarget)!\pv{height}!270:(\tikztotarget)$)
    and ($(\tikztostart)!1-\pv{pos}!(\tikztotarget)!\pv{height}!270:(\tikztotarget)$)
    .. (\tikztotarget)\tikztonodes}},
    settings/.code={\tikzset{quiver/.cd,#1}
        \def\pv##1{\pgfkeysvalueof{/tikz/quiver/##1}}},
    quiver/.cd,pos/.initial=0.35,height/.initial=0}

\tikzset{tail reversed/.code={\pgfsetarrowsstart{tikzcd to}}}
\tikzset{2tail/.code={\pgfsetarrowsstart{Implies[reversed]}}}
\tikzset{2tail reversed/.code={\pgfsetarrowsstart{Implies}}}
\tikzset{no body/.style={/tikz/dash pattern=on 0 off 1mm}}
    
\title{The Reduced Phase Space of $N=1, D=4$ Supergravity in the BV-BFV formalism}

\author{Alberto S. Cattaneo }
\author{Filippo Fila-Robattino}

\renewcommand\footnotemark{}

\date{}                     
\thanks{We acknowledge partial support of the SNF Grant No. 200021\_227719 and of the Simons Collaboration on Global Categorical Symmetries. This research was (partly) supported by the NCCR SwissMAP, funded by the Swiss National Science Foundation. This article is based upon work from COST Action 21109 CaLISTA, supported by COST (European Cooperation in Science and Technology)(www.cost.eu), MSCA-2021-SE-01-101086123 CaLIGOLA, and MSCA-DN CaLiForNIA -101119552. FFR acknowledges funding from the EU project Caligola HORIZON-MSCA-2021-SE-01, Project ID: 101086123.}
\begin{document}

\begin{abstract}
    This paper describes the reduced phase space of $N=1$, $D=4$ supergravity in the fully off-shell Palatini--Cartan formalism. This is achieved through the KT construction, allowing an explicit description of first-class constraints on the boundary. The corresponding BFV description is obtained, and its relation with the BV one in the bulk is described by employing the BV pushforward in the particular example of a cylindrical spacetime.
\end{abstract}
\maketitle
\tableofcontents
\section{Introduction}
Supergravity occupies a central role in modern theoretical physics, standing at the intersection of gauge theory, general relativity, and supersymmetry. Among the many formulations of supergravity, the $N=1$, $D=4$ model holds a special status as the simplest realistic incarnation, coupling a spin-$\frac{3}{2}$ gravitino to Palatini–Cartan gravity.

In this paper, we study the geometry of the boundary structure of the theory and, in particular, the reduced phase space, in relation with the BFV and BV formalisms.

Historically, boundaries have been interesting to the physics community insofar as they are Cauchy surfaces $\Sigma$ on which one can study the Hamiltonian description of the theory. The original methods developed by Dirac \cite{Dir58} involve the analysis of first and second-class constraints, which define the phase space of the theory.

In the context of this paper, we start by consider the Lagrangian field theory in the bulk of spacetime, which we assume to have a non-degenerate boundary $\Sigma$, on which the Euler-Lagrange equations split into evolution equations and constraints. The latter define a submanifold of the space of boundary fields, which can be realized as a (infinite-dimensional) symplectic manifold thanks to the Kijowski--Tulczjiew \cite{KT79} construction. 
When the constraints form a first-class set, i.e., generate a Poisson ideal with respect to the symplectic structure, they define a coisotropic submanifold. Furthermore, their Hamiltonian vector fields can be interpreted as generators of the symmetries of the theory.

In relations to the supergravity theory at hand, the constraints are nothing but the restrictions of the Einstein, torsion, and Rarita--Schwinger equations to the boundary, and are implemented via the use of Lagrange multipliers, which take the roles of gauge parameters. Indeed, the Hamiltonian vector fields of the above constraints, which are shown to be first-class, generate respectively the diffeomorphisms, the internal Lorentz transformations and the local supersymmetry transformations. In this regard, supersymmetry emerges naturally as the gauge symmetry associated to the gravitino, in a way not dissimilar to standard (non-fermionic) symmetries. 

The coisotropic submanifold defined as the constrained set is the space of initial conditions of the theory, but as such it is not a symplectic manifold. Indeed, it includes gauge-equivalent boundary field configurations, which are precisely the ones related by the flow of the Hamiltonian vector fields, reinterpreted as gauge transformations. In general, the reduced phase space is obtained by taking the appropriate coisotropic reduction, which amounts to quotienting with respect to the action of the Hamiltonian vector fields. In favorable cases, this turns out to be the Marsden--Weinstein \cite{MW74} reduction, however, in the case at hand (and in gravity in general, see e.g. \cite{CCS2020}, \cite{blohmann2021hamiltonianliealgebroids}) it is not the case.

The solution to this problem is provided by the BFV formalism \cite{BV3,Scht08,Stasheff1997}, which allows to cohomologically resolve the reduced phase space, by regarding the algebra of functions on it as the degree-0 cohomology of a certain nilpotent operator $Q$ of odd degree, defined on an enlarged space of fields which is endowed with the structure of a symplectic supermanifold.

The BFV formalism is closely related to its bulk counterpart, the BV formalism, introduced by Batalin, Fradkin and Vilkovisky\cite{BV1,BV2, BV3} to treat the quantization of gauge theories. In particular, BV and BFV procedures generalise the Fadeev--Popov and BRST constructions on manifolds with boundary and in a way compatible with cutting and gluing, which provides an important step towards the axiomatization of quantum field theories. 

In many cases, the presence of a boundary alters the BV description of a theory in the bulk, in the sense that the BV axioms (this scheme, called BV-BFV, has been studied in \cite{CMR2012,CMR2}) are satisfied only up to boundary terms, which turn out to define the data of a BFV theory. In the case studied in this paper, however, this is not the case, as there are obstruction to the existence of a symplectic structure on the boundary BFV fields compatible with the bulk BV fields, very much like it is the case in PC gravity \cite{CS2017}, therefore only obtaining a pre-BFV theory. Indeed the BV structure of supergravity, computed in \cite{CFR25}, fails to induce a BFV one on the boundary. On top of it, while the first-class condition guarantees its existence, the hands-on computation of the BFV action is immensely complicated by the fact that supersymmetry closes only on-shell, hence introducing higher order terms in the action, presenting a cumbersome technical challenge.

The way out of this standstill is provided by the BV pushforward, a technique that allows to eliminate extra auxiliary degrees of freedom within the BV formalism, e.g. integrating out heavy modes. In the case of gravity and supergravity, the idea is to get rid of the degrees of freedom responsible for the obstruction in the BV-BFV compatibility of the theory. In particular, we show that the original BV theory is equivalent (in the BV sense) to ''reduced" one, where a gauge-invariant set of constraints is imposed. This equivalence is established explicitly, starting from the work in \cite{C25}, via the BV pushforward, with the caveat of having to work on a cylinder, where spacetime is of the form $M=I\times \Sigma$. The situation is summed up in the following diagram
\[\begin{tikzcd}
	F && {F_\Sigma} \\
	{\mathfrak{F}} && {\mathfrak{F}_\Sigma} \\
	{\mathfrak{F}^r} & {} & {\mathfrak{F}_\Sigma^r}
	\arrow["KT", from=1-1, to=1-3]
	\arrow["BV"', from=1-1, to=2-1]
	\arrow["BFV", dashed, from=1-3, to=2-3]
	\arrow["{BV-BFV}", dashed, from=2-1, to=2-3]
	\arrow["{BV pf}"', two heads, from=2-1, to=3-1]
	\arrow["\simeq", tail reversed, from=2-3, to=3-3]
	\arrow["{BV-BFV}", from=3-1, to=3-3]
\end{tikzcd}\]
where $\mathfrak{F}$ and $\mathfrak{F}_\Sigma$ indicate respectively the BV and BFV theories, while $F$ and $F_\Sigma$ the classical ones. Furthermore, the apex ${}^r$ indicates the reduced theory, while the dashed lines indicate the failure of the construction. 

BV-BFV compatibility is one of the necessary requirements for the quantisation of BV theories with boundary,  and it is therefore important to establish it for theories that one wishes to provide a perturbative quantization of. We achieve this for supergravity by showing that the reduced BV theory exists and is indeed BFV extendible, in such a way that the resulting BFV theory is equivalent (there exists a symplectomorphism relating the two theories) to the one resolving the reduced phase. 

\subsection*{Structure of the paper}
 In Section \ref{sec: Review of previous results} we summarise the previous results on the BV description of supergravity, and present the main construction for the BV-BFV formalism. 

 In Section \ref{sec: RPS} we review the Kijowski--Tulczijew construction in field theory and apply it to the case of supergravity, showing that the constraints obtained are first-class, computing explicitly their Poisson brackets. The basis for the BFV description of the theory are posed. 

 Section \ref{sec: BFV extension} contains a brief review of the BV pushforward both in the general case and in the case of PC gravity, where a couple of redefinitions are implemented to make the coming computations more suitable for the case at hand. In the second part of the section, the BV pushforward for supergravity is explicitly computed, yielding a BV-BFV compatible theory.

Lastly, in Section \ref{sec: BFV SG}, the BFV description of supergravity is established, albeit only implicitly, by means of the 1D AKSZ construction, which is summarised in \ref{sec: AKSZ theory}.

Throughout the paper we use a vast body of technical results, regrouped in Appendix \ref{app: tools}.
\subsection*{Acknowledgements} The authors thank Giovanni Canepa for the useful comments and discussion. 

\section{Review of previous results}\label{sec: Review of previous results}
In the following section, we provide a short overview on the results obtained in \cite{CFR25}. We start by recalling some generalities on the BV formalism for field theories.

Let $F_M$ be the space of classical fields, usually obtained as the space of sections of a vector bundle over the spacetime manifold $M$. When gauge symmetries are present, it is convenient to enlarge the space $F_M$ to a $\mathbb{Z}$--graded symplectic supermanifold $\mathcal{F}_M$, whose symplectic form is of degree -1. Furthermore, the classical symmetries of the theory will be encoded in a degree 1 vector field $Q$, such that $Q^2=\frac{1}{2}[Q,Q]=0$, i.e. $Q$ is cohomological. This data is regrouped in the following definition.

\begin{definition}
    A BV manifold on $M$ is given by the collection $(\mathcal{F}_M,\mathcal{S}_M,Q,\varpi_M)$, where $(\mathcal{F}_M,\varpi_M)$ is a $\mathbb{Z}$-graded manifold endowed with a -1-symplectic form $\varpi_M$,  $\mathcal{S}_M$ is a degree 0 functional (the BV action) and $Q$ is a degree 1 vector field such that 
    \begin{itemize}
        \item $\iota_{Q}\varpi_M=\delta\mathcal{S}_M$, i.e. $Q$ is the Hamiltonian vector field of $\mathcal{S}_M$;
        \item $Q^2=\frac{1}{2}[Q,Q]=0$, i.e. $Q$ is cohomological.
    \end{itemize}
\end{definition}

\begin{remark}
    Typically, the BV action extends the classical action $S_M^0$ of the theory, in such a way to contain the symmetries of the fields and the antifields. 
    Furthermore, the condition $Q^2=0$ is equivalent to the classical master equation 
    \begin{equation}
        (S,S)=\iota_Q\iota_Q \varpi_M.
    \end{equation}
\end{remark}

In favorable cases, the symmetries of the classical action $S_M^0$ form a distribution $D \subset \mathfrak{X}(F_M)$, which might correspond to the infinitasimal action of the Lie algebra of a Lie group. If this happens, the BV formalism simplifies to the BRST prescription (see \cite{Mnev2017}). However, in the general case, one only requires that the distribution is involutive on the Euler-Lagrange locus $EL_M := \{\phi \in F_M \mid \delta S \vert_\phi = 0\}$ of the theory. 

The simplest BV manifold that can be constructed at this point is $\mathcal{F}_M := T^*[-1] D[1]$, where the vector fields representing symmetries are promoted to degree one fields (the ghosts), and the odd cotangent fibers define the antifields and antighosts with degrees -1 and -2, respectively. These graded field spaces can then be endowed with the canonical $-1$-symplectic form, defined on a $-1$-shifted cotangent bundle. In this framework, let $\Phi = (\Phi^\alpha)$ represent a multiplet (comprising fields and ghosts) in $D[1]$, and $\Phi^\Finv = (\Phi^\Finv_\alpha)$ be its canonical conjugate (containing antifields and antighosts) in the fiber of $T^*[-1] D[1]$. Let $Q_0$ denote the vector field encoding the symmetries of the theory (i.e. such that $Q_0(S_M^0)=0$). Then the simplest putative BV action is obtained as a linear functional in the antifields/ghosts, with
\begin{align*}
    &\varpi_M = \int_M \delta \Phi^\Finv_\alpha \delta \Phi^\alpha, \qquad \qquad \mathcal{S}_M = S_M^0 + \int_M \Phi^\Finv_\alpha Q_0(\Phi^\alpha).
\end{align*}
However, it is often the case that a functional linear in the antifields is not enough to obtain a BV action satifying the CME, hence the necessity to introduce higher order terms, obtaining
    \begin{equation*}
        \mathcal{S}_M = S_M^0 + \int_M \Phi^\Finv_\alpha Q_0(\Phi^\alpha) + \sum_{k=2}^{p}\frac{1}{k!}M^{\alpha_1\cdots\alpha_k}(\Phi)\Phi^\Finv_{\alpha_1}\cdots\Phi^\Finv_{\alpha_k}.
    \end{equation*}
The number $p$ is called the rank of the BV action and, in the relevant cases, it is finite. Indeed each coeffiecient $M^{\alpha_1\cdots\alpha_k}(\Phi)$ is obtained recursively from the lower order ones, and one can prove this procedure is finite.

\subsection{The BV action of $N=1,D=4$ supergravity}\label{subsec: BV Sugra}
In this section, we see how the previous discussion applies to the case of $N=1,D=4$ supergravity. We start by defining the classical space of fields, and then provide the classical action. 

First of all, we assume $M$ to be a pseudo-riemannian spin $4$-manifold, endowed with a principal Spin$(3,1)$--bundle $P_{\text{Spin}}\rightarrow M$.  Consider a 4-dimensional real vector space $V$ equipped with a Lorentzian metric $\eta$ of signature $(-,+,+,+)$, then, without loss of generality, we take $\eta = \text{diag}(-1,1,1,1)$ as the Minkowski metric. The corresponding associated bundle, referred to as the "Minkowski bundle," is defined as $\mathcal{V} := P_{\text{Spin}} \times_\Lambda V$, where $\Lambda$ represents the spin-1 representation of $\text{Spin}(3,1)$.

For future convenience, assuming $M$ to be a manifold with boundary $\partial M=\colon\Sigma$, we define the following spaces\footnote{Indicating by $\mathcal{V}$ both the vector bundles $\mathcal{V}\rightarrow M$ and $\mathcal{V}|_\Sigma\rightarrow\Sigma$.}
        \begin{equation*}
            \Omega^{(k,l)}:=\Omega^k(M,\wedge^l \mathcal{V}) \qquad \Omega_\partial^{(k,l)}:=\Omega^k(\Sigma,\wedge^l \mathcal{V}),
        \end{equation*}
    and maps 
        \begin{eqnarray}
            W_{k}^{ (i,j)}: \Omega^{i,j}  \longrightarrow &\Omega^{i+k,j+k}  \\
            X  \longmapsto  &  \frac{1}{k!}\underbrace{e \wedge \dots \wedge e}_{k-times}\wedge X  \nonumber\\
            W_{k}^{ \partial, (i,j)}: \Omega_{\partial}^{i,j}  \longrightarrow & \Omega_{\partial}^{i+k,j+k}  \\
            X  \longmapsto &  \frac{1}{k!} \underbrace{e \wedge \dots \wedge e}_{k-times}\wedge X\nonumber
        \end{eqnarray}
The classical space of fields is then given by
    \begin{equation*}
        F_M=\Omega^{(1,1)}_{n.d.} \otimes \mathcal{A}_M \otimes \Omega^{1}(M,\Pi \mathbb{S}_M),
    \end{equation*}
where the classical fields consist of the vielbein $e \in \Omega^{(1,1)}_{n.d.}$,\footnote{The vielbein is defined to be a linear isomorphism of the tangent bundle with the minkowsky bundle, hence a non-degenerate element of $\Omega^{(1,1)}$. Given the reference metric $\eta$ on $\mathcal{V}$, one can induce a metric on $M$ by $g:=e^*(\eta)$, in components $g_{\mu\nu}=e_\mu^a e_\nu^a \eta_{ab}$} the spin connection $\omega \in \mathcal{A}_M$, and the gravitino $\psi \in \Omega^{1}(M, \Pi \mathbb{S}_M)$\footnote{We define $\mathbb{S}_M$ as the subspace of Majorana spinors, i.e. the set of Dirac spinors on $M$ such that $\bar{\psi}=\psi^t C$, where $C$ is the charge conjugation matrix. See \cite{Canepa:2024rib,FR25} for more details.}.

The classical action reads
    \begin{equation}
        S^0_M=\int_M \frac{e^2}{2}F_\omega + \frac{1}{3!}e \bar{\psi}\gamma^3 d_\omega \psi,
    \end{equation}
where $\gamma=\gamma^a v_a$ is a constant section of $\mathcal{V}\otimes\mathcal{C}(V)$, where $\mathcal{C}(V)$ is the Clifford algebra of $V$, locally represented by complex $4\times 4$ matrices\footnote{The gamma matrices $\{\gamma_a\}$, $a=0,\cdots,3$ are precisely the Dirac gamma matrices, subject to the following relation
    \begin{equation*}
        \{\gamma_a,\gamma_b\}=-2\eta_{ab}\mathbb{1}
    \end{equation*}
    and generate the whole algebra of complex $4\times 4$ matrices.}, and $\{v_a\}$ is a local basis of $\mathcal{V}$. Furthermore, $d_\omega\psi=d\psi +[\omega,\psi]$, where $[\omega,\cdot]$ generically denotes the action of the spin connection (as a section of the Lie algebra $\mathfrak{spin}(3,1)=\mathfrak{so}(3,1)\simeq \wedge^2 V$) via the correct representation. In this case, the spin connection acts on spinors via the gamma representation, which sends an element $v_a\wedge v_b$ of the basis of $ \wedge^2 V $ to $-\frac{1}{4}\gamma_{ab}$, with $\gamma_{ab}:=\gamma_{[ab]}$. We then obtain
        \begin{equation}
            d_\omega\psi=d\psi-\frac{1}{4}\omega^{ab}\gamma_{ab}\psi,
        \end{equation}
    providing the correct spin transformation of the Majorana spinor\footnote{One can also show that $\gamma$ is covariantly constant, i.e. $d_\omega\gamma=0$, for any spin connection $\omega$. Indeed by definition $d\gamma=0$, hence $d_\omega\gamma=[\omega,\gamma]$. At this point, since $\gamma$ has values both in $V$ and $\mathcal{C}(V)$, an element of $\wedge^2 V\simeq \mathfrak{so}(3,1)=\mathfrak{spin}(3,1)$ acts separately on the two factors, respectively via the vector representation and the gamma representation, obtaining
        \begin{equation*}
        [\omega,\gamma]=[\omega,\gamma]_V+[\omega,\gamma]_S=\omega^{ab}\gamma_b v_a -\frac{1}{4}\omega_{ab}(\gamma^{ab}\gamma^c-\gamma^c\gamma^{ab})v_c=0.
        \end{equation*}
        where in the last step we used twice the anticommutation relation
        \[
        \{\gamma^a,\gamma^b\}=-2\eta^{ab}.
        \]
        .}.

Finally, the BV space of fields is given by 
    \begin{equation*}
        \mathcal{F}_{\mathrm{SG}}=T^*[-1]\big(\Omega^{(1,1)}_{n.d.} \otimes \mathcal{A}_M \otimes \Omega^{1}(M,\Pi \mathbb{S}_M) \otimes \Omega^{(0,2)}[1] \times \mathfrak{X}[1](M) \otimes \Gamma[1](M,\Pi \mathbb{S}_M) \big),
    \end{equation*}
where
\begin{itemize}
    \item The ghost fields, which can be thought of as (odd) gauge parameters, include $c \in \Omega^{(0,2)}[1] = \Gamma[1](M, \wedge^2 \mathcal{V})$ for internal Lorentz transformations, $\xi \in \mathfrak{X}[1](M)$ for diffeomorphisms, and $\chi \in \Gamma[1](M, \Pi \mathbb{S}_M)$ for local supersymmetry.
    \item The antifields are given by $e^\Finv \in \Omega^{(3,3)}[-1]$, $\omega^\Finv \in \Omega^{(3,2)}[-1]$, and $\psi^\Finv \in \Omega^{(3,4)}[-1](M, \Pi \mathbb{S}_M)$. The corresponding antighosts are $c^\Finv \in \Omega^{(4,2)}[-2]$, $\xi^\Finv \in \Omega^1(M)[-2] \otimes \Omega^{(4,4)}$, and $\chi^\Finv \in \Omega^{(4,4)}[-2](M, \Pi \mathbb{S}_M)$.
\end{itemize}
The $-1$-symplectic form is expressed as:
\begin{equation}
    \varpi_{\mathrm{SG}} = \int_M \delta e \delta e^\Finv + \delta \omega \delta \omega^\Finv + i \delta \bar{\psi} \delta \psi^\Finv + \delta c \delta c^\Finv + \iota_{\delta \xi} \delta \xi^\Finv + i \delta \bar{\chi} \delta \chi^\Finv.
\end{equation}
Before showing the full form of the BV action, some redefinition is in order. Let $\ugam:=\gamma_\mu dx^\mu$, then we set
    \begin{equation}\label{eq: redefinition of antifields}
        \omega^\Finv=e\check{\omega},\qquad   c^\Finv=\frac{e^2}{2}\check{c}, \qquad \psi_\Finv=\frac{1}{3!}\gamma^3\ugam\psi^0_\Finv.
    \end{equation}
It was proven in \cite{FR25} and \cite{Canepa:2024rib} that the above maps sending the newly defined fields to the original ones are isomorphisms, hence providing a well-given redefinition. 

Furthermore, lemmata \ref{lem: splitting (2,1) spin} and \ref{lem: splitting (3,1)} allow to define maps
    \begin{align*}
            &\alpha\colon\Omega^{(3,1)}(\Pi \mathbb{S}_M)\rightarrow \Omega^{(1,0)}(\Pi \mathbb{S}_M) &\beta\colon\Omega^{(3,1)}(\Pi \mathbb{S}_M)\rightarrow\ker{(\gamma^3_{(3,1)})}\\
            &\kappa\colon\Omega^{(2,1)}(\Pi \mathbb{S}_M)\rightarrow\Omega^{(1,0)}(\Pi \mathbb{S}_M) & \varkappa\colon \Omega^{(2,1)}(\Pi \mathbb{S}_M)\rightarrow\ker{(\ugam\gamma^3_{(2,1)})}
    \end{align*}
such that for all $\theta\in\Omega^{(3,1)} $ and $\sigma\in\Omega^{(2,1)}$ one has
    \begin{align*}
       & \theta= i e \ugam\alpha(\theta) + \beta(\theta), \qquad\sigma = e \kappa(\sigma) + \varkappa(\sigma),
    \end{align*}
where $(\ugam\gamma^3)_{(2,1)}\colon\Omega^{(2,1)}(\Pi \mathbb{S_M})\rightarrow \Omega^{(3,4)}(\Pi \mathbb{S_M})\colon\beta\mapsto\ugam\gamma^3\beta$.  

With these tools we can finally define the full BV action, which turns out to be of rank 2, indeed $\mathcal{S}_{\mathrm{SG}}=S^0_{\mathrm{SG}}+{s}^1_{\mathrm{SG}} + {s}^2_{\mathrm{SG}}$, where ${s}^1_{\mathrm{SG}}$ and ${s}^2_{\mathrm{SG}}$ are respectively the linear and quadratic parts in the antifields. Explicitly, one finds 
    \begin{align}
        \nonumber{s}^1_{\mathrm{SG}}&=\int_M -(L_\xi^\omega e - [c,e] + \bar{\chi}\gamma\psi) e^\Finv + \left(e\iota_\xi F_\omega - ed_\omega c -\frac{1}{3!}\bar{\chi}\gamma^3 d_\omega\psi \right)\check{\omega}  \\
        &\qquad -i (L_\xi^\omega \bar{\psi} - [c,\bar{\psi}] -d_\omega \bar{\chi})\psi_\Finv + \left(\frac{1}{2}\iota_\xi\iota_\xi F_\omega -\frac{1}{2} [c,c] + \iota_\xi \delta_\chi\omega  \right)c^\Finv   \\
        \nonumber&\qquad +\frac{1}{2}(\iota_{[\xi,\xi]} + \iota_\varphi)\xi^\Finv  - i\left(L_\xi^\omega\bar{\chi} - [c,\bar{\chi}]  -\frac{1}{2}\iota_\varphi \bar{\psi} \right)\chi_\Finv,
    \end{align}
and, defining the combination of fields $k^\Finv:=\omega^\Finv - \iota_\xi c^\Finv$ and $k^\Finv= e \check{k}$, we have 
    \begin{equation}\label{eq: quadratic BV action}
        \begin{split}
            s_2=\int_M &\frac{1}{2}e^\Finv \vf\check{k} +\frac{1}{4}\left(\frac{1}{2}\bar{\psi}^\Finv_0 + \alpha(\check{k}\bar{\psi})\ugam - \frac{1}{2}\check{c}\bar{\chi}\right)\vf\psi^\Finv\\
            & - \frac{i}{4\cdot 3!}\left(\bar{\psi}^\Finv_0\ugam + \alpha(\check{k}\bar{\psi} )\ugam  \right)\gamma^3 \chi <e,\bar{\chi}[\check{k},\gamma]\psi> \\
            &+\frac{i}{8\cdot 3!}\left( \alpha(\check{k}\bar{\psi})\ugam - i \check{c}\bar{\chi} \right)\gamma^3 \vf(\check{k}\psi) + \frac{1}{8\cdot 3!} \bar{\psi}^\Finv_0\ugam \gamma^3 \chi <e,\bar{\chi}\ugam^2\psi^\Finv_0>\\
            &-\frac{1}{32}\left(i\bar{\psi}^\Finv  \chi + \frac{1}{3!}\check{k}\bar{\psi}\gamma^3 \chi\right)\bar{\chi}\igam\igam([\check{k},\gamma]\psi) +\frac{1}{32}\bar{\psi}^\Finv\chi \bar{\chi}\igam\igam(\ugam^2 \psi^\Finv_0).
        \end{split}
        \end{equation}
where $\hat{\gamma}:=\gamma^\mu\partial_\mu=e^\mu_a \gamma^a \partial_\mu$, $\varphi=\bar{\chi}\hat{\gamma}\chi$ and the map $<e,->$ is defined via the inverse vielbein as 
    \begin{align*}
        <e,->\colon\Omega^{(i,j)}&\longrightarrow\Omega^{(i-1,j+1)}\\
        \sigma&\longmapsto v_a\eta^{ab}e^\mu_b\iota_{\partial_\mu}\sigma.
    \end{align*}    
The full cohomological vector field $Q_{\mathrm{SG}}$ is given by $Q_{\mathrm{SG}}=Q_{\mathrm{SG}}^0+\mathbb{q}$, where $Q^0_{\mathrm{SG}}$ is just the sum of the the infinitesimal gauge transformations\footnote{In particular, the supersymmetry transformation for the connection $\delta_\chi\omega$ is defined implicitly by
\[
e\delta_\chi\omega=\frac{1}{3!}\bar{\chi}\gamma^3d_\omega\psi.
\]
Such definition is well given because $W_e^{(1,2)}\colon \alpha \mapsto e\wedge \alpha$ is an isomorphism
}
    \begin{align}\label{eq: Q_0 fields}
        \nonumber& Q_{\mathrm{SG}}^0 e = L_\xi^\omega e - [c,e] +  \bar{\chi}\gamma\psi & Q_{\mathrm{SG}}^0 \omega = \iota_\xi F_\omega - d_\omega c + \delta_\chi \omega \\
        &Q_{\mathrm{SG}}^0 \psi= L_\xi^\omega \psi - [c,\psi] - d_\omega \chi & Q_{\mathrm{SG}}^0 \xi = \frac{1}{2}[\xi,\xi] + \frac{1}{2}\varphi \\
        \nonumber& Q_{\mathrm{SG}}^0 c = \frac{1}{2}( \iota_\xi\iota_\xi F_\omega -[c.c] ) + \iota_\xi \delta_\chi\omega  & Q_{\mathrm{SG}}^0 \chi = L_\xi^\omega \chi - [c,\chi] - \frac{1}{2}\iota_\varphi \psi,
    \end{align}
and
\begin{align}\label{eq: q fields}
        \nonumber &\mathbb{q}_e =  \frac{1}{2}\iota_\varphi\check{\omega} -\frac{1}{2}\iota_\varphi\check{c}\iota_\xi e -\frac{1}{4}\iota_\varphi(e \iota_\xi \check{c} ) \\
         \nonumber&e \mathbb{q}_\omega =   \frac{1}{2} \vf e^\Finv +\frac{i}{4\cdot 3!} \vf(\bar{\psi^0_\Finv}\ugam)\gamma^3 \psi +\frac{i}{4\cdot 3! }\bar{\psi}\gamma^3\vf\left(\ugam\alpha(\check{\omega}\psi )\right) - \frac{1}{8\cdot 3!} \vf \check{c} \bar{\chi}\gamma^3 \psi - \frac{1}{8\cdot 3!} \iota_\xi \check{c} \bar{\psi}\gamma^3 \vf\psi \\
         \nonumber&\qquad \hspace{3mm}   -\frac{i}{4\cdot 3!} \bar{\psi}\gamma^3 \vf\left( \ugam\alpha(\check{c}\iota_\xi e \psi) \right)  + \frac{1}{2\cdot 3!}\bar{\psi}\gamma^3\chi\kappa\left[ <e, \bar{\chi}\left( -\frac{i}{2}\ugam^2\psi_0^\Finv - [\check{\omega},\gamma]\psi - \frac{1}{2}\ugam\iota_\xi\check{c}\psi - \iota_\xi\ugam \check{c}\psi \right) > \right]\\
         \nonumber&\qquad \hspace{3mm} + \frac{1}{16\cdot 3!}\bar{\psi}\gamma^3\chi \bar{\chi}\igam\igam\left(  -\frac{i}{2}\ugam^2\psi_0^\Finv - [\check{\omega},\gamma]\psi - \frac{1}{2}\ugam\iota_\xi\check{c}\psi - \iota_\xi\ugam \check{c}\psi  \right)\\
        \nonumber&\mathbb{q}_\psi = \frac{i}{4}\vf(\ugam \psi^0_\Finv) -\frac{i}{4}\vf\left( \ugam\alpha( \check{\omega}\psi )  \right)   - \frac{i}{4}\vf\left( \ugam\alpha( \check{c}\iota_\xi e \psi )   \right) + \frac{1}{8}\vf\check{c} \chi - \frac{1}{8} \vf(\iota_\xi\check{c}\psi )\\
        \nonumber& \hspace{7mm} +\frac{i}{4}\chi\kappa\left( <\bar{e}, \bar{\chi}\ugam^2\psi^0_\Finv +i \bar{\chi}[\check{\omega}-\frac{i}{2}\iota_\xi \check{c} e + \iota_\xi e \check{c}]> \right)+\frac{1}{16}\chi\bar{\chi}\igam\igam(\ugam^2\psi^0_\Finv +i [\check{\omega}-\frac{i}{2}\iota_\xi \check{c} e + \iota_\xi e \check{c}])\\
        &\frac{e^2}{2} \mathbb{q}_c = -\frac{i}{8}\bar{\chi}\vf\psi_\Finv -\frac{i}{8\cdot 3!}\vf(\check{\omega }\bar{\chi}\gamma^3 \psi) -\frac{1}{2}\iota_\xi e \vf e^\Finv+ \frac{1}{4}\iota_\xi( e \vf e^\Finv)- \frac{i}{4\cdot 3!}\vf(\bar{\psi}^0_\Finv \ugam) \gamma^3 \iota_\xi e \psi  \\
        \nonumber&\qquad \hspace{3mm} + \frac{i}{4\cdot 3!}\vf\left(\alpha(\check{\omega}\bar{\psi}) \ugam\right) \gamma^3 \iota_\xi e \psi -\frac{i}{8}\iota_\xi(\bar{\psi} \vf\psi_\Finv) - \frac{1}{8\cdot 3!}\iota_\xi(\check{\omega}\bar{\psi}\gamma^3 \vf\psi)\\
       \nonumber &\qquad \hspace{3mm}+\frac{1}{4\cdot3!} \iota_\xi\left(\bar{\psi}\gamma^3\chi<e, \bar{\chi}([\check{\omega},\gamma]\psi + i\ugam^2\psi^0_\Finv) >  \right)-\frac{1}{2\cdot 3!}\iota_\xi e \bar{\psi}\gamma^3\chi\kappa\left(<e, \bar{\chi}([\check{\omega},\gamma]\psi + i\ugam^2\psi^0_\Finv) > \right)\\
        \nonumber&\qquad \hspace{3mm} + \frac{1}{32\cdot 3!}\iota_\xi e \bar{\psi}\gamma^3\chi \bar{\chi}\igam\igam([\check{\omega},\gamma]\psi + i\ugam^2\psi^0_\Finv) - \frac{1}{32\cdot3!} e \iota_\xi\left(\bar{\psi}\gamma^3\chi \bar{\chi}\igam\igam([\check{\omega},\gamma]\psi + i\ugam^2\psi^0_\Finv)\right),
    \end{align}
while $\mathbb{q}_\chi=\mathbb{q}_\xi=0.$

\subsection{The BV--BFV formalism on manifolds with boundary}\label{subsec: BV BFV formalism}
Assume now that the space--time manifold $M$ admits a boundaty $\Sigma:=\partial M$. In this case, it is possible that the variation of the BV action gives rise to a boundary term, i.e. the condition $\delta\mathcal{S}=\iota_Q \varpi_{BV}$ is not satisfied. In this case, if one still has that $[Q,Q]=0$, $(\mathcal{F}_M,\mathcal{S}_M,Q,\varpi_M)$ is called a broken BV-manifold.

    \begin{definition}
        Given a manifold $\Sigma$, a BFV manifold on $\Sigma$ is a tuple $(\mathcal{F}_\Sigma,\mathcal{S}_\Sigma,Q_\Sigma,\varpi_\Sigma)$, where $\mathcal{F}_\Sigma$ is a 0-symplectic graded manifold, $\mathcal{S}_\Sigma$ is a functional of degree 1 on $\mathcal{F}_\Sigma$ such that its Hamiltonian vector field $Q_\Sigma$ is cohomological. In particular, this implies that $\mathcal{S}_\Sigma$ satisfies the CME $(\mathcal{S}_\Sigma,\mathcal{S}_\Sigma)_\Sigma=0$.

        If the symplectic form $\varpi_\Sigma$ is exact, then $(\mathcal{F}_\Sigma,\mathcal{S}_\Sigma,Q_\Sigma,\varpi_\Sigma)$ is called an exact BFV manifold. 
    \end{definition}
The above definition is closely related to that of a BV manifold, where only the grading of the symplectic form and of the action is shifted by one. In the context of field theory, this is related to the boundary $\Sigma$ having codimension one, hence introducing a shift in degree of the fields, as it will be cleared by the following definition.

\begin{definition}
    An exact BV-BFV pair is given by the data $(\mathcal{F}_M,\mathcal{S}_M,Q_M,\varpi_M;\pi)$, where $(\mathcal{F}_M,\mathcal{S}_M,Q_M,\varpi_M)$ is a broken BV manifold and $\pi:\mathcal{F}_M\rightarrow \mathcal{F}_\Sigma$ is a surjective submersion, such that
        \begin{align}
           & \iota_{Q_M}\varpi_M=\delta\mathcal{S}_M + \pi^*\vartheta_\Sigma,
           & d\pi Q_M = Q_\Sigma
        \end{align}
    together with an exact BFV manifold $(\mathcal{F}_\Sigma=\pi(\mathcal{F}_M),\mathcal{S}_\Sigma,Q_\Sigma,\delta\vartheta_\Sigma)$  
\end{definition}

As expected, $\vartheta$ will be identified with the boundary term arising from the variation of the BV action, definining a variational 1--form on the space of boundary fields $\mathcal{F}_\Sigma$, while its variation $\varpi_\Sigma:= \delta\vartheta$ defines a closed 2-form on $\mathcal{F}_\Sigma$, which, in the case where it is non-degenerate\footnote{In general, $\delta\vartheta$ will have a non-trivial kernel, and the corresponding quotient needs to be performed, assuming the appropriate regularity of the kernel.} defines a 0--graded symplectic form on $\mathcal{F}_\Sigma$. 

\begin{theorem}\cite{CMR2012b}
    With the above assumptions, the cohomological vector field $Q$ induces a cohomological vector field $Q_\Sigma$ on $\mathcal{F}_\Sigma$, which is the Hamiltonian vector field for the functional $\mathcal{S}_\Sigma$.
\end{theorem}

We have then obtained a theoretical framework with which we can describe theories with non-trivial symmetries, generalizing the BRST procedure and compatible with the presence of a boundary. 

In the following section, we will study the boundary structure of $N=1, D=4$ supergravity, proving the existence of a BFV structure and relating it to the algebra of functions (i.e. the observables of the theory) on the reduced phase space.

\section{The reduced phase space of $\mathcal{N}=1$, $D=4$ SUGRA}\label{sec: RPS}

In the context of field theory, the phase space is related to the space of Cauchy data of the theory and can be endowed with the structure of symplectic manifold. It can happen (and will happen, in the case of supergravity) that such phase space only admits a closed degenerate two form, which is therefore only pre-symplectic. If the kernel of such two-form defines a regular distribution, it is possible to perform a reduction and obtain the reduced phase space.

\subsection{The Kijowski--Tulczijew construction}

We employ here a rather geometric construction by Kijowski and Tulczijew \cite{KT79} (see also \cite{CMR2012b, CMR2012} and \cite{Cattaneo:2023wxd} that allows us to obtain the reduced phase space (RPS) in a fashion compatible with the BFV formalism, in order to obtain the algebra of observables on reduced coisotropic spaces.

We assume $M$ to be a manifold with boundary $\partial M= \Sigma$. A classical Lagrangian field theory on $M$ is specified by a space of fields $F_M$ and an action functional $S_M$. The variation of the action gives the Euler--Lagrange form and a boundary term, such as \begin{equation*} \delta S_M = el_M + \int_{\partial M} a_M, 
    \end{equation*}
where $a_M$ is a local form depending on the fields and their jets at $\partial M$. Letting $EL_M$ be the zero-locus of $el_M$, we notice that, defining $\alpha_M:=\int_{\partial M} a_M$,
    \begin{itemize}
        \item $\alpha_M$ is $\delta$--exact on $EL_M$, as $\alpha_M = \delta S_M|_{EL_M}$
        \item defining $\varpi_M:= \delta \alpha_M$, we notice $\varpi_M|_{EL_M}=0 $ and that $\varpi_M$ is invariant under the following transformation of the Lagrangian $\mathrm{L}_M$
            \begin{align*}
                & \mathrm{L}_M\longmapsto \tilde{\mathrm{L}}_M := \mathrm{L}_M + \phi_M\\
                &\alpha_M \longmapsto\tilde{\alpha}_M:= \alpha_M + \delta \Phi_M,
            \end{align*}
        where $\Phi_M:=\int_{\partial M} \phi_M$ is a boundary term.\footnote{What this tells us is that $\alpha_M$ can be regarded as a one-form connection on a line--bundle over $F_M$, and the property above is just a form of gauge--invariance of the curvature two-form. Indeed, when one considers $e^{i S_M} $ as a section of the line--bundle over $F_M$, $\mathrm{L}_M  + \phi $ is just the action of the gauge transformation $e^{i \Phi_M }$. }    
    \end{itemize}
Now, by definition $\varpi_M$ is in general closed but not non-degenerate, i.e. it might present a non-trivial kernel $\ker{\varpi_M}=\{ \mathbb{X}\in\mathfrak{X}(F_M)\hspace{1mm}|\hspace{1mm} \iota_{\mathbb{X}}\varpi_M=0 \} $. Assuming a certain degree of regularity\footnote{In particular, assuming that $\ker{\varpi_M}$ defines a regular distribution (which is automatically involutive since $\varpi_M$ is closed), and assuming that it is integrable (Frobenius theorem does not hold on the nose on Frechet manifolds). } we obtain a quotient $\pi\colon F_M\rightarrow F_\Sigma:= F_M/\ker{\varpi_M}$ which is called the geometric phase space and which, in general, might be singular. It is therefore convenient to proceed by stages, first performing a reduction which always yields a smooth quotient, and then further reducing the latter. 

We start by noticing that vector fields on $F_M$ that preserve the fields at the boundary are by definition in the kernel of $\varpi_M$, since $\varpi_M$ only depends on the values of the fields (and their jets) and the points on $\Sigma$. This allows us to define the space of pre--boundary fields $\tilde{F}_\Sigma$ as the leaf space of the distribution of such vector fields, which in turn defines a surjective submersion
    \begin{equation*}
        \tilde{\pi}: F_M \rightarrow \tilde{F}_\Sigma
    \end{equation*}
defined by restricting the fields and their transversal jets to the boundary. Furthermore, $\tilde{\pi}$ uniquely induces the forms $\tilde{\alpha}_\Sigma$ and $\tilde{\varpi}_\Sigma$ on $\tilde{F}_\Sigma$. In particular, $\tilde{\varpi}_\Sigma= \delta \tilde{\alpha}_\Sigma$ is still a closed two--form and 
    \begin{equation*}
        \delta S_M = el_M + \tilde{\pi}^*\tilde{\alpha}_\Sigma. 
    \end{equation*}
It is also convenient to define the subspace $\tilde{L}_M:=\tilde{\pi}(EL_M)$ of pre--boundary fields  that can be extended to a solution of the E--L equations in the bulk. Such space is isotropic with respect to $\tilde{\varpi}_\Sigma$, since as before $\tilde{\varpi}_\Sigma|_{\tilde{L}_M}=0 $.\footnote{In most cases, $\tilde{L}_M$ is a submanifold. Here we assume that it is the case}

As it happens, $\tilde{\varpi}_\Sigma$ is still degenerate, leading to another quotient $p:\tilde{F}_\Sigma \rightarrow F_\Sigma$, which recovers the geometric phase space of the theory. If $F_\Sigma$ is smooth, then $\pi= p \circ \tilde{\pi}$ is a surjective submersion.  

It is then clear that $(F_\Sigma, \varpi_\Sigma)$ defines a symplectic manifold, but as such it is not yet the physical phase space of the theory, as the latter is seen as the set of "Cauchy data" of the theory. To better understand this statement, we notice that the Euler--Lagrange equations split into evolution equations that contain derivatives transversal to the boundary and constraints which only contain derivatives of the fields in the directions tangential to the boundary. The constraints need to be imposed on the space of pre--boundary fields, which usually enlarges the kernel of the pre--symplectic form. The corresponding reduction leads to the reduced phase space. 

To be precise, one works with the cylindrical manifold $M_\epsilon:= \Sigma \times [0,\epsilon]$, for some positive $\epsilon$. The boundary $\partial M$ is then given by $(\Sigma\times \{0\})\sqcup (\Sigma\times\{\epsilon\})$, while $\tilde{L}_{M_\epsilon} $ can be seen as a relation\footnote{A relation between two sets $A$ and $B$ is a subset of $A\times B$.} between $\tilde{F}_\Sigma\simeq \tilde{F}_{\Sigma\times \{0\}} $ and $\tilde{F}_{\Sigma\times \{\epsilon\} } $. The space of Cauchy data $\tilde{C}_\Sigma$ is then defined to be the subset of pre--boundary fields at $\Sigma$ that can be extended to solutions to the E--L equations in a cylindrical neighborhood of $\Sigma$, i.e.
    \begin{equation*}
        \tilde{C}_\Sigma:=\{ c\in \tilde{F}_\Sigma\simeq \tilde{F}_{\Sigma\times\{0\}} \hspace{1mm}|\hspace{1mm} \exists u\in \tilde{F}_{\Sigma\times \{\epsilon\} } \text{ s.t. } (c,u)\in\tilde{L}_{M_\epsilon}  \}.
    \end{equation*}
The induced 2--form $\tilde{\varpi}^C_\Sigma$ is generally degenerate on $\tilde{C}_\Sigma$, and the quotient $\underline{C}_\Sigma$ is finally the reduced phase space of the theory. Such space is often non--smooth, but in the context of field theory one is interested in the algebra of functions on it, i.e. the physical observables of the theory, or even better in a cohomological resolution. Under certain assumptions (first-class constraints), such a resolution can be obtained within the BFV formalism. 

\subsubsection{BFV data from the reduced phase space}\label{sec: BFV data from the reduced phase space}
To understand how the BFV formalism is helpful in the cohomological resolution of the reduced phase space, we start by studying the finite dimensional setting.
Here $(F,\omega)$ is a finite dimensional symplectic manifold, with functions $\phi_\alpha\in\mathcal{C^1}(F)$, $\alpha=1,\cdots,n$, whose differentials are independent and such that their zero locus  $C$ defines a coisotropic submanifold, i.e. such that there exist functions $f_{\alpha\beta}^\gamma\in\mathcal{C^1}(F)$ satisfying  
    \begin{equation*}
        \{\phi_\alpha,\phi_\beta\}=f_{\alpha\beta}^\gamma \phi_\gamma.
    \end{equation*}
If we let $\mathcal{I}$ be the ideal generated by the functions $\phi_\alpha$'s, then the functions on $C$ are simply given by $\mathcal{C^1}(F)/\mathcal{I}$, as functions differing by a combination of the $\phi_\alpha$'s will coincide on $C$. Now, letting $X_\alpha$'s be the Hamiltonian vector fields associated to the $\phi_\alpha$'s, they span the characteristic foliation, hence the coisotropic reduction $\underline{C}$ is given as $C/\{X_\alpha\} $.\footnote{Indeed the kernel of the restriciton  to $C$ of the symplectic form $\omega_C$ is spanned precisely by the $X_\alpha$'s, as $\iota_{X_\alpha}\omega=\delta\phi_\alpha=0$ on $C$.} Furthermore, if $\underline{C}$ is smooth, then 
    \begin{equation*}
        \mathcal{C^1}(\underline{C})\simeq\left(\mathcal{C^1}(F)/\mathcal{I}\right)^{(X_1\cdots X_n)},
    \end{equation*}
meaning that the functions on $\underline{C}$ are equivalent to the $X_\alpha$--invariant functions on $C$.

Now, we can introduce odd coordinates $c^\alpha$ (the ghosts) and $c^\Finv_\alpha$ (the ghost momenta) respectively of degree 1 and -1 (seen as coordinates on $T^*R^n[-1]$ , extending $F$ to a graded symplectic manifold $F \times T^*R^n[-1]$ with 0--symplectic form given by $\omega + dc^\alpha dc^\Finv_\alpha$. We can furthermore introduce a cohomological vector field $Q$ on $F \times T^*R^n[-1]$ given by $Q(f)=c^\alpha X_\alpha(f)$, $Q(c^\Finv_\alpha)=\phi_\alpha$ and $Q(c^\alpha)=0$, for all $f\in\mathcal{C^1}(F)$. As it turns out, this is the Hamiltonian vector field of $S=c^\alpha \phi_\alpha$, and its degree zero cohomology gives exactly the functions on $\underline{C}$, as $H^0_Q\simeq\{ f\in\mathcal{C^1}(F)\hspace{1mm}\vert\hspace{1mm} X_\alpha(f)=0\}/\mathcal{I}$. 
\begin{remark}
It can happen that $Q$ defined as above is not cohomological, i.e. it does not square to zero. However, a result from Stasheff \cite{Stasheff1997} proves that one can always deform the Hamiltonian $S$ in such a way that $Q$ is cohomological.    
\end{remark}
In practice one can always start by defining
    \begin{equation}\label{eq: BFV action}
        \mathcal{S}_\Sigma=\int_\Sigma c^\alpha\phi_\alpha + \frac{1}{2}f^{\gamma}_{\alpha\beta}c^\Finv_\gamma c^\alpha c ^\beta + R,
    \end{equation}
where $R$ is determined degree by degree by requiring $\{S,S\}=0$.
    \begin{theorem}[\cite{BV3, Stasheff1997,Scht08}]\label{thm: CMR12}
        Let $(F,\omega)$ be a symplectic manifold and $C$ a coisotropic submanifold, then there exist a BFV manifold $(\mathcal{F},Q,S,\varpi)$ whose body is given by $F$ and such that $$H^0_Q\simeq \mathcal{C^1}(\underline{C}).$$
    \end{theorem}

\subsection{Constraint analysis of $\mathcal{N}=1$, $D=4$ SUGRA}
We briefly recall the form of the action of $\mathcal{N}=1, D=4$ supergravity:
    \begin{equation}
        S_{\mathrm{SG}}=\int_M \frac{e^2}{2}F_\omega + \frac{1}{3!}e \bar{\psi}\gamma^3 d_\omega \psi,
    \end{equation}
while the equations of motion are
        \begin{align}
            &\label{eq:Einstein bdr}eF_\omega + \frac{1}{3!}\bar{\psi}\gamma^3 d_\omega \psi=0\\
            &\label{eq:Torsion bdr}d_\omega e - \frac{1}{2}\bar{\psi}\gamma\psi=0\\
            &\label{eq:Dirac bdr}\frac{1}{3}\left(e d_\omega \bar{\psi}\gamma^3 + \frac{1}{2}d_\omega e \bar{\psi}\gamma^3\right)=0.
        \end{align}
        
    The boundary term in the variation of the action depends only on the value of the fields at the boundary. In particular, we consider only those coframes defining non-degenerate metrics on the boundary.\footnote{Specifically, we require $g_{ij}^\partial:=(e_i,e_j)$ to be non denegenerate on $\Sigma$, i.e. either time--like or space--like.} The reason for this restriction will become clear in the following Section, but, as it is an open condition, it does not define a constraint on the space of bulk and (pre-)boundary fields $\tilde{F}_\Sigma:=\Omega_{n.d.}^1(\Sigma,\mathcal{V}) \times \mathcal{A}_\Sigma \times \Omega^1(\Sigma,\mathbb{S}_M)$.

    We have
    \begin{equation*}
        \delta S_{\mathrm{SG}}=\int_M \text{EL}_M - \int_\Sigma \frac{e^2}{2}\delta\omega +\frac{1}{3!}e\bar{\psi}\gamma^3 \delta\psi,
    \end{equation*}
    hence obtaining
        \begin{equation}\label{eq: pre-symplectic form bdr}
            \tilde{\varpi}^\partial_{\mathrm{SG}}=\int_\Sigma e\delta e\delta \omega + \frac{1}{3!}\bar{\psi}\gamma^3 \delta \psi \delta e + \frac{1}{3!}e \delta \bar{\psi}\gamma^3 \delta \psi.
        \end{equation}
    \subsubsection{First reduction and structural constraint}\label{subsubsec: first reduction and struct constraint}

    As it turns out, $\tilde{\varpi}^\partial_{\mathrm{SG}}$ is closed but not non--degenerate, its kernel is given by $\Ker{\tilde{\varpi}^\partial_{\mathrm{SG}}}=\{ \int_\Sigma \mathbb{X}_\omega \frac{\delta}{\delta \omega} \in \mathfrak{X}(\tilde{F}_\Sigma) \hspace{1mm}\vert \hspace{1mm} e \mathbb{X}_\omega=0, \hspace{1mm} \mathbb{X}_\omega\in\Omega^{(1,2)}_\partial \}$.\footnote{To be precise, any tangent vector field $\mathbb{X}=\int_\Sigma\mathbb{X}_e \frac{\delta }{\delta e} + \mathbb{X}_\omega \frac{\delta }{\delta \omega} + \mathbb{X}_\psi \frac{\delta }{\delta \psi}$ is in the kernel of $\tilde{\varpi}^\partial_{\mathrm{SG}}$ iff $e\mathbb{X}_e=0$, $e\gamma^3 \mathbb{X}_\psi=0$  and $e\mathbb{X}_\omega=0$, but the first two conditions imply $\mathbb{X}_e=0$ and $\mathbb{X}_\psi=0$.} In other words, any vector field acting on the boundary connections as $\omega\mapsto \omega + v$, such that $ev=0$, is in the kernel of the pre--symplectic form. We then define the geometric phase space as the quotient of the space of preboundary fields with respect to the action of the vector fields in $\Ker{\tilde{\varpi}^\partial_{\mathrm{SG}}}$ (i.e. consider the charachteristic foliation of this distribution), obtaining
        \begin{equation*}
            F_{\mathrm{SG}}^\partial:=\big(\Omega_{n.d.}^1(\Sigma,\mathcal{V}) \times \mathcal{A}_\Sigma \times \Omega^1(\Sigma,\mathbb{S}_M)\big)/\ker(\tilde{\varpi}^\partial_{\mathrm{SG}})        \end{equation*}
    where $\mathcal{A}_{\Sigma,\mathrm{red}}:=\mathcal{A}_{\Sigma}/\{ \omega\sim\omega+v,\hspace{1mm}ev=0 \}$.

    At this point, the constraints are simply obtained by restricting the equations of motion \eqref{eq:Einstein bdr}, \eqref{eq:Torsion bdr} and \eqref{eq:Dirac bdr} to the boundary, defining functions on $\tilde{F}_\Sigma$. However as pointed out in \cite{CCS2020}, the constraints are generally not invariant under the distribution defined by $\ker(\tilde{\varpi}^\partial_{\mathrm{SG}})$, \footnote{One can check that, for any $v\in\Omega^{(1,2)}_\partial$ such that $ev=0$, equations \eqref{eq:Einstein bdr} and \eqref{eq:Dirac bdr} are invariant under $\omega\mapsto \omega + v$, after applying \eqref{eq:Torsion bdr}, hence they will only depend on $[\omega]\in\mathcal{A}_{\Sigma,\mathrm{red}}$.
    In particular, one sees
    \begin{align*}
        \delta_v \left(e F_{\omega+v}+\frac{1}{3!}\bar{\psi}\gamma^3d_{\omega+v}\psi\right)=-e d_\omega v - \frac{1}{2}e\bar{\psi}\gamma\psi v=-d_\omega(ev)+e\left( d_\omega e -\frac{1}{2}\bar{\psi}\gamma\psi \right)v\approx 0,
    \end{align*}
    where the symbol $\approx$ is used to indicate an equality modulo equations of motion, i.e. an equality holding on--shell. We also see
    \begin{align*}
        \delta_v\left( e\gamma^3 d_\omega\psi -\frac{1}{2}d_\omega e \gamma^3 \psi \right)&= -e\gamma^3[v,\psi] - \frac{1}{2}[v,e]\gamma^3\psi\overset{\eqref{id: action of omega}}{=}-3ev \gamma\psi -\frac{1}{2}e[v,\gamma^3]_V\psi - \frac{1}{2}v [e,\gamma^3]\psi=0,
    \end{align*}
    having used $ev=0$ and $e[v,\gamma^3]_V=[ev,\gamma^3]-v[e,\gamma^3]$.
    } hence they cannot be naively extended to functions on $F_\Sigma$. In order to do so, we cleverly fix a representative of $[\omega]\in\mathcal{A}_{\Sigma,\mathrm{red}}$ (i.e. choose a $v$-section), such that it imposes the non-invariant part of the constraint. 

    \begin{remark}
        Notice that, in the bulk, the torsion equation $d_\omega e - \frac{1}{2}\bar{\psi}\gamma\psi=0$ is equivalent to $e \left( d_\omega e - \frac{1}{2}\bar{\psi}\gamma\psi \right)=0$, but it is not the case when $e$ is the vielbein restricted to the boundary, as $W_1^{\partial(2,1)}$ is not an isomorphism and in particular not injective. Indeed one finds that $e \left( d_\omega e - \frac{1}{2}\bar{\psi}\gamma\psi \right)$ has $6$ local components, and is invariant under the action of $\ker(\tilde{\varpi}^\partial_{\mathrm{SG}})$, while the remaining part of $d_\omega e - \frac{1}{2}\bar{\psi}\gamma\psi $ has another 6 local components, which will be used to fix $v\in\ker(W_e^{\partial (1,2)})$. 
    \end{remark}
    \begin{definition}
        In the following we fix a nowhere vanishing section $\epsilon_n\in\Gamma(\Sigma,\mathcal{V})$ and restrict the space of pre-boundary fields to the open subspace where $e$ and $\epsilon_n$ form a basis of $\mathcal V$ at    every point point.\footnote{If $\Sigma$ is space-like, we can make a global choice of $\epsilon_n$ time-like.    Otherwise, the choice is local on patch in $\Omega^{1,1}_{\partial,n.d.}$} Also note that $\epsilon_n$ is lindependent of the fields, so $\delta_\textrm{fields}\epsilon_n=0$.
        \end{definition}
    \begin{theorem}\label{thm: fix rep of omega bdry}
        Assume the metric $g^\partial$ induced by the boundary vielbein $e$ is non-degenerate. Then for any $\tilde{\omega}\in\Omega^{(1,2)}_\partial$ there exists a unique decomposition $\tilde{\omega}=\omega+ v$ such that 
        \begin{equation}\label{constr: classical constraint omega bdry}
            ev=0 \qquad \mathrm{and}\qquad \epsilon_n \left( d_\omega e - \frac{1}{2}\bar{\psi}\gamma\psi \right)= e \sigma,
        \end{equation}
        for some $\sigma\in\Omega_\partial^{(1,1)} $.
        Furthermore, the constraint $d_\omega e -\frac{1}{2}\bar{\psi}\gamma\psi=0$ splits as 
        \begin{equation}\label{eq: struct constr}
            d_\omega e -\frac{1}{2}\bar{\psi}\gamma\psi=0\quad \Leftrightarrow \quad \begin{cases}
                e \left(d_\omega e-\frac{1}{2}\bar{\psi}\gamma\psi\right) =0\\
                \epsilon_n \left( d_\omega e - \frac{1}{2}\bar{\psi}\gamma\psi \right)\in \mathrm{Im}(W_1^{\partial(1,1)} )
            \end{cases}
        \end{equation}
        We call $ \epsilon_n \left( d_\omega e - \frac{1}{2}\bar{\psi}\gamma\psi \right)= e \sigma$ structural constraint and $e \left(d_\omega e-\frac{1}{2}\bar{\psi}\gamma\psi\right) =0$ invariant constraint.
    \end{theorem}
    \begin{proof}
        We start by noticing that the splitting of the torsion constraints into structural and invariant part is a simple consequence of  \ref{lem: splitting (2,1)}. Now, from Lemma \ref{lem: splitting (2,2)}, we know that there exist $\sigma\in\Omega_\partial^{(1,1)} $ and $v\in\Ker{W_1^{\partial(1,2)}}$ such that 
        \begin{equation*}
             \epsilon_n \left( d_{\tilde{\omega}} e - \frac{1}{2}\bar{\psi}\gamma\psi \right)= e \sigma + \epsilon_n[v,e].
        \end{equation*}
        Then one fixes $\omega:=\tilde{\omega}-v$, obtaining the desired result. The uniqueness is simply shown by assuming there exist different splittings $\tilde{\omega}=\omega_1+v_1=\omega_2+v_2$ as above, implying $[e,v_1-v_2]\in\mathrm{Im}(W_1^{\partial(1,1)} )$, which, by Lemma \ref{lem: splitting (2,2)} and \ref{lem: splitting (2,1)} shows $v_1=v_2$.
    \end{proof}

    \subsubsection{Constraints and the first-class condition}

    Now we can finally define the constraints on $F_{\mathrm{SG}}^\partial$ simply as the restriction of \eqref{eq:Dirac bdr} and \eqref{eq:Einstein bdr} to the boundary plus the invariant torsion constraint. In order to readily have them as functionals over $F_{\mathrm{SG}}^\partial$, we make use of Lagrange multipliers\footnote{In view of the BFV description of the theory in the following chapter, we shift the degree of the Lagrange multiplier by one, as they will later represent the ghosts of the theory.} $\mu\in\Omega^{(1,1)}_\partial[1]$, $c\in\Omega_\partial^{(0,2)}[1]$ and $\chi\in\Gamma(\mathbb{S}_M|_\Sigma)[1]$, obtaining
        \begin{align*}
            J_\mu&=\int_\Sigma \mu\left( eF_\omega + \frac{1}{3!}\bar{\psi}\gamma^3 d_\omega \psi \right)\\
            L_c&=\int_\Sigma c\left( e d_\omega e - \frac{1}{2}e\bar{\psi}\gamma\psi \right)\\
            M_\chi&=\int_\Sigma\frac{1}{3}\bar{\chi}\left(e \gamma^3 d_\omega \psi -\frac{1}{2}d_\omega e  \gamma^3 \psi\right).
        \end{align*}

    \begin{remark}
        As we will see later, the Hamiltonian vector fields are related to the gauge symmetries of the theory (i.e. they define the infinitesimal gauge transformations). In particular, $L_c$ generates the internal Lorentz symmetry, $M_\chi$ is the generator of the supersymmetry and $J_\mu$ generates the diffeomorphism symmetry. The last statement can be refined once one notices that, since $\{e_1,e_2,e_3,\epsilon_n\}$ defines a local basis of $\mathcal{V}$, it is possible to split $\mu=\lambda \epsilon_n + \iota_\xi e$, with $\lambda\in \mathcal{C}^\infty(\Sigma)[1]$ and $\xi\in \mathfrak{X}[1](\Sigma)$. Then $\xi $ and $\lambda $ can be interpreted respectively as the gauge parameters associated to the tangential and transversal diffeomorphisms with respect to $\Sigma$. The constraint $J_\mu$ splits into 
            \begin{align*}
                P_\xi&=\int_\Sigma \frac{1}{2}\iota_\xi(e^2)F_\omega + \frac{1}{3!}\iota_\xi e \bar{\psi}\gamma^3 d_\omega \psi  \quad \mathrm{and}\quad
                H_\lambda=\int_\Sigma \lambda \epsilon_n \left( eF_\omega + \frac{1}{3!}\bar{\psi}\gamma^3 d_\omega \psi \right).
            \end{align*}
    \end{remark}

    \begin{remark}\label{rem: redefinition of constraints}
        Notice that $L_c$ can be rewritten in a nicer form, in particular one finds\footnote{Here we used the fact that
        \begin{equation*}
            -\frac{1}{2}ce\bar{\psi}\gamma\psi=\frac{i}{2\cdot 3!} e \left(  \bar{\psi}\gamma^3 [c,\psi] + [c,\psi]\gamma^3\psi \right)=\frac{i}{3!}\bar{\psi}\gamma^3 [c,\psi].
        \end{equation*}}  
            \begin{align*}
                &L_c=\int_\Sigma c ed_\omega e + \frac{1}{3!}e\bar{\psi}\gamma^3 [c,\psi].
            \end{align*}
        To further simplify the computations, we introduce a reference connection $\omega_0$ and use Cartan magic formula to define, for any field $\phi$, the Lie derivative along $\xi$ with respect to $\omega_0$ as
            \begin{equation*}
                \mathrm{L}_\xi^{\omega_0}\phi:=[\iota_\xi,d_{\omega_0}]\phi=\iota_\xi d_{\omega_0}\phi - d_{\omega_0}\iota_\xi\phi.
            \end{equation*}
        Then, to make the dependence of $P_\xi$ on the newly defined Lie derivative apparent and to make the Hamiltonian vector field of $P_\xi$ well defined, we take into consideration the following redefinition:\footnote{Redefining the constraint set as a $\mathcal{C}^\infty(F_{\mathrm{SG}}^\Sigma)$-linear combination of the original constraints does not change the zero-locus of such constraints, which ultimately is what we will be interested in.} $P_\xi \rightarrow P_\xi + L_{\iota_\xi (\omega-\omega_0)} + M_{\iota_\xi\psi}$, yielding
            \begin{equation*}
                P_\xi =\int_\Sigma \frac{1}{2}\iota_\xi e^2 F_\omega + \iota_\xi(\omega-\omega_0) e d_\omega e - \frac{1}{3!}e \bar{\psi}\gamma^3 \mathrm{L}_\xi^{\omega_0}\psi.
            \end{equation*}
        Lastly, it will be convenient to rewrite $M_\chi$ in the following form, obtained by integrating by parts
            \begin{equation*}
                M_\chi=\int_\Sigma \frac{1}{3!}e ( d_\omega \bar{\chi}\gamma^3 \psi + \bar{\chi}\gamma^3 d_\omega\psi )
            \end{equation*}

        The full list of constraints then reads
            \begin{align*}
                &L_c=\int_\Sigma c ed_\omega e + \frac{1}{3!}e\bar{\psi}\gamma^3 [c,\psi],\\
                &P_\xi =\int_\Sigma \frac{1}{2}\iota_\xi e^2 F_\omega + \iota_\xi(\omega-\omega_0) e d_\omega e - \frac{1}{3!}e \bar{\psi}\gamma^3 \mathrm{L}_\xi^{\omega_0}\psi,\\
                &M_\chi=\int_\Sigma \frac{1}{3!}e ( d_\omega \bar{\chi}\gamma^3 \psi + \bar{\chi}\gamma^3 d_\omega\psi ).
            \end{align*}

        \end{remark}
    \begin{theorem}\label{thm: PB of constraints}
        Let $g^\partial$ be non-degenerate on $\Sigma$. Then the functions $P_\xi,H_\lambda,L_c$ and $M_\chi$ form a first-class set of constraints,  defining a coisotropic submanifold as their zero-locus. In particular
        \begin{align}\label{eq: Poisson brackets}
            \nonumber\{L_c,L_c\}&=-\frac{1}{2}L_{[c,c]} &   \{L_c,M_\chi\}&=M_{[c,\chi]}\\
            \nonumber\{L_c,P_\xi\}&=L_{\mathrm{L}_\xi^{\omega_0}c} & \{P_\xi,M_\chi\}&=-M_{\mathrm{L}_\xi^{\omega_0}\chi}\\
            \nonumber\{P_\xi,P_\xi\}&= \frac{1}{2}P_{[\xi.\xi]}- \frac{1}{2} L_{\iota_\xi\iota_\xi F_{\omega_0}} & \{H_\lambda,H_\lambda\}&=0\\
            \nonumber\{L_c,  H_{\lambda}\} & = -P_{\zeta}+L_{\iota_\zeta(\omega-\omega_0)} + M_{\iota_\zeta\psi}-H_{\zeta^n} %- \frac{i}{3!}L_{\alpha(\iota_X \psi,\psi)}
            &\{M_\chi,H_\lambda\}&=0 \\
            \nonumber\{P_{\xi},H_{\lambda}\} &=  P_\vartheta -L_{\iota_\vartheta(\omega-\omega_0)} -M_{\iota_\vartheta \psi}  + H_{\vartheta^{(n)}} \\%+ \frac{i}{3!}L_{\alpha(\iota_Y,\psi,\psi)}
            \{M_\chi,M_\chi\}&=\frac{1}{2}P_\varphi - \frac{1}{2}L_{\iota_\varphi (\omega-\omega_0)} - \frac{1}{2}M_{\iota_\varphi\psi}+ \frac{1}{2}H_{\varphi^{(n)}} 
        \end{align}
        where, setting $\{x^i\}$ local coordinates on $\Sigma$, one has $\zeta=e^i_a[c,\lambda \epsilon_n]^a\partial_i$, $\vartheta=e^i_a(\mathrm{L}_\xi^{\omega_0}(\lambda \epsilon_n))^a\partial_i$, $\varphi=e^i_a\bar{\chi}\gamma^a \chi \partial_i$, while $\zeta^{(n)}=[c,\lambda \epsilon_n]^n$, $\vartheta^{(n)}=(\mathrm{L}_\xi^{\omega_0}(\lambda \epsilon_n))^n$, $\varphi^{(n)}=\bar{\chi}\gamma^n \chi$.
        %\begin{align*}
        %    \frac{1}{3!}\bar{\chi}\gamma^3 d_\omega\chi&=e\alpha^\partial(\chi,d_\omega\chi)+\epsilon_n \beta^\partial(\chi,d_\omega\chi)\\
        %    &\epsilon_n\mathbb{M}_\omega=e\alpha^\partial(\epsilon_n\mathbb{M}_\omega)+\epsilon_n \beta^\partial(\epsilon_n\mathbb{M}_\omega),
        %\end{align*}
        %where $\mathbb{M}_\omega$ is the component of the Hamiltonian vector field $M_\chi$ along $\omega$. 
        %Furthermore, we have that
            %\begin{align*}
             %  L_{\alpha(\chi,\omega,\chi)}=\int_\Sigma \bar{\chi}\gamma^3 d_\omega \chi d_\omega e,% L_{\alpha(\iota_R\psi,\psi)}=\int_\Sigma \iota_R\bar{\psi}\gamma^3\psi d_\omega e, \qquad &, \qquad L_{\alpha(\chi,\omega,\psi)}=\int_\Sigma \epsilon_n \mathbb{M}_\omega\left( d_\omega e -\frac{i}{2}\bar{\psi}\gamma\psi \right)
            %\end{align*}
    \end{theorem}

    \begin{proof}
        We begin by computing  the Hamiltonian vector fields of the constraints, defined by $\iota_{\mathbb{X}_f}\varpi_{\mathrm{SG}}^\partial=\delta f$.
        
        We see
        \begin{equation}
            \iota_{\mathbb{X}}\varpi_{\mathrm{SG}}^\partial= \int_\Sigma e \mathbb{X}_e \delta \omega +\left( e \mathbb{X}_\omega +\frac{1}{3!}\mathbb{X}_{\bar{\psi}}\gamma^3 \psi \right)\delta e + \frac{1}{3}\left( \frac{1}{2}\mathbb{X}_e \bar{\psi}\gamma^3 + e\mathbb{X}_{\bar{\psi}} \gamma^3 \right)\delta \psi .
        \end{equation}
        \begin{align*}
            \delta L_c&=\int_\Sigma [c,e]e \delta \omega + e d_\omega c  \delta e + \frac{1}{3!}\bar{\psi}\gamma^3 [c,\psi]\delta e + \frac{1}{3!}e \left( 
            \delta\bar{\psi}\gamma^3 [c,\psi] + \bar{\psi}\gamma^3[c,\delta\psi] \right)\\
            &=\int_\Sigma [c,e]e \delta \omega + \left(  e d_\omega c  + \frac{1}{3!}\bar{\psi}\gamma^3 [c,\psi]\right) \delta e + \frac{1}{3!}e \left( \delta\bar{\psi}\gamma^3 [c,\psi] +[c,\bar{\psi}] \gamma^3 \delta\psi - \bar{\psi}[c,\gamma^3]\delta\psi  \right)\\
            &=\int_\Sigma [c,e]e \delta \omega + \left( e d_\omega c  + \frac{1}{3!}\bar{\psi}\gamma^3 [c,\psi]\right) \delta e +  \frac{1}{3}e \left( [c,\bar{\psi}] \gamma^3 \delta\psi - \frac{1}{2}\bar{\psi}[c,\gamma^3]_V\delta\psi  \right)\\
            &=\int_\Sigma [c,e]e \delta \omega + \left( e d_\omega c  + \frac{1}{3!}\bar{\psi}\gamma^3 [c,\psi]\right) \delta e +  \frac{1}{3} \left(  \frac{1}{2}[c,e]\bar{\psi}\gamma^3 + e[c,\bar{\psi}] \gamma^3     \right)\delta\psi
        \end{align*}
            In the second step we used the fact that an element in $\wedge^2 V$ acts on $\gamma$ both in the spinor representation and in the vector one, and they cancel each other out. Explicitly, using the fact that $\wedge^2 V\simeq \mathfrak{so(3,1)}\simeq \mathfrak{spin}(3,1)$, a basis for it is given by $\{-\frac{1}{4}\gamma_{[a}\gamma_{b]}\}$ or equivalently by $\{v_a\wedge v_b\}$. Now, since $\gamma=\gamma^a v_a$ has values in $\mathcal{C}(V)\otimes V$, when acting with $c$ we have $[c,\gamma]=[c,\gamma]_S + [c,\gamma]_V=0$, in fact\footnote{ We indicate the graded commutator by double square brackets $\llbracket A, B\rrbracket=AB -{(-1)}^{|A||B|}BA$.  }
                \begin{align*}
                    [c,\gamma]_S:=\frac{1}{4} \llbracket [\gamma,[\gamma,c]], \gamma \rrbracket  =-\frac{1}{4}(c^{ab}\gamma_a\gamma_b \gamma + \gamma c^{ab}\gamma_a\gamma_b) =  -c^{ab} \gamma_a v_b=-[c,\gamma]_V.
                \end{align*}
            Now, since $\bar{\chi}\gamma^N\psi$ has no spinor indices (for any two arbitrary spinors $\chi$ and $\psi$), $[c,\bar{\chi}\gamma^N\psi]=(-1)^{|\chi|}\bar{\chi}[c,\gamma^N]_V\psi$, but at the same time the Leibniz rule for $[c,\cdot]$ holds, we find
                \begin{equation}\label{eq: Leibniz spinors}
                    [c,\bar{\chi}\gamma^N\psi]=(-1)^{|\chi|}\bar{\chi}[c,\gamma^N]_V\psi=[c,\chi]\gamma^N \psi - (-1)^{|\chi|}\bar{\chi}\gamma^N [c,\psi].
                \end{equation}
        
        \begin{align*}
            \delta P_\xi&=\int_\Sigma -\mathrm{L}_\xi^{\omega_0} e \delta \omega - \left( \iota_\xi F_{\omega_0} + \mathrm{L}_\xi^{\omega_0}(\omega-\omega_0) +\frac{1}{3!}\bar{\psi}\gamma^3 \mathrm{L}_\xi^{\omega_0}\psi \right)\delta e -\frac{1}{3!}e\left( \delta\bar{\psi}\gamma^3 \mathrm{L}_\xi^{\omega_0}\psi + \bar{\psi}\gamma^3 \mathrm{L}_\xi^{\omega_0}\delta\psi \right)\\
            &=\int_\Sigma -\mathrm{L}_\xi^{\omega_0} e \delta \omega - \left( \iota_\xi F_{\omega_0} + \mathrm{L}_\xi^{\omega_0}(\omega-\omega_0) +\frac{1}{3!}\bar{\psi}\gamma^3 \mathrm{L}_\xi^{\omega_0}\psi \right)\delta e -\frac{1}{3}\left( e \mathrm{L}_\xi^{\omega_0}\bar{\psi}\gamma^3 + \frac{1}{2}\mathrm{L}_\xi^{\omega_0}  e\bar{\psi}\gamma^3 \right)\delta\psi;
        \end{align*}
        \begin{align*}
            \delta H_\lambda &=\int_\Sigma d_\omega(\lambda \epsilon_n e ) \delta \omega + \lambda \epsilon_n  F_\omega \delta e +\frac{1}{3!}\lambda \epsilon_n (\delta \bar{\psi} \gamma^3 d_\omega \psi  - \bar{\psi} \gamma^3[\delta\omega,\psi] + \bar{\psi} \gamma^3 d_\omega\delta\psi) \\
            &=\int_\Sigma \left(d_\omega(\lambda \epsilon_n e ) -\frac{1}{2}\lambda \epsilon_n \bar{\psi}\gamma\psi \right)\delta \omega + \lambda \epsilon_n  F_\omega \delta e + \frac{1}{3}\left( \lambda \epsilon_n d_\omega \bar{\psi}\gamma^3 + \frac{1}{2}d_\omega(\lambda \epsilon_n) \bar{\psi}\gamma^3 \right)\delta\psi;
        \end{align*}
        \begin{align*}
            \delta M_\chi &=\int_\Sigma \frac{1}{3!}\delta e (d_\omega \bar{\chi}\gamma^3 \psi + \bar{\chi}\gamma^3 d_\omega \psi) + \frac{1}{3!}e([\delta \omega,\bar{\chi}]\gamma^3 \psi - \bar{\chi}\gamma^3[\delta\omega,\psi])+\frac{1}{3}\delta \bar{\psi}\left( e\gamma^3 d_\omega\chi -\frac{1}{2}d_\omega e \gamma^3 \chi \right) \\
            &= \int_\Sigma \frac{1}{3!} (d_\omega \bar{\chi}\gamma^3 \psi + \bar{\chi}\gamma^3 d_\omega \psi) \delta e - i e \bar{\chi}\gamma\psi \delta \omega +\frac{1}{3}\delta \bar{\psi}\left( e\gamma^3 d_\omega\chi -\frac{1}{2}d_\omega e \gamma^3 \chi \right)
        \end{align*}
        This allows to extract the following vector fields\footnote{Strictly speaking, in the computation of $\mathbb{M}_\psi$ one would have a term proportional to $\mathbb{M}_e\bar{\psi}\gamma^3\delta\psi=- \bar{\psi}\gamma\chi\bar{\psi}\gamma^3\delta\psi$, but this one vanishes because of \ref{lem: Fierz}.}
            \begin{align}\label{eq: Ham v.f.}
                \nonumber&\mathbb{L}_e=[c,e] &\mathbb{L}_\omega=d_\omega c + \mathbb{V}_L 
                &&\mathbb{L}_\psi=[c,\psi]\\
                \nonumber&\mathbb{P}_e= -\mathrm{L}_\xi^{\omega_0}e  &\mathbb{P}_\omega=-\iota_\xi F_{\omega_0} - \mathrm{L}_\xi^{\omega_0} (\omega-\omega_0) + \mathbb{V}_P &&\mathbb{P}_\psi=  -\mathrm{L}_\xi^{\omega_0} \psi\\
                \nonumber&\mathbb{H}_e= d_\omega(\lambda \epsilon_n) + \lambda \sigma & e\mathbb{H}_\omega=\lambda \epsilon_n F_\omega - \frac{1}{3!}\mathbb{H}_{\bar{\psi}}\gamma^3\psi && e\gamma^3 \mathbb{H}_\psi= \lambda \epsilon_n \gamma^3 d_\omega \psi + \frac{1}{2}\lambda \sigma \gamma^3 \psi\\
                &\mathbb{M}_e= - \bar{\chi}\gamma\psi &e\mathbb{M}_\omega=\frac{1}{3!}(d_\omega\bar{\chi}\gamma^3\psi + \bar{\chi}\gamma^3 d_\omega \psi) - \frac{1}{3!}\bar{\psi}\gamma^3 \mathbb{M}_\psi &&e\gamma^3\mathbb{M}_\psi=e\gamma^3 d_\omega \chi - \frac{1}{2}d_\omega e \gamma^3 \chi
            \end{align}
        
            Notice that the components along $\omega$ of the Hamiltonian vector fields are defined up to a term $\mathbb{V}\in\Ker{W_e^{\partial(1,2)}}$, which is fixed by requiring that $\mathbb{X}_\omega$ preserves the structural constraint \eqref{eq: struct constr}. In most of the future calculations we will only need $e\mathbb{X}_\omega$, but one can in any case prove \cite{CCS2020} that elements in $\Omega^{(2,3)}_\partial$ are in the image of $W_e^{\partial(1,2)}$, hence such $\mathbb{X}_\omega$ always exists.

            The discussion is analogous in the computation of $\mathbb{H_\psi}$ and $\mathbb{M}_\psi$, indeed by considering $e\gamma^3 \mathbb{X}_\psi$, one sees
                \begin{equation*}
                    e\gamma^3\mathbb{X}_\psi = e^a \gamma^{bcd}\mathbb{X}_\psi \epsilon_{abcd} \mathrm{Vol}_V \overset{\eqref{id: gamma5gammac}}{=} i \gamma^5 [e,\gamma]\mathbb{X}_\psi.
                \end{equation*}
            Now, since $\Omega_\partial^{(1,0)}(\Pi \mathbb{S}_M) $ and $\Omega_\partial^{(2,0)}(\Pi \mathbb{S}_M)$ have the same local dimension, showing $e\gamma^3$ is injective proves that it is also an isomorphism, but that amounts to show that 
                \begin{equation*}
                    [e,\gamma]\mathbb{X}_\psi=0 \quad \Rightarrow \quad \mathbb{X}_\psi=0,
                \end{equation*}
            or in other words that $\gamma_{[i}\mathbb{X}_{\psi,j]}=0$ implies $\mathbb{X}_{\psi,j}=0$, which is immediately verified by solving the system of three equations.\footnote{We defined $\gamma_i:=\gamma_a e^a_i$, which is still an invertible matrix.}\footnote{This computation also explicitly shows that the kernel of the pre-symplectic form \eqref{eq: pre-symplectic form bdr} does not contain any term of the kind $\mathbb{X}_\psi\frac{\delta}{\delta\psi}$.} One can then define $\mathbb{M}^e_\psi$ such that 
                \begin{equation}\label{eq: M_e psi}
                    e\gamma^3 \mathbb{M}_\psi^e :=-\frac{1}{2}d_\omega e \gamma^3 \chi \overset{(\ref{lem: Fierz})}{=}-\frac{1}{2}\left(d_\omega e - \frac{1}{2}\bar{\psi}\gamma\psi\right)\gamma^3\chi.
                \end{equation}

    The rest of the proof, which amounts to showing that the constraints form a first-class set, is found in Appendix \ref{app: PB constraints}.
    \end{proof}
    \subsection{Towards a BFV description}\label{subsec: BFV attempt}
   Having proved that the constraints form a first-class set, under the assumption of a regular boundary, Theorem \ref{thm: CMR12} tells us that there must exist a BFV structure on ${F}_{\mathrm{SG}}^\Sigma$. Indeed, one can consider the bundle
            \begin{equation*}
                \mathcal{F}^\Sigma_{\mathrm{SG}} \rightarrow \Omega^{(1,1)}_{\partial,n.d.}\times \Omega^1_\partial(\mathbb{S}_{M})
            \end{equation*}
        with local trivialization on an open $\mathcal{U}_\Sigma\subset \Omega^{(1,1)}_{\partial,n.d.}\times \Omega^1_\partial(\mathbb{S}_{M})$ 
            \begin{equation*}
                \mathcal{U}_\Sigma\times \mathcal{A}_{\Sigma,\mathrm{red}} \times T^*(\underset{(c, k^\dagger)}{\Omega_\partial^{(0,2)}[1]} \oplus \underset{(\xi,\zeta^\dagger)}{\mathfrak{X}[1](\Sigma)}   \oplus \underset{(\lambda,\lambda^\dagger)}{\mathcal{C}^{\infty}[1](\Sigma)} \oplus \underset{(\chi,\theta_\dagger)}{\Gamma[1](\Pi\mathbb{S}_M})  )
            \end{equation*}
        where   $e\in  \Omega^{(1,1)}_{\partial,n.d.}$, $\omega \in  \mathcal{A}_{\Sigma,\mathrm{red}}$ and $\Omega^1_\partial(S_{\Sigma,m})$, while the antighosts of $c,\xi$ and $\chi$ are denoted respectively by $k^\Finv\in\Omega^{(3,2)}_\partial[-1] $, $\zeta^\dagger\in\Omega_\partial^{(1,0)}[-1]\otimes\Omega_\partial^{(3,4)} $ and $\theta_\Finv\in\Omega^{(3,4)}[-1](\Pi\mathbb{S}_M)$. In particular, generalizing \cite{CCS2020}, one defines $\mathcal{A}_{\Sigma,\mathrm{red}}$ as the space of connections modeled over $\Omega_\partial^{(1,2)} $ satisfying a modified version of structural constraint \eqref{constr: classical constraint omega bdry}, called the \textbf{BFV structural constraint}
            \begin{equation*}
                \epsilon_n\left( d_\omega e -\frac{1}{2}\bar{\psi} \gamma\psi \right) + \left( \mathrm{L}_\xi^{\omega_0}(\epsilon_n)^i - [c,\epsilon_n]^i\chi\right)k^\dag_i + \cdots= e\sigma,
            \end{equation*}
        where $\cdots$ regroups possible extra terms to ensure the constraint is invariant under the action of the cohomological vector field $Q^\partial_{\mathrm{SG}}$.
            
        The canonical symplectic form is defined as
            \begin{equation}
                \varpi_{\mathrm{SG}}^\Sigma=\int_\Sigma e \delta e \delta\omega + \frac{1}{3!}\bar{\psi}\gamma^3\delta\psi \delta e + \frac{1}{3}e\delta\bar{\psi}\gamma^3 \delta\psi + \delta c \delta k^\Finv + \iota_{\delta\xi}\delta\zeta^\Finv + \delta\lambda\delta\lambda^\Finv + i\delta\bar{\chi}\delta\theta_\Finv.
            \end{equation}

        Now one can, as a first attempt, define the tentative BFV action from the structure of the constraints as
            \begin{equation}\label{eq: BFV bdry action}
            \begin{split}
                \mathcal{S}_{\mathrm{SG}}^\Sigma:=&L_c+ M_\chi + P_\xi + H_\lambda \\
                &+  \int_\Sigma \left(\frac{1}{2}[c,c] + \frac{1}{2}\iota_\xi\iota_\xi F_{\omega_0}  - \mathrm{L}_\xi^{\omega_0}c \right)k^\Finv +i(\mathrm{L}_\xi^{\omega_0}\bar{\chi} - [c,\bar{\chi}])\theta_\Finv \\
                &\qquad -\frac{1}{2}\iota_{[\xi,\xi]}\zeta^\Finv + \left([c,\lambda \epsilon_n]^n  - \mathrm{L}_\xi^{\omega_0}(\lambda \epsilon_n)^n + \frac{1}{2}\bar{\chi}\gamma^n\chi \right)\lambda^\Finv\\
                &\qquad +\left( [c,\lambda \epsilon_n]^j - \mathrm{L}_\xi^{\omega_0}(\lambda \epsilon_n)^j + \frac{1}{2}\bar{\chi}\gamma^j\chi \right)(\zeta^\dag_j - (\omega-\omega_0)_j k^\Finv -i\bar{\psi}_j \theta_\Finv )\\
            \end{split}
            \end{equation}
        In order to obtain a BFV structure, one needs to show that $\{\mathcal{S}_{\mathrm{SG}}^\Sigma,\mathcal{S}_{\mathrm{SG}}^\Sigma\}_{BFV}=0$. However, doing so is a  computationally hard task, which, when carried out fully\footnote{We spare the reader of the cumbersome details of this computation.} does not yield the desired result, hence $\mathcal{S}_{\mathrm{SG}}^\Sigma$ needs to be complemented with terms of higher order in the antighosts.  
       
        \subsubsection{Introducing new variables}
        We know from diagram $\eqref{diag: prop e bdry}$ that $W_e^{\partial,(2,1)}$ is surjective, therefore it is possible to rewrite $k^\Finv=e\tilde{k}$, for (more than) a $\tilde{k}\in\Omega_\partial^{(2,1)}$.\footnote{This is due to the fact that $W_e^{\partial,(2,1)}$ is not injective, unlike its counterpart in the bulk. Such non-uniqueness in the definition of a preimage of $W_e^{\partial,(2,1)}$ is one of the obstructions to finding an extended BV/BFV formalism for gravity and supergravity.} Considering the field redefiniton with $\tilde{k}$ as the new field, we immediately notice that the new symplectic form contains a term $ e\delta\tilde{k}\delta c$ leading to a degeneracy, in particular
            \begin{equation*}
                \Ker{\varpi_{\mathrm{SG}}^\Sigma}=\left\{\mathbb{X}_{\tilde{k}}\frac{\delta}{\delta\tilde{k}},\hspace{1mm}\mathbb{X}_{\tilde{k}}\in\Omega^{(2,1)}_\partial\hspace{1mm}\bigg\vert\hspace{1mm}e\mathbb{X}_{\tilde{k}}=0\right\}.
            \end{equation*}
        To obtain a well defined symplectic form one needs to consider $[\tilde{k}]\in\Omega^{(2,1)}_\partial/\Ker{W_e^{\partial,(2,1)}}$. However, we can cleverly fix a representative thanks to the following.  
    \begin{theorem}\label{thm: fix rep of kdag}
        For all $\tilde{k}\in\Omega^{(2,1)}_\partial$ there exist a unique decomposition
            \begin{equation*}
                \tilde{k}=\check{k}+ r
            \end{equation*}
        with $\check{k},r\in\Omega^{(2,1)}_\partial$ such that 
            \begin{equation}\label{eq: constr k bdry}
                er=0, \qquad \epsilon_n \check{k}=e\check{a},
            \end{equation}
        for some $\check{a}\in\Omega_\partial^{(1,1)}$. Furthermore, the field $\check{k}$ in the decomposition above only depends on the equivalence class $[\tilde{k}]\in\Omega_{\partial,\text{red}}^{(2,1)} $.
    \end{theorem}
    \begin{remark}
        Notice that, as an immediate consequence of the first statement we obtain $[\tilde{k}]=[\check{k}]$.
    \end{remark}
    \begin{proof}
        The decomposition is a direct consequence of Lemma \ref{lem:split cdag}. 
        Now consider $\tilde{k}_1,\tilde{k}_2\in[\tilde{k}]$, then by definition $\tilde{k}_1-\tilde{k}_2=r'\in \Ker{W_e^{\partial(1,2)}}  $. By \ref{lem:split cdag}, $\tilde{k}_1=\check{k_1}+r_1  $ and $\tilde{k}_2=\check{k_2}+r_2$ such that $\epsilon_n\check{k}_1,\epsilon_n\check{k}_2\in \text{Im}(W_e^{\partial,(1,1)})$ and $e r_1= er_2=0$, hence
            \begin{equation*}
                \begin{cases}
                    \check{k}_1-\check{k}_2=r_2-r_1-r' \in\Ker{W_e^{\partial(1,2)}}\\
                    \epsilon_n(\check{k}_1-\check{k}_2)\in \text{Im}(W_e^{\partial,(1,1)}),
                \end{cases}    
            \end{equation*}
        which implies, by Lemma \ref{lem: splitting (2,1)}, that $\check{k}_1=\check{k}_2$.    
    \end{proof}
        At this point, we recall from Remark \ref{rem: redefinition of constraints} that we altered the original constraint set (to an equivalent one) by redefining $P_\xi$. In particular, in order to obtain nicer Hamiltonian vector fields, we had
        \begin{equation*}
            P_\xi\mapsto P_\xi - L_{\iota_\xi(\omega-\omega_0)} - M_{\iota_\xi\psi}.
        \end{equation*}
        Following \cite{CCS2020} it is convenient to change variables in order to get rid of the redundancies. We introduce 
            \begin{equation}
               c'=c+\iota_\xi(\omega-\omega_0) \qquad \chi'=\chi+\iota_\xi \psi \qquad \zeta_\bullet^{\Finv'}=\zeta_\bullet^\Finv - (\omega-\omega_0)_\bullet k^\Finv -i \bar{\psi}_\bullet\theta_\Finv.
            \end{equation}
        Lastly, we can define the new variable $y^\Finv\in\Omega_\partial^{(3,3)}[-1]$ such that $e_i y^\Finv=\zeta^{\Finv'}_i $ and $\epsilon_n y^\Finv= \lambda^\Finv$, which, combined with the field redefinition $k^\Finv= e\check{k}$, yields (omitting the $'$ apex)
        \begin{eqnarray}
            &\begin{split}\label{eq: BFV linear action new variables}
                \mathcal{S}_{\mathrm{SG}}^\Sigma&=\int_\Sigma(\iota_\xi e + \lambda\epsilon_n )\left(eF_\omega + \frac{1}{3!}\bar{\psi}\gamma^3 d_\omega\psi \right)+ c \left( e d_\omega e - \frac{1}{2}\bar{\psi}\gamma\psi \right)\\
                &\qquad +\frac{1}{3}\bar{\chi}\left(e\gamma^3d_\omega\psi -\frac{1}{2}d_\omega e \gamma^3\psi \right)+\left( \frac{1}{2}[c,c] -\frac{1}{2}\iota_\xi\iota_\xi F_\omega -\mathrm{L}_\xi^\omega c \right)e\check{k}\\
                &\qquad +\check{k}\frac{1}{3!}(\bar{\chi}-\iota_\xi\bar{\psi})\gamma^3d_\omega \chi  - \check{a}\frac{1}{3!}(\bar{\chi}-\iota_\xi\bar{\psi})\gamma^3d_\omega\psi  - \frac{1}{2}\iota_{[\xi,\xi]}e y^\Finv - i \iota_\xi\big(\mathrm{L}_\xi^\omega \bar{\psi} - [c,\psi]\big)\theta_\Finv\\
                &\qquad + i \big( \mathrm{L}_\xi^\omega\bar{\chi} - [c,\bar{\chi}] \big)\theta_\Finv + \left([c,\lambda\epsilon_n] - \mathrm{L}_\xi^\omega(\lambda \epsilon_n) +\frac{1}{2}\bar{\chi}'\gamma\chi -\iota_\xi(\bar{\chi}\gamma\psi) \right)y^\Finv
                \end{split} \\  
            &\begin{split}\label{eq: BFV SG symplectic form}
                \varpi_{\mathrm{SG}}^\Sigma&=\int_\Sigma e \delta e \delta\omega + \frac{1}{3!}\bar{\psi}\gamma^3\delta\psi \delta e + \frac{1}{3}e\delta\bar{\psi}\gamma^3 \delta\psi + \delta c' \delta (e\check{k})  + \delta\omega\delta\big(\iota_\xi (e \check{k})\big)\\
                &\qquad + i\delta\bar{\chi}\delta\theta_\Finv + i\delta\bar{\psi}\delta(\iota_\xi\theta_\Finv) -\delta \big(\iota_{\delta\xi}(e)y^\Finv\big) - \delta\lambda\epsilon_n\delta y^\Finv.
            \end{split}    
        \end{eqnarray}
    This is a good starting point but, as remarked before, it is not yet the full BFV action. In order to obtain it, one way would be to extract the Hamiltonian vector field $Q_{\mathrm{SG}}^\partial$ of $\mathcal{S}_{\mathrm{SG}}^\Sigma$ and compute $(Q^\partial_{\mathrm{SG}})^2$, which would allow us to algorithmically produce the extra terms in the action. Such method, while theoretically feasible, provides many challenges. The other way to obtain a BFV action is to induce it from the BV structure in the bulk. In the following section, we explain how to obtain it from the latter option, using the example of PC gravity as a starting point.

\section{The BV pushforward of $\mathcal{N}=1, D=4$ supergravity}\label{sec: BFV extension}
As stated in section \ref{subsec: BFV attempt}, we know from the study of the phase space that a BFV description for supergravity must exist. At the same time, we know that the BV action in the bulk is of rank 2, hence one could expect the same behavior for the BFV action. This section is devoted to studying the compatibility of the BV and BFV structures in the sense of \ref{subsec: BV BFV formalism}, i.e. whether there exists a BFV action induced by the BV one such that it recovers and extends the one found in \ref{subsec: BFV attempt}. We start by reviewing some previous results for pure gravity.

\subsection{Obstructions in the free theory}
Consider the free theory of gravity in the first order formalism, i.e. Palatini-Cartan gravity. The BV-BFV formulation of PC gravity has been studied in \cite{CS2017}, where the authors found that the complete free theory in the bulk is not BV-BFV extendible, i.e. the BV data $(\mathcal{F}_{PC},\mathcal{S}_{PC},Q_{PC},\varpi_{PC})$ in the bulk do not give rise to a BV-BFV theory. In particular, one has
    \begin{equation*}
        \delta \mathcal{S}_{PC}=\iota_{Q_{PC}}\varpi_{PC} + \pi_{PC}^*(\tilde{\vartheta}^\partial_{PC}),
    \end{equation*}
where $\pi_{PC}:\mathcal{F}_{PC} \rightarrow \tilde{\mathcal{F}}_{PC}^\Sigma$ is the surjective submersion mapping the space of bulk (BV) fields to the space of pre-boundary (pre-BFV) fields. The main obstruction is given by the fact that the pre-symplectic form $\tilde{\varpi}_{PC}^\partial:=\delta\tilde{\vartheta}^\partial_{PC} $ on $\tilde{\mathcal{F}}_{PC} $ has  singular kernel. Specifically, letting $\tilde{\omega}\in\Omega_\partial^{(1,2)} $ be the induced spin connection on the boundary, we know from the discussion in Section \ref{subsubsec: first reduction and struct constraint} (and from the study of the reduced phase space of PC gravity in \cite{CCS2020}) that we can split $\tilde{\omega}=\omega + v$, with $v\in\ker(W_1^{\partial(1,2)}) $ and $\omega$ satisfying the structural constraint. It is a matter of computations to see that the singularity in the kernel of the pre-symplectic form $\tilde{\varpi}_{PC}^\partial$ is given by terms proportional to $\delta v$. This suggests the idea of constraining the BV theory in the bulk such that the induced connection on the boundary is precisely $\omega$ satisfying the structural constraint, discarding the terms proportional to $v$. However, in order to have a well-defined BV symplectic form, one also needs to fix some of the components of the canonical anti--field $\omega^\Finv$, which otherwise would carry too many degrees of freedom. 

The correct constraints to be imposed in the space of BV PC fields were identified in \cite{Graksz21}, employing the AKSZ \cite{AKSZ} construction on the BFV PC structure found in \cite{CCS2020} in the case of cylindrical manifold $M=I\times \Sigma$, where $I=[0,1]$.\footnote{For a more detailed introduction on the 1-D AKSZ construction, we refer to Appendix \ref{subsec: AKSZ theory}} The authors in \cite{Graksz21} also showed that the constrained BV theory is equivalent to the the uncostrained one in the BV sense\footnote{This amounts to asking that there exists a symplectomorphism of the two theories such that the actions are related by the pullback via the symplectomorphism.} However, since we currently do not have a complete BFV theory of supegravity, but rather we want to find one, this method cannot be suitable to obtain a compatible BV/BFV theory in our case. In order to overcome this issue, we employ a construction which allows to ''integrate out" certain degrees of freedom of the spin connection and of its antifield in such a way that the resulting theory  retains the properties of BV (i.e. the restricted action satisfies the CME) and is equivalent to the full one. This is the content of the next section.

\subsection{The BV-Pushforward in Field Theories}
Before delving into the definition of BV pushforward, it is better to recall the object we are dealing with. We start by a quick overview on the BV integral and the Quantum Master Equation (QME) in finite dimension. We refer to \cite{CMR2, Mnev2017} for a complete discussion.

We start by considering a $\mathbb{Z}$--graded manifold $\mathcal{F}$ endowed with a $-1$--symplectic form $\varpi$. We denote by Dens$^{\frac{1}{2}}(\mathcal{F}) $ the space of half--densities on $\mathcal{F}$. A theorem by \cite{Khudaverdian2000SemidensitiesOO} \cite{Severa06} allows to define the following.
    \begin{definition}
        The BV laplacian $\Delta$ is a degree 1 coboundary operator, acting on Dens$^{\frac{1}{2}}(\mathcal{F})$. In a Darboux chart $(x^i,\xi_i)$, it reads
            \begin{equation*}
                \Delta:=\sum_i \frac{\partial}{\partial x^i}\frac{\partial}{\partial \xi_i}.
            \end{equation*}
        Furthermore, if $\mu^{\frac{1}{2}}$ is a nowhere vanishing element of Dens$^{\frac{1}{2}}(\mathcal{F})$, we can construct a $\mu$--dependent BV laplacian $\Delta_\mu$ acting on functions on $\mathcal{F}$ as $\mu^{\frac{1}{2}}\Delta_\mu f =\Delta(\mu^{\frac{1}{2}} f)\in\mathcal{C^1}(\mathcal{F}).$

        Letting $\mathcal{L}\subset \mathcal{F}$ be a Lagrangian submanifold, we define the BV integral to be the following composition
            \begin{equation*}
                \mathrm{Dens}^{\frac{1}{2}}(\mathcal{F})\overset{\cdot|_\mathcal{L}}{\longrightarrow}\mathrm{Dens}^{\frac{1}{2}}(\mathcal{L})\overset{\int_{\mathcal{L}} }{\longrightarrow}\mathbb{c}\qquad \xi\longmapsto\int_{\mathcal{L}}\xi|_{\mathcal{L}}.
            \end{equation*}
    \end{definition}
    The main theorem of BV \cite{BV1,BV2} shows that for all half densities $\xi$ on $\mathcal{F}$ and for all Lagrangian submanifolds $\mathcal{L}$, 
        \begin{equation*}
            \int_{\mathcal{L}}\Delta \xi =0.
        \end{equation*}
    Furthermore, if $\xi$ satisfies $\Delta\xi=0$, and $\mathcal{L}_t$ is a smooth family of Lagrangians parametrized by $t\in[0,1]$, we have    
        \begin{equation*}
            \frac{d}{dt}\int_{\mathcal{L}_t}\xi=0 \quad \Rightarrow \int_{\mathcal{L}_0} \xi= \int_{\mathcal{L}_1} \xi.
        \end{equation*}
\begin{remark}
In the context of field theory, the choice of a Lagrangian submanifold corresponds to gauge fixing, while the
invariance under the choice of Lagrangian submanifold corresponds to independence on gauge fixing.  The relevant object is the "path integral measure" $\mu e^{\frac{i}{\hbar}\mathcal{S}})=0\quad $, where $\mathcal{S}$ is the BV action. The quantum master equation is then given by the condition
    \begin{equation*}
        \Delta_\mu(e^{\frac{i}{\hbar}\mathcal{S}})=0\quad \Leftrightarrow \quad \frac{1}{2}(\mathcal{S},\mathcal{S})-i\hbar \Delta\mathcal{S}=0,
    \end{equation*}
which provides a modification to the CME.\footnote{If we assume we can expand $\mathcal{S}$ in powers of $\hbar$, we see that at order zero we retrieve the CME.}    
\end{remark}
\begin{definition}
    Assume that $\mathcal{F}$ factorises as a direct product of -1--symplectic graded manifolds $\mathcal{F}=\mathcal{F}'\times \mathcal{F}''$, with $\varpi=\varpi'+\varpi''$. Then
        \begin{equation*}
            \mathrm{Dens}^{\frac{1}{2}}(\mathcal{F})=\mathrm{Dens}^{\frac{1}{2}}(\mathcal{F'})\otimes\mathrm{Dens}^{\frac{1}{2}}(\mathcal{F''}).
        \end{equation*}
    The BV pushforward is a map between half densities $\mathrm{Dens}^{\frac{1}{2}}(\mathcal{F})\rightarrow\mathrm{Dens}^{\frac{1}{2}}(\mathcal{F'})$ defined by BV integration over the second factor. In particular, choosing a Lagrangian submanifold of $\mathcal{F}''$, the BV pushforward is given by the map
    \begin{equation*}
        \mathcal{P}_{\mathcal{L}}\colon\mathrm{Dens}^{\frac{1}{2}}(\mathcal{F})\overset{\mathrm{id\otimes\int_{\mathcal{L}}}}{\longrightarrow}\mathrm{Dens}^{\frac{1}{2}}(\mathcal{F'}),
    \end{equation*}
    sending a half density $\phi'\otimes\phi''$ to $\mathcal{P}_{\mathcal{L}}(\phi'\otimes\phi''):=\phi'\int_{\mathcal{L}}\phi''$.
\end{definition}
\begin{remark}
    In the context of field theory, the BV pushforward is a consistent way to eliminate "heavy modes", resulting in an effective theory. The main feature of this construction is that the effective action obtained after integrating out the heavy modes (that lie in $\mathcal{F}''$) still satisfies the QME, as showed in the following theorem.
\end{remark}
\begin{theorem}
    Letting $\Delta$, $\Delta'$ and $\Delta''$ be the canonical BV Laplacians respectively on $\mathcal{F}$, $\mathcal{F}'$ and $\mathcal{F}''$, we have:
    \begin{enumerate}
        \item $\mathcal{P}_{\mathcal{L}}$ is a chain map, i.e.
            \begin{equation*}
                \int_{\mathcal{L}}\Delta\xi=\Delta'\int_{\mathcal{L}}\xi.
            \end{equation*}
        \item if $\mathcal{L}_0$ and $\mathcal{L}_1$ are homotopic Lagrangian submanifolds in $\mathcal{F}''$, then
            \begin{equation*}
               \frac{d}{dt}\int_{\mathcal{L}_t}\xi 
            \end{equation*}
        is $\Delta'$--exact.
        \item if $\mathcal{S}\in\mathcal{C^1}(\mathcal{F})\llbracket\hbar\rrbracket$ satisfies the QME, defining $\mathcal{S}'\in\mathcal{C^1}(\mathcal{F}')\llbracket\hbar\rrbracket$ such that $$(\mu')^{\frac{1}{2}}e^{\frac{i}{\hbar}\mathcal{S}'}=\mathcal{P}_{\mathcal{L}}\left(\mu^{\frac{1}{2}}e^{\frac{i}{\hbar}\mathcal{S}}\right), $$ then $\mathcal{S}'$ satisfies the QME. 
    \end{enumerate}
\end{theorem}
In the case we are interested in, the space of fields does not factor as a product of odd symplectic manifolds, but rather it requires a more general construction.

In the following, we will need a generalization of the above construction to the case where $\mathcal{F}$ is only locally the product of two odd symplectic supermanifolds. Indeed, we assume $\pi\colon \mathcal{F}\rightarrow \mathcal{F}'$ to be a fiber bundle of odd symplectic supermanifold, where the fiber is locally also an odd symplectic supermanifold. 

We assume there is a bundle isomorphism
% https://q.uiver.app/#q=WzAsNSxbMCwwLCJcXG1hdGhjYWx7Rn0iXSxbMiwwLCJcXHRpbGRle1xcbWF0aGNhbHtGfX0nXFx0aW1lcyBcXHRpbGRle1xcbWF0aGNhbHtGfX0nJyJdLFswLDIsIlxcbWF0aGNhbHtGfSciXSxbMiwyLCJcXHRpbGRle1xcbWF0aGNhbHtGfX0nIl0sWzIsMV0sWzAsMiwiXFxwaSIsMl0sWzEsMywiXFx0aWxkZXtcXHBpfSJdLFswLDEsIlxcUGhpIl0sWzIsMywiXFxwaGkiXV0=
\[\begin{tikzcd}
	{\mathcal{F}} && {\tilde{\mathcal{F}}'\times \tilde{\mathcal{F}}''} \\
	&& {} \\
	{\mathcal{F}'} && {\tilde{\mathcal{F}}'}
	\arrow["\Phi", from=1-1, to=1-3]
	\arrow["\pi"', from=1-1, to=3-1]
	\arrow["{\tilde{\pi}}", from=1-3, to=3-3]
	\arrow["\phi", from=3-1, to=3-3]
\end{tikzcd}\]
where $\tilde{\mathcal{F}}'\times \tilde{\mathcal{F}}'' $ is a product of odd symplectic supermanifolds and both $\Phi$ and $\phi$ are symplectomorphisms. Choosing a Lagrangian submanifold $\tilde{\mathcal{L}}$ of $\tilde{\mathcal{F}}''$ defines the BV pushforward $\mathcal{P}_{\tilde{\mathcal{L}}}$. Then the composition
    \[
    (\phi^{-1})_*\circ\mathcal{P}_{\tilde{\mathcal{L}}}\circ\Phi_*
    \]
satisfies the relevant properties of the BV pushforward, and is precisely what we will be working on.
\begin{remark}
    The above construction is a special case of the so-called BV hedgehog, described in \cite{CMR2}.
\end{remark}

\subsection{The BV pushforward of PC gravity}
The construction has been applied to the case of Palatini--Cartan gravity on cylindrical manifolds in \cite{C25,canepa20254dpalatinicartangravityhamiltonian}, where it was shown that the resulting BV theory is precisely the one which is induced by the AKSZ construction\footnote{See Appendix \ref{subsec: AKSZ theory} for a short introduction} of the BFV PC theory \cite{Graksz21}. 

From now, we indicate any bulk field $\phi$ with the bold character $\bm{\phi}$.
\begin{definition}[\cite{CS2017}]
    The full BV PC theory is given by the quadruple $\mathfrak{F}_{PC}:=(\mathcal{F}_{PC}, \varpi_{PC}, Q_{PC}, \mathcal{S}_{PC})$, where $\mathcal{F}_{PC} := T^*[-1] F_{PC}$ and 
    \begin{equation*}
        {F}_{PC}=\Omega^{(1,1)}_{n.d.} \oplus \mathcal{A}_M  \oplus \Omega^{(0,2)}[1] \oplus \mathfrak{X}[1](M) \ni (\bm{e},\bm{\omega},\bm{c},\bm{\xi}).
    \end{equation*}
    The canonical symplectic form is given by
    \begin{equation*}
        \varpi_{PC} = \int_M \delta \bm{e} \delta \bm{e}^\Finv + \delta\bm{\omega} \delta \bm{\omega}^\Finv + \delta {\bm{c}} \delta {\bm{c}}^\Finv + \iota_{\delta {\bm{\xi}}} \delta {\bm{\xi}}^\Finv,
    \end{equation*}
    and the action is written as
    \begin{align}\label{eq: BV PC action}
        \mathcal{S}_{PC} &= \int_M \frac{\bm{e}^2}{2} F_{\bm{\omega}} - (L_{\bm{\xi}}^{\bm{\omega}} \bm{e} - [{\bm{c}},\bm{e}]) \bm{e}^\Finv + (\iota_{\bm{\xi}} F_{\bm{\omega}} - d_{\bm{\omega}} {\bm{c}})\bm{\omega}^\Finv \\
        &\nonumber\quad + \frac{1}{2} (\iota_{\bm{\xi}} \iota_{\bm{\xi}} F_{\bm{\omega}} - [{\bm{c}},{\bm{c}}]) {\bm{c}}^\Finv + \frac{1}{2} \iota_{[{\bm{\xi}},{\bm{\xi}}]} {\bm{\xi}}^\Finv.
    \end{align}
    Finally, the cohomological vector field acting on the fields and ghosts is given by
    \begin{align*}
        & Q_{PC} \bm{e} = L_{\bm{\xi}}^{\bm{\omega}} \bm{e} - [{\bm{c}}, \bm{e}], & Q_{PC} {\bm{\omega}} = \iota_{\bm{\xi}} F_{\bm{\omega}} - d_{\bm{\omega}} {\bm{c}}, \\
        & Q_{PC} \bm{c} = \frac{1}{2} (\iota_{\bm{\xi}} \iota_{\bm{\xi}} F_{\bm{\omega}} - [{\bm{c}}, {\bm{c}}]), & Q_{PC} {\bm{\xi}} = \frac{1}{2} [{\bm{\xi}}, {\bm{\xi}}].
    \end{align*}
\end{definition}

Now, letting $M=I\times \Sigma$, if $\bm{\phi}\in\Omega^k(I\times \Sigma)$,
    \begin{equation*}
        \bm{\phi}=\tilde{\phi}+\un{\tilde{\phi}_n}, \qquad\text{with}\quad \tilde{\phi}\in\Omega^k(\Sigma)\otimes\mathcal{C^1}(I),\quad \underline{\tilde{\phi}_n}\in\Omega^{k-1}(\Sigma)\otimes\Omega^1(I),
    \end{equation*}
assuming $x^n$ to be the coordinate along $I$, then $\un{\tilde{\phi_n}}=\tilde{\phi}_n dx^n$, with $\tilde{\phi}_n\in\mathcal{C^1}(I)\otimes\Omega^{k-1}(\Sigma).$
In the same way a vector field $\bm{\zeta}\in\mathfrak{X}(I\times\Sigma)$ is going to be split as
    \begin{equation*}
        \bm{\zeta}=\tilde{\zeta} + \ol{\tilde{\zeta}}^n, \qquad \mathrm{with} \quad \tilde{\zeta}\in\mathfrak{X}(\Sigma)\otimes\mathcal{C^1}(I),\quad  \overline{\tilde{\zeta}}^n\in \mathcal{C^1}(\Sigma)\otimes\mathfrak{X}(I),
    \end{equation*}
with $\ol{\tilde{\zeta}}^n=\tilde{\zeta}^n\partial_n$. 

\begin{remark}
    Notice that the coframe is going to be split as $\bm{e}=\tilde{e}+\underline{\tilde{e}_n}$, and since each component of $\bm{e}$ is an independent basis vector of $\mathcal{V}$,\footnote{That is the requirement that the boundary component $\tilde{e}$ are non--degenerate.} then, considering $\epsilon_n$ as in the previous sections,\footnote{i.e. such that $\{\tilde{e}_i,\epsilon_n\}$ is a local basis of $\mathcal{V}$ and such that $\partial_n\epsilon_n=0=\delta \epsilon_n$} we have that $\underline{\tilde{e}_n}$ is expressed as the following linear combination
    \begin{equation*}
        \underline{\tilde{e}_n}=\underline{z}^i \tilde{e}_i + \underline{\mu}\epsilon_n = \iota_{\underline{z}}\tilde{e} + \underline{\mu}\epsilon_n,
    \end{equation*}
where $\underline{z}\in\Omega^1(I)\otimes\mathfrak{X}(\Sigma) $ and $\underline{\mu}\in\Omega^1(I)\otimes\mathcal{C^1}(\Sigma)$. 

\end{remark}

\begin{definition}\cite{C25}\label{def: reduced PC theory}
    The restricted BV PC theory is given by data $\mathfrak{F}_{PC}^r:=(\mathcal{F}^r_{PC}, \varpi^r_{PC}, Q^r_{PC}, \mathcal{S}^r_{PC})$, where 
    \begin{itemize}
        \item The restricted space of BV fields is given by the subspace of $\mathcal{F}_{PC}$ satisfying the following structural constraints 
            \begin{align}
                \label{eq: BV PC constr omega dag}&\underline{\mathfrak{W}}^\Finv:=\underline{\tilde{\omega}}_n^\Finv - \iota_{\underline{z}}\tilde{\omega}^\Finv - \iota_{\tilde{\xi}}\underline{\tilde{c}}_n^\Finv + \iota_{\underline{z}}{\tilde{c}}_n^\Finv\tilde{\xi}^n\in\mathrm{Im}(W_{\tilde{e}}^{(1,1)})\\
                \label{eq: BV PC constr omega}&\epsilon_n d_{\tilde{\omega}}\tilde{e} - \epsilon_n W_{\tilde{e}}^{-1}(\underline{\mathfrak{W}}) d\tilde{\xi}^n + \iota_{\hat{X}}(\tilde{\omega}_n^\Finv - \tilde{c}_n^\Finv\tilde{\xi}^n)\in\mathrm{Im}(W_{\tilde{e}}^{(1,1)}),
            \end{align}
        where $W_{\tilde{e}^k}^{(i,j)}\colon\Omega^{(i,j)}\rightarrow\Omega^{(i+k,j+k)}\colon \alpha\mapsto \tilde{e}^k \wedge \alpha$ shares the same properties of $W_k^{\partial,(i,j)} $ and
            \begin{equation}
                \tilde{X}=\mathrm{L}_{\tilde{\xi}}^{\tilde{\omega}}(\epsilon_n) - d_{\tilde{\omega}_n}(\epsilon_n)\tilde{\xi}^n - [\tilde{c},\epsilon_n]; \qquad \hat{X}=\tilde{e}^i_a \tilde{X}^a\partial_i
            \end{equation}
        \item $\varpi^r_{PC}:=\varpi_{PC}(I\times \Sigma)|_{\mathcal{F}^r_{PC}} $;
        \item $\mathcal{S}^r_{PC}:=\mathcal{S}_{PC}|_{\mathcal{F}^r_{PC}}  $;
        \item $Q^r_{PC}=Q_{PC}$.
    \end{itemize}
\end{definition}

It is worth investigating the source of the constraints \eqref{eq: BV PC constr omega dag} and \eqref{eq: BV PC constr omega} in a more direct and systematic way, as we will want to generalize them to the case of supergravity. 

First of all, we see that the BFV-PC theory found in \cite{CCS2020} is also obtained by "turning off" the gravitino interaction in \eqref{eq: BFV action}, yielding\footnote{assuming all the fields in the expression below to be defined on the boundary.}
    \begin{eqnarray}
        &\begin{split}
            \mathcal{S}_{PC}^\partial&=\int_\Sigma(\iota_\xi e + \lambda\epsilon_n )eF_\omega + c  e d_\omega e +\left( \frac{1}{2}[c,c] -\frac{1}{2}\iota_\xi\iota_\xi F_\omega -\mathrm{L}_\xi^\omega c \right)k^\Finv\\
            &\qquad   - \frac{1}{2}\iota_{[\xi,\xi]}e y^\Finv  + \left([c,\lambda\epsilon_n] - \mathrm{L}_\xi^\omega(\lambda \epsilon_n)  \right)y^\Finv
            \end{split} \\  
        &\begin{split}\label{BFV PC sympl form}
            \varpi_{PC}^\partial&=\int_\Sigma e \delta e \delta\omega  + \delta c \delta k^\Finv  + \delta\omega\delta\big(\iota_\xi (k^\Finv)\big) -\delta \big(\iota_{\delta\xi}(e)y^\Finv\big) - \delta\lambda\epsilon_n\delta y^\Finv.
        \end{split}    
    \end{eqnarray}    
This was obtained in \cite{CCS2020} with the same procedure as in Section \ref{sec: BFV data from the reduced phase space}, but without the need to introduce any second-rank terms.     
Furthermore we know that, in order to have a well-defined phase space, we need to carefully fix some components of the boundary connection $\omega$, in such a way that the term $e\wedge\delta e\wedge \delta\omega$ in the symplectic form does not give rise to any degeneracy. This is the content of Theorem \ref{thm: fix rep of omega bdry} (or equivalently of Theorem 33 in \cite{CCS2020}). We might therefore be tempted to impose the same constraint for the bulk fields, i.e.
    \begin{equation*}
        \epsilon_nd_{\tilde{\omega}}\tilde{e}\in\mathrm{Im}(W_{\tilde{e}}^{(1,1)}),
    \end{equation*}
which amounts to imposing only some parts of the condition $d_{\bm{\omega}}\bm{e
}=0$. Unfortunately, such constraint is not invariant under the action of $Q_{PC}$, hence not defining a suitable restricted BV theory. As it turns out, the correct one is given by \eqref{eq: BV PC constr omega}. In particular, a direct (and immediate) generalization of Lemma \ref{lem: splitting (2,2)} tells us that there must exist unique $\sigma\in\mathcal{C^1}(I)\otimes\Omega_\partial^{(1,1)} $ and $\rho\in\Ker{W_{\tilde{e}}^{(1,2)}}$ such that
    \begin{equation*}
        \epsilon_n d_{\tilde{\omega}}\tilde{e} - \epsilon_n W_{\tilde{e}}^{-1}(\underline{\mathfrak{W}}) d\tilde{\xi}^n + \iota_{\hat{X}}(\tilde{\omega}_n^\Finv - \tilde{c}_n^\Finv\tilde{\xi}^n) = \tilde{e}\sigma + \epsilon_n[\rho,\tilde{e}].
    \end{equation*}
In the same way, we can generalize Theorem \ref{thm: fix rep of omega bdry} and see that we can split   $\tilde{\omega}$ as $\tilde{\omega}=\hat{\omega}+\tilde{v}$, with $\tilde{v}\in\Ker{W_{\tilde{e}}^{(2,1)}}$ and $\hat{\omega}$ satisfying
    \begin{equation*}
        \epsilon_n d_{\hat{\omega}}\tilde{e} - \epsilon_n W_{\tilde{e}}^{-1}(\underline{\mathfrak{W}}) d\tilde{\xi}^n + \iota_{\hat{X}}(\tilde{\omega}_n^\Finv - \tilde{c}_n^\Finv\tilde{\xi}^n)\in\mathrm{Im}(W_{\tilde{e}}^{(1,1)}).
    \end{equation*}
Then, since $\epsilon_n d_{\tilde{\omega}}\tilde{e}=\epsilon_n d_{\hat{\omega}}\tilde{e} +\epsilon_n[\tilde{v},\tilde{e}]$. Constraint \eqref{eq: BV PC constr omega} is satisfied if and only if $\rho=\tilde{v}=0$, implying that only $\hat{\omega}$ survives in $\mathcal{F}_{PC}^r$, taking the role of the "reduced boundary connection", which (thanks to Theorem \ref{thm: fix rep of omega bdry}) can be seen as a representative living in $\mathcal{A}_{\mathrm{red}}(I\times\Sigma):=\mathcal{A}(I\times\Sigma)/\ker(W_{\tilde{e}}^{(1,2)})$.

The other structural constraint \eqref{eq: BV PC constr omega dag} is interpreted as fixing some components of the canonical antifield $\underline{\tilde{\omega}}^\Finv_n$\footnote{Indeed in the symplectic form the relevant term is $\delta\tilde{\omega}\delta\underline{\tilde{\omega}}^\Finv_n$.}. In particular, since $\underline{\tilde{\omega}}^\Finv_n\in\Omega^1(I)\otimes\Omega_\partial^{(2,2)} $, we can apply Lemma \ref{lem: splitting (2,2)} to show there exists a splitting  
    \begin{equation*}
        \underline{\omega}^\Finv_n=\tilde{e}\underline{\rho}^\Finv + \epsilon_n[\tilde{e},\underline{\sigma}^\Finv].
    \end{equation*}
Then $\underline{\rho}^\Finv$ is exactly in the "dual" of $\mathcal{C^1}(I)\otimes\Omega^{(1,2)}_\partial/\Ker{W_{\tilde{e}}^{(1,2)}} $, providing a perfect candidate for the antifield of $\tilde{\omega}$ in $\mathcal{F}_{PC}^r$. However, the constraint $\underline{\omega}^\Finv_n \in \mathrm{Im}(W_{\tilde{e}}^{(1,1)})$ is not $Q$--invariant, and \eqref{eq: BV PC constr omega dag} turns out to be the correct one. 

In order to shed some more light into the choice of constraints, we consider the "more covariant" combination of fields $\bm{\omega}^\Finv-\iota_{\bm{\xi}}\bm{c}^\Finv$, as such expression appears repeatedly in computations. Specifically, when computing the variation of the BV-PC action \eqref{eq: BV PC action}, one obtains a total derivative term $\vartheta_{PC}$, whose variation (having used Stokes' theorem on $I\times\Sigma$) produces a term on the boundary
    \begin{equation*}
        \delta\tilde{\vartheta}_{PC}^\partial=\int_\Sigma\cdots+ \delta\tilde{\omega}\delta\big(\iota_{\tilde{\xi}}(\tilde{\omega}^\Finv - \tilde{c}_n^\Finv\tilde{\xi}^n)\big)+\cdots.
    \end{equation*}
Confronting it with the BFV symplectic form \eqref{BFV PC sympl form}, one notices that the expression $\tilde{\omega}^\Finv - \tilde{c}_n^\Finv\tilde{\xi}^n$ is a good candidate for the boundary field $k^\Finv$, representing the antighost of $ c\in\Omega_\partial^{(0,2)}[1] $. The same expression also appears inside \eqref{eq: BV PC constr omega dag}.

At this point we know from diagram \eqref{diag: prop e bulk} and from the discussion in Section \ref{subsec: BV Sugra}  that we can redefine $\bm{\omega}^\Finv = \bm{e}\check{\bm{\omega}}$ and $\bm{c}^\Finv=\frac{\bm{e}^2}{2}\check{\bm{c}}$, so we have $\bm{\omega}^\Finv-\iota_{\bm{\xi}}\bm{c}^\Finv=\bm{e}\left(\check{\bm{\omega}} -\iota_{\bm{\xi}}\bm{e}\check{\bm{c} } - \frac{1}{2}\bm{e}\iota_{\bm{\xi}}\check{\bm{c}} \right) $, and since $\bm{\omega}^\Finv=\tilde{\omega}^\Finv+ \underline{\tilde{\omega}}^\Finv_n$ and $\bm{c}^\Finv=\underline{\tilde{c}}_n^\Finv$, unpacking the expression yields 
    \begin{align*}
      \bm{\omega}^\Finv-\iota_{\bm{\xi}}\bm{c}^\Finv=&\tilde{\omega}^\Finv - \tilde{c}^{\Finv}_n\tilde{\xi}^n + \underline{\tilde{\omega}}_n^\Finv - \iota_{\tilde{\xi}}\underline{\tilde{c} }^\Finv_n \\
       \bm{e}\left(\check{\bm{\omega}}-\iota_{\bm{\xi}}e\check{\bm{c}}-\frac{\bm{e}}{2}\iota_{\bm{\xi}}\check{\bm{c}}\right)=& \tilde{e}\left( \tilde{\check{\omega}} + \underline{\tilde{\check{\omega}}}_n -\iota_{\tilde{\xi}}\tilde{e}\tilde{\check{c}} -\tilde{e}_n\tilde{\xi}^n\tilde{\check{c}} - \frac{\tilde{e}}{2}( \iota_{\tilde{\xi}}\tilde{\check{c}} - \tilde{\check{c}}_n\tilde{\xi}^n ) -\iota_{\tilde{\xi}}\tilde{e}\underline{\tilde{\check{c}}}_n - \tilde{e}_n\tilde{\xi}^n\underline{\tilde{\check{c}}}_n \right)  \\
      &+ \underline{\tilde{e}}_n\left(\tilde{\check{\omega}}- \iota_{\tilde{\xi}}\tilde{e}\tilde{\check{c}} -\tilde{e}_n\tilde{\xi}^n\tilde{\check{c}} - \frac{\tilde{e}}{2}( \iota_{\tilde{\xi}}\tilde{\check{c}} - \tilde{\check{c}}_n\tilde{\xi}^n )  \right).
    \end{align*}
By inspection (discarding the terms proportional to $dx^n$) one could then infer 
    \begin{equation}\label{eq: definition brdy omegadag - cdag}
        \tilde{\omega}^\Finv - \tilde{c}^{\Finv}_n\tilde{\xi}^n=\tilde{e}\left( \tilde{\check{\omega}} -\iota_{\tilde{\xi}}\tilde{e}\tilde{\check{c}} -\tilde{e}_n\tilde{\xi}^n\tilde{\check{c}} - \frac{\tilde{e}}{2}( \iota_{\tilde{\xi}}\tilde{\check{c}} - \tilde{\check{c}}_n\tilde{\xi}^n )\right)=:\tilde{e}\tilde{\check{k}},
    \end{equation} 
having defined $\bm{k}^\Finv:=\bm{\omega}^\Finv -\iota_{\bm{\xi}}\bm{c}^\Finv = \bm{e}\bm{\check{k}}=\tilde{e}\tilde{\check{k}} + \underline{\tilde{e}}_n\tilde{\check{k}} +\tilde{e}\underline{\tilde{\check{k}}}_n $. However, we know from diagram $\ref{diag: prop e bdry}$ that $W_{\tilde{e}}^{(2,1)} $ is surjective but not injective, hence the right hand side of \eqref{eq: definition brdy omegadag - cdag} is defined up to a term in $\ker{W_{\tilde{e}}^{(2,1)}}$.\footnote{Defining $\bm{k}^\Finv=\tilde{k}^\Finv+\underline{\tilde{k}}_n^\Finv$, we have that $\underline{\tilde{k}}_n^\Finv=\underline{\tilde{e}}_n\tilde{\check{k}} +\tilde{e}\underline{\tilde{\check{k}}}_n$, which is ill-defined since $\tilde{\check{k}}$ is unique only up to elements in $\Ker{W_{\tilde{e}}^{(2,1)}}$. Such ambiguity is resolved by imposing the structural constrain \eqref{eq: BV PC constr omega dag}, as is shown in the next proposition and in Remark \ref{rem: K dag n}} In order to fix the representative of the equivalence class in $\mathcal{C^1}(I)\otimes\big(\Omega_\partial^{(2,1)}/\ker{W_{\tilde{e}}^{(2,1)}}\big) $, we can generalize Theorem \ref{thm: fix rep of kdag} to see that we have to impose the constraint
    \begin{equation*}
        \epsilon_n\tilde{\check{k}}=\epsilon_n\left(\tilde{\check{\omega}} -\iota_{\tilde{\xi}} \tilde{e}\tilde{\check{c}} - \tilde{e}_n\tilde{\check{c}}\tilde{\xi}^n- \frac{e}{2}( \iota_{\tilde{\xi}}\tilde{\check{c}} - \tilde{\check{c}}_n\tilde{\xi}^n ) \right)\in\mathrm{Im}(W_{\tilde{e}}^{(1,1)}),
    \end{equation*}
which, thanks to the following proposition, turns out to be equivalent to $\eqref{eq: BV PC constr omega dag}$. 
    \begin{proposition}\label{prop: equivalence of BV PC r constraints}
        Constraint \eqref{eq: BV PC constr omega dag} is equivalent to 
        \begin{equation}\label{eq: new BV PC constr omegadag}
            \epsilon_n\tilde{\check{k}}\in\mathrm{Im}(W_{\tilde{e}}^{(1,1)}).
        \end{equation}
        Furthermore, constraint \eqref{eq: BV PC constr omega} is obtained by applying the cohomological vector field $Q_{PC} $ to \eqref{eq: new BV PC constr omegadag}.
    \end{proposition}
    \begin{remark}
        As an immediate corollary, since $Q_{PC}^2=0$, one obtains that the structural constraints on $\mathcal{F}_{PC}^r$ are invariant with respect to the action of $Q_{PC} $.
    \end{remark}
    \begin{proof}
        The proof is found in Appendix \ref{proof: equiv BVPCr}
    \end{proof}

Having fixed such constraint, we recall there exist $\underline{\tilde{\tau}}^\Finv\in\Omega^1(I)\otimes\Omega_\partial^{(1,1)}[-1] $ and $\underline{\tilde{\mu}}^\Finv\in\Ker{W_{\tilde{e}}^{(1,2)}}[-1] $ such that
    \begin{equation}
        \underline{\mathfrak{W}}^\Finv=\tilde{e}\underline{\tilde{\tau}}^\Finv + \epsilon_n[\underline{\tilde{\mu}}^\Finv,\tilde{e}].
    \end{equation}
Constraint \eqref{eq: new BV PC constr omegadag} becomes $\underline{\mathfrak{W}}^\Finv=\tilde{e}\underline{\tilde{\tau}}^\Finv$, telling us that $\underline{\tilde{\tau}}^\Finv$ is isomorphic to a field in $\mathcal{F}^r_{PC}$ and $\underline{\tilde{\mu}}^\Finv$ defines its complement in $\mathcal{F}_{PC}$.

Explicitly, setting $\underline{\tilde{\omega}}^\Finv_n=\underline{\hat{\omega}}_n^\Finv + \underline{\tilde{v}}^\Finv$, we can rewrite
    \begin{eqnarray}
        &\label{eq: def omega dag hat}\underline{\hat{\omega}}_n :=\tilde{e}\underline{\tilde{\tau}}^\Finv +\iota_{\underline{z}}\tilde{\omega}^\Finv + \iota_{\tilde{\xi}}\underline{\tilde{c}}^\Finv_n - \iota_{\underline{z}}\tilde{c}_n^\Finv\tilde{\xi}^n  \\
        &\underline{\tilde{v}}^\Finv= \epsilon_n[\underline{\tilde{\mu}}^\Finv,\tilde{e}].
    \end{eqnarray}
    \begin{remark}\label{rem: K dag n}
    Imposing \eqref{eq: new BV PC constr omegadag} is equivalent to setting $\underline{\mu}^\Finv=0$, implying $\underline{\tilde{v}}^\Finv=0$, which is consistent with the fact that the other constraint \eqref{eq: BV PC constr omega} fixes $\tilde{v}=0$. Furthermore, from \eqref{eq: def tau dag a check} and \eqref{eq: def omega dag hat} we see that $\underline{\tilde{\omega}}^\Finv_n - \iota_{\tilde{\xi}}\underline{\tilde{c}}^\Finv_n=\tilde{e}\big(\underline{\tilde{k}}_n + \underline{\mu}\tilde{\check{a}}\big) + \iota_{\underline{z}}\tilde{e}\tilde{\check{k}} + \underline{\tilde{v}}^\Finv$, which, after imposing \eqref{eq: new BV PC constr omegadag}, becomes 
        \begin{align*}
            \underline{\tilde{\omega}}^\Finv_n - \iota_{\tilde{\xi}}\underline{\tilde{c}}^\Finv_n = \tilde{e}\underline{\tilde{\check{k}}}_n + \underline{\tilde{e}}_n \tilde{\check{k}}, 
        \end{align*}
    which solves any possible ambiguity in the definition of $\underline{\tilde{k}}^\Finv_n$.    
\end{remark}
Lastly, we can explicitly write the symplectic form on $\mathcal{F}_{PC}^r$ as
    \begin{align*}
        \varpi_{PC}^r&=\int_{I\times \Sigma} \delta\tilde{e}\delta\underline{\tilde{e}}^\Finv_n + \delta\underline{\tilde{e}}_n\delta\tilde{e}^\Finv + \delta\hat{\omega}\delta\underline{\hat{\omega}}^\Finv_n + \delta\underline{\tilde{\omega}}_n\delta\tilde{\omega}^\Finv
        + \delta\tilde{c}\delta\underline{\tilde{c}}^\Finv_n + \iota_{\delta\tilde{\xi}}\delta\underline{\tilde{\xi}}^\Finv + \delta\underline{\tilde{\xi}}^\Finv_n\delta\tilde{\xi}^n.
    \end{align*}

\begin{proposition}\cite{C25}\label{thm: PC sympl restr}
    The full BV PC theory $\mathfrak{F}_{PC}$ is symplectomorphic to the following $-1$--symplectic graded manifold
    \begin{equation*}
        \mathfrak{F}_{PC}^H:=\left(\mathcal{F}_{PC},\varpi_{PC}^r + \int_{I\times \Sigma}\delta\tilde{v}\delta\underline{\tilde{v}}^\Finv\right), \quad\text{with}\quad \tilde{v}\in\Ker{W_{\tilde{e}}^{(1,2)}}\quad \text{and} \quad \underline{\tilde{v}}^\Finv=\epsilon_n[\underline{\tilde{\mu}}^\Finv,\tilde{e}],
    \end{equation*} for some $\underline{\tilde{\mu}}^\Finv\in\mathcal{C^1}(I)\otimes\Omega_\partial^{(1,2)}[-1]$. Furthermore, fixing a reference coframe $e_0$ and considering $\mathcal{F}_f:=T^*[-1](\Ker{W_{{e}_0}^{(1,2)}})$ together with its canonical symplectic form $\varpi_f$, there is a surjective submersion $\pi\colon\mathfrak{F}_{PC}^H\rightarrow\mathfrak{F}_{PC}^r$ such that the quadruple $$(\mathfrak{F}_{PC}^H,\mathfrak{F}_{PC}^r,\mathfrak{F}_f,\pi)$$
    is a BV hedgehog.
\end{proposition}
\begin{theorem}\cite{C25}
    The restricted BV PC theory $\mathfrak{F}_{PC}^r$ is the BV pushforward of $\mathfrak{F}_{PC}^H$ obtained by integrating along the Lagrangian submanifold $\mathcal{L}_f:=\{(\tilde{v},\underline{\tilde{v}}^\Finv)\in\mathcal{F }_f\hspace{1mm}\vert\hspace{1mm}\underline{\tilde{v}}^\Finv=0 \}\subset \mathcal{F }_f$.
\end{theorem}
\begin{remark}
    In particular, the above theorem hold because the symplectomorphism $\Phi$ of Theorem \ref{thm: PC sympl restr} is such that
    \begin{equation*}
        \Phi^*\left( \mathcal{S}_{PC}^r + \int_{I\times\Sigma}\frac{1}{2}\underline{\tilde{e}}_n \tilde{e} [\tilde{v},\tilde{v}] \right)=\mathcal{S}_{PC}\big\vert_{\mathcal{L}_f},
    \end{equation*}
    with $\int_{I\times\Sigma}\frac{1}{2}\underline{\tilde{e}}_n \tilde{e} [\tilde{v},\tilde{v}]$ taking the role of a Gaussian integral\footnote{Indeed it can be proved \cite{C25,canepa20254dpalatinicartangravityhamiltonian} that  the quadratic form $\int_{I\times\Sigma}\frac{1}{2}\underline{\tilde{e}}_n \tilde{e} [\tilde{v},\tilde{v}]$ is non--degenerate.} inside $\int_{\mathcal{L}_f}e^{\frac{i}{\hbar}\mathcal{S}_{PC}}$, contributing to the partition function as multiplicative factor (depending on $e$).
\end{remark}
The details of such constructions are provided in the next section.

\subsection{The BV-Pushforward of $N=1, D=4$ supergravity}
In this section we repeat the same procedure above in the case of supergravity, obtaining the reduced BV theory of $N=1, D=4$ supergravity as the BV pushforward of the supergravity hedgehog. 

\begin{definition}
    Letting $M=I\times \Sigma$, the reduced $\mathcal{N}=1, D=4$ supergravity theory is given by $\mathfrak{F}_{\mathrm{SG}}^r:=(\mathcal{F}_{\mathrm{SG}}^r,\varpi_{\mathrm{SG}}^r,Q_{\mathrm{SG}}^r,\mathcal{S}_{\mathrm{SG}}^r)$, where 
    \begin{itemize}
        \item $\mathcal{F}_{\mathrm{SG}}^r$ is given by the the subspace of $\mathcal{F}_{\mathrm{SG}}(I\times\Sigma)$ satisfying the following constraints 
            \begin{eqnarray}
                &\epsilon_n\left(\tilde{\check{\omega}} -\iota_{\tilde{\xi}} \tilde{e}\tilde{\check{c}} - \tilde{e}_n\tilde{\check{c}}\tilde{\xi}^n- \frac{e}{2}( \iota_{\tilde{\xi}}\tilde{\check{c}} - \tilde{\check{c}}_n\tilde{\xi}^n ) \right)=\tilde{e}\tilde{\check{a}}\\
                &\label{constr: BV Sugra omega}Q_{\mathrm{SG}}\left[\epsilon_n\left(\tilde{\check{\omega}} -\iota_{\tilde{\xi}} \tilde{e}\tilde{\check{c}} - \tilde{e}_n\tilde{\check{c}}\tilde{\xi}^n- \frac{e}{2}( \iota_{\tilde{\xi}}\tilde{\check{c}} - \tilde{\check{c}}_n\tilde{\xi}^n ) \right)-\tilde{e}\tilde{\check{a}}\right]=0,
            \end{eqnarray}
        for some $\tilde{\check{a}}\in\mathcal{C^1}(I)\otimes\Omega^{(1,1)}_\partial[-1]$;    
        \item $\varpi^r_{\mathrm{SG}}:=\varpi_{\mathrm{SG}}(I\times \Sigma)|_{\mathcal{F}^r_{\mathrm{SG}}} $;
        \item $\mathcal{S}^r_{\mathrm{SG}}:=\mathcal{S}_{\mathrm{SG}}|_{\mathcal{F}^r_{\mathrm{SG}}}  $;
        \item $Q^r_{\mathrm{SG}}=Q_{\mathrm{SG}}$.
    \end{itemize}
\end{definition}
\begin{remark}
    Carrying out some computations, \eqref{constr: BV Sugra omega} can be expressed as
        \begin{align}\label{constr: BV SG red omega}
            &\epsilon_n \left(d_{\tilde{\omega}}\tilde{e}-\frac{1}{2}\bar{\tilde{\psi}}\gamma\tilde{\psi}\right)  - \epsilon_n \underline{\tilde{\tau}}^\Finv d\tilde{\xi}^n + \iota_{\hat{X}}\tilde{k}^\Finv +\mathfrak{Q} = \tilde{e}\sigma,
            %&\nonumber\mathfrak{Q} = \epsilon_n\left( i\bar{\tilde{\psi}}^\Finv_0\tilde{\ugam}\gamma\tilde{\chi} - \frac{i}{2}e \left( \bar{\tilde{\psi}}^\Finv_0,\hat{\gamma}\right)\gamma\tilde{\ugam}^2\chi +\frac{\tilde{e}}{4}\bar{\chi}\gamma\iota_{\hat{\gamma}}\left([\tilde{\check{k}},\gamma]\tilde{\psi}\right) +\frac{\tilde{e}}{4}<\tilde{e},[\tilde{\check{k}},\gamma]\tilde{\psi}> -\frac{1}{2}\bar{\tilde{\chi}}\gamma\tilde{\chi}\tilde{\check{c}} \right) + \tilde{e}(\cdots ), 
        \end{align}
    where terms $\mathfrak{Q}$ are left implicitly defined by applying the terms in $Q_{\mathrm{SG}}$ depending on the $\tilde{\psi}$ and $\tilde{\chi}$ to $\epsilon_n\check{k}$. Furthermore, we can also define $x^\Finv\in\mathcal{C^1}(I)\otimes\Omega^{(1,1)}_\partial$ and $y^\Finv\in\mathcal{C^1}(I)\otimes\Omega^{(1,2)}_\partial$ such that 
        \begin{equation}\label{eq: def x e y}
            \mathfrak{Q}=\tilde{e} x^\Finv + \epsilon_n[\tilde{e},y^\Finv].
        \end{equation}
    Lastly, we have
        \begin{equation*}
           \tilde{X}:=\mathrm{L}_{\tilde{\xi}}^{\tilde{\omega}}(\epsilon_n) - d_{\tilde{\omega}_n}(\epsilon_n)\tilde{\xi}^n - [\tilde{c},\epsilon_n] , \qquad \hat{X}=\tilde{e}^i_a \tilde{X}^a\partial_i.
        \end{equation*}
\end{remark}
\begin{remark}
    As of now, we do not know if such reduced theory is a genuine BV theory, since we did not check explicitly that $(\mathcal{S}^r_{\mathrm{SG}},\mathcal{S}^r_{\mathrm{SG}})=0$ holds. However, showing that one can recover the reduced theory as the BV pushfoward of the supergravity hedgehog will automatically assure that the CME holds.
\end{remark}
The first goal is to find a symplectomorphism between $\mathfrak{F}_{\mathrm{SG}}=(\mathcal{F}_{\mathrm{SG}}(I\times\Sigma),\varpi_{\mathrm{SG}}(I\times\Sigma))$ and $\mathfrak{F}_{\mathrm{SG}}^H=\left(\mathcal{F}_{\mathrm{SG}}(I\times\Sigma),\varpi_{\mathrm{SG}}^r+\int_{I\times\Sigma}\delta\tilde{v}\delta\underline{\tilde{v}}^\Finv\right)$. To do so, we follow \cite{C25,canepa20254dpalatinicartangravityhamiltonian}, splitting the symplectomorphism in two steps.
\begin{lemma}
    There exists a symplectomorphism $\phi_1\colon\mathcal{F}_{\mathrm{SG}}(I\times \Sigma)\rightarrow\mathcal{F}_{\mathrm{SG}}(I\times\Sigma)$ such that 
    \begin{equation*}
        \phi_1^*\left(\varpi_{\mathrm{SG}}^r+\int_{I\times\Sigma} \delta\tilde{v}\delta\underline{\tilde{v}}^\Finv + \delta\hat{\omega}\delta\underline{\tilde{v}}^\Finv \right)=\varpi_{\mathrm{SG}}(I\times \Sigma).
    \end{equation*}
    Explicitly, we have that all the fields are preserved by the symplectomorphism\footnote{In particular we also have $\phi_1^*(\tilde{v})=\tilde{v}$, $\phi_1^*(\hat{\omega})=\hat{\omega}$, $\phi_1^*(\underline{\tilde{\tau}}^\Finv)=\underline{\tilde{\tau}}^\Finv$ and $\phi_1^*(\underline{\tilde{\mu}}^\Finv)=\underline{\tilde{\mu}}^\Finv$.} except 
    \begin{align*}
        &\phi_1^*(\tilde{e}^\Finv)=\tilde{e}^\Finv+\tilde{v}\tilde{\check{k}} &\phi_1^*(\underline{\tilde{e}}_n^\Finv)=\underline{\tilde{e}}^\Finv_n - \tilde{v}\underline{\tilde{\check{k}}}_n - \iota_{\underline{z}}\tilde{v}\tilde{\check{k}}\\
        &\phi_1^*(\tilde{c})=\tilde{c} -\iota_{\tilde{\xi}}\tilde{v} +\iota_z\tilde{v}\tilde{\xi}^n \qquad \phi_1^*(\underline{\tilde{\xi}}^\Finv)=\underline{\tilde{\xi}}^\Finv- \tilde{v}\underline{\tilde{c}}^\Finv_n & \qquad \phi_1^*(\underline{\tilde{\xi}}^\Finv_n)=\underline{\tilde{\xi}}^\Finv_n +\iota_{z}\tilde{v}\underline{\tilde{c}}^\Finv_n \qquad    \phi_1^*(\underline{\tilde{\omega}}_n)=\underline{\tilde{\omega}}_n +\iota_{\underline{z}}\tilde{v}
    \end{align*}
\end{lemma}

\begin{proof}
    The proof can be copied, mutatis mutandis, from \cite{C25}, since $\varpi_{\mathrm{SG}}=\varpi_{PC}+\int_M i \delta\bm{\bar{\psi}}\delta{\bm{\psi}}^\Finv + i \delta\bar{\bm{\chi}}\delta\bm{\chi}^\Finv$, and only the field inside $\varpi_{PC}$ transform under $\phi_1.$
\end{proof}
\begin{lemma}
    There exists a symplectomorphism $\phi_2\colon\mathcal{F}_{\mathrm{SG}}(I\times \Sigma)\rightarrow\mathcal{F}_{\mathrm{SG}}(I\times\Sigma)$ such that 
    \begin{equation*}
        \phi_2^*\left(\varpi_{\mathrm{SG}}^r+\int_{I\times\Sigma} \delta\tilde{v}\delta\underline{\tilde{v}}^\Finv  \right)=\varpi_{\mathrm{SG}}^r+\int_{I\times\Sigma} \delta\tilde{v}\delta\underline{\tilde{v}}^\Finv + \delta\hat{\omega}\delta\underline{\tilde{v}}^\Finv .
    \end{equation*}
    In particular, defining $\alpha\in\mathcal{C^1}(I)\otimes\Omega^{(1,1)}_\partial$ and $\beta\in\mathcal{C^1}(I)\otimes\Omega^{(1,2)}_\partial$ such that
        \begin{equation}\label{eq: def nu}
            d\tilde{\xi}^n \tilde{\mu}^\Finv = \tilde{e}\alpha + \epsilon_n[\tilde{e,\beta}],
        \end{equation}
    then $\epsilon_n d\tilde{\xi}^n \tilde{\mu}^\Finv= \epsilon_n \tilde{e}\alpha\eqqcolon \tilde{e}\nu$, where $\nu=\epsilon_n \alpha\in\mathcal{C^1}(I)\otimes\Omega^{(1,2)}_\partial$. 
    
    The action of $\phi_2$ is given by 
        \begin{align*}
            %&\phi_2^*(\underline{\tilde{e}}^\Finv_n)=\underline{\tilde{e}}^\Finv_n + d_{\tilde{\omega}}(\epsilon_n\underline{\tilde{\mu}}^\Finv) + \sigma\underline{\tilde{\mu}}^\Finv+ \iota_{\tilde{X}}(\tilde{\check{k}}\underline{\tilde{\mu}}^\Finv) - \nu \underline{\tilde{\check{k}}}_n-\iota_{\underline{z}}\nu \tilde{\check{k}} - x^\Finv \underline{\tilde{\mu}}^\Finv + [\epsilon_n y^\Finv,\underline{\tilde{y}}^\Finv]\\
            & \phi_2^*(\tilde{e}^\Finv)= \tilde{e}^\Finv + \nu \tilde{\check{k}} &  \phi_2^*(\tilde{v})= \tilde{v} +y^\Finv \\
            &\phi_2^*(\tilde{\omega}_n)=\tilde{\omega}_n + \iota_z \nu + \iota_{\tilde{X}}\tilde{\mu}^\Finv & \phi_2^*(\hat{\omega})=\hat{\omega} + \nu\\
            &\phi_2^*(\tilde{c})=\tilde{c} + \iota_{\tilde{X}}\tilde{\mu}^\Finv\tilde{\xi}^n + \iota_{\tilde{\xi}}\nu + \iota_{z}\nu \tilde{\xi}^n &\phi_2^*(\tilde{c}^\Finv_n)= \tilde{c}^\Finv_n + [\epsilon_n,\tilde{\check{k}}\tilde{\mu}^\Finv]\\
            &\phi_2^*(\tilde{\omega}^\Finv)=\tilde{\omega}^\Finv + [\epsilon_n, \tilde{\xi}^n \tilde{\check{k}}\tilde{\mu}^\Finv] & \phi_2^*(\tilde{\omega}_n)=\tilde{\omega}_n + [\epsilon_n,\iota_{\xi}(\tilde{\check{k}}\tilde{\mu}^\Finv)]\\
            &\phi_2^*(\tilde{\phi}^\Finv_n)=\tilde{\phi}^\Finv_n - i \epsilon_n \gamma\tilde{\psi}\tilde{\mu}^\Finv,
        \end{align*}
        \begin{align*}
            &\phi_2^*(\underline{\tilde{e}}^\Finv_n)=\underline{\tilde{e}}^\Finv_n + d_{\tilde{\omega}}(\epsilon_n\underline{\tilde{\mu}}^\Finv) + \sigma\underline{\tilde{\mu}}^\Finv+ \iota_{\tilde{X}}(\tilde{\check{k}}\underline{\tilde{\mu}}^\Finv) - \nu \underline{\tilde{\check{k}}}_n-\iota_{\underline{z}}\nu \tilde{\check{k}} - x^\Finv \underline{\tilde{\mu}}^\Finv + [\epsilon_n y^\Finv,\underline{\tilde{y}}^\Finv]\\
            &\phi_2^*(\tilde{\xi}^\Finv_n)=\tilde{\xi}^\Finv_n + \iota_{\tilde{X}}\tilde{c}^\Finv_n \tilde{\mu}^\Finv + d(\epsilon_n \tau^\Finv \tilde{\mu}^\Finv) + \iota_z \tilde{c}^\Finv_n \nu + \iota_z \nu [\epsilon_n, \tilde{\check{k}}\tilde{\mu}^\Finv]+(d_{\tilde{\omega}_n}\epsilon_n)\tilde{\check{k}}\tilde{\mu}^\Finv + \iota_{\tilde{X}}\tilde{\mu}^\Finv[\epsilon_n,\tilde{\check{k}}\tilde{\mu}^\Finv]\\
            &\phi_2^*(\tilde{\xi}^\Finv)=\tilde{\xi}^\Finv + \tilde{c}^\Finv_n \nu + d_{\hat{\omega}}\epsilon_n (\tilde{\check{k}}\tilde{\mu}^\Finv) + \nu [\epsilon_n, \tilde{\check{k}}\tilde{\mu}^\Finv].
        \end{align*}
\end{lemma}
\begin{proof}
    Once again, we can refer to \cite{C25,canepa20254dpalatinicartangravityhamiltonian} and the proof therein. In particular, defining $\phi_2=\phi_2^{PC}+\phi_3$, where $\phi_2^{PC}$ is the symplectomorphism appearing in \cite{C25}, we have that $\phi_3=0$ on all fields except
        \begin{align*}
            &\phi_3^*(\tilde{v})=y^\Finv \qquad \quad \phi_3^*(\tilde{\psi}^\Finv_n)=-i\epsilon_n \gamma\tilde{\psi}\tilde{\mu}^\Finv\qquad \quad \phi_3^*(\tilde{e}_n^\Finv)=x^\Finv\tilde{\mu}^\Finv + [\epsilon_n y^\Finv,\tilde{\mu}^\Finv]
        \end{align*}
    where we recall $\mathfrak{Q}=\tilde{e}x^\Finv + \epsilon_n[\tilde{e},y^\Finv]$. Then, by \cite{C25}, we have 
        \begin{equation*}
            \phi_2^*\left(\varpi_{PC}^r +\int_{I\times\Sigma}\delta\tilde{v}\delta\underline{\tilde{v}}^\Finv \right) = \varpi^r_{PC}+ \int_{I\times\Sigma} \delta\tilde{v}\delta\underline{\tilde{v}}^\Finv + \delta\hat{\omega}\delta\underline{\tilde{v}}^\Finv + \delta\tilde{\bar{\psi}}\delta(\epsilon_n \gamma\psi \tilde{\mu}^\Finv) + \delta\mathfrak{Q}\delta\tilde{\mu}^\Finv,
        \end{equation*}
    where we have taken into account that the structural constraint that was used in \cite{C25} has been changed to take the gravitino interaction into consideration. 

    The term $\epsilon_n\delta \tilde{\psi}\gamma\tilde{\psi}\tilde{\mu}^\Finv$ is balanced by the term $\phi_3^*\left(\int_{I\times\Sigma}\delta\tilde{\psi}\delta\underline{\tilde{\psi}}_{n,\Finv}\right)=\phi_3^*(\varphi^r_{\mathrm{SG}}-\varpi^r_{PC})$ while 
        \begin{align*}
            \phi^*_3\left(\varpi_{PC}^r +\int_{I\times\Sigma}\delta v \delta \underline{\tilde{v}}^\Finv\right)=\int_{I\times\Sigma}\delta\tilde{e} \delta(x^\Finv\tilde{\mu}^\Finv + [\epsilon_n y^\Finv,\tilde{\mu}^\Finv]) + \delta y^\Finv \delta(\epsilon_n[\tilde{e},\tilde{\mu}^\Finv])= \int_{I\times\Sigma}-\delta\mathfrak{Q}\delta\tilde{\mu}^\Finv,
        \end{align*}
    balancing exactly the remaining term and showing the Lemma.\footnote{We remark we have used the property that $\tilde{e}\tilde{\mu}^\Finv=0$.} 
\end{proof}
\begin{proposition}\label{prop: split sg action}
    Let $\mathfrak{F}^H_{\mathrm{SG}}$ be the BV theory given by
    \begin{equation*}
        (\mathcal{F}^H_{\mathrm{SG}}, \varpi^H_{\mathrm{SG}},\mathcal{S}^H_{\mathrm{SG}}),
    \end{equation*}
    where
    \begin{equation*}
        \varpi_{\mathrm{SG}}^H=\varpi^r_{\mathrm{SG}}+\int_{I\times \Sigma}\delta\tilde{v}\delta\underline{\tilde{v}}^\Finv
    \end{equation*}
    and 
    \begin{equation}
        \mathcal{S}^H_{\mathrm{SG}}=S^r_{\mathrm{SG}}+\int_{I\times \Sigma}\frac{1}{2}\underline{\tilde{e}}_n \tilde{e}[\tilde{v}-y^\Finv,\tilde{v}-y^\Finv] + g(\tilde{v}^\Finv),
    \end{equation}
    with 
    \begin{equation}
        g(\tilde{v}^\Finv)=f(v^\Finv) + \left(\delta_\chi (\tilde{v} + \nu) + \mathbb{q}(\tilde{v}+ \nu)\right)\underline{\tilde{v}}^\Finv,
    \end{equation}
    having defined $\delta_\chi$ as the supersymmetry transformation and $\mathbb{q}$ as the Hamiltonian vector field of the rank--2 part of the BV action defined in \eqref{eq: quadratic BV action} and $f(\tilde{v}^\Finv)$ as in \cite{C25}.
    
    Then, letting $\Phi\coloneqq\phi_1\circ\phi_2$, we have $$\Phi^*(\mathcal{S}^H_{\mathrm{SG}})=\mathcal{S}_{\mathrm{SG}}.$$
\end{proposition}
The proposition will follow immediately from the following Lemmata.
\begin{lemma}\label{lem: phi_2 Sg}
    The symplectomorphism $\phi_2$ is such that
        \begin{equation*}
            \phi_2^*\left( \mathcal{S}^r_{\mathrm{SG}}+\int_{I\times\Sigma} \frac{1}{2}\underline{\tilde{e}}_n \tilde{e}[\tilde{v}-y^\Finv,\tilde{v}-y^\Finv] + g(\tilde{v}^\Finv)\right)=\mathcal{S}^r_{\mathrm{SG}}+\int_{I\times\Sigma} \frac{1}{2}\underline{\tilde{e}}_n \tilde{e}[\tilde{v},\tilde{v}] + h(\tilde{v}^\Finv),
        \end{equation*}
    where $h(\tilde{v}^\Finv)=g(\tilde{v}^\Finv)+ (\iota_{\tilde{\xi}}F_{\hat{\omega}} + F_{\tilde{\omega}_n}\tilde{\xi}^n + d_{\hat{\omega}}\tilde{c}+ \delta_\chi \hat{\omega}+ \mathbb{q}\hat{\omega})\tilde{v}^\Finv $.\footnote{Notice that $ h(\tilde{v}^\Finv)=Q_{\mathrm{SG}}(\hat{\omega}+\tilde{v}) \tilde{v}^\Finv$}
\end{lemma}

\begin{lemma}\label{lem: phi1 SG}
    The symplectomorphism $\phi_1$ is such that
        \begin{equation*}
            \phi_1^*\left( \mathcal{S}^r_{\mathrm{SG}}+\int_{I\times\Sigma} \frac{1}{2}\underline{\tilde{e}}_n \tilde{e}[\tilde{v},\tilde{v}] + h(\tilde{v}^\Finv)\right)=\mathcal{S}_{\mathrm{SG}}
        \end{equation*}
\end{lemma}
\begin{proof}
    The proofs of the above lemmata are found in Appendix \ref{proofs of BVPFSG}
\end{proof}

\begin{remark}
    So far, we have showed that the BV theory $\mathfrak{F}^H_{\mathrm{SG}}$ is BV-equivalent to the full BV theory. In particular, this implies that the BV action is equivalent to the reduced BV action with the addition of the terms depending on $\tilde{v}^\Finv$ and $\tilde{v}$. Ultimately, we are interested in integrating out the field $\tilde{v}$, which is responsible for the singularity of the pre-symplectic form induced on the boundary by the full BV action. Therefore, obtaining a decoupling of $S_r$ from the dynamics of $\tilde{v}$ is fundamental. 

    We start by noticing that we have a fiber bundle $\mathcal{F}_{\mathrm{SG}}^H\rightarrow\mathcal{F}_{\mathrm{SG}}^r$ whose fiber is locally given by $\mathcal{F}_f\coloneq T^*[1](\ker{W_{\tilde{e}_0}^{(1,2}})$, for a reference non--degenerate coframe $e_0$. Furthermore we obtain a BV bundle
    \[\begin{tikzcd}
	{\mathcal{F}_{\mathrm{SG}}^H} && {\tilde{\mathcal{F}}_{\mathrm{SG}}^H} \\
	\\
	{} & {\mathcal{F}_{\mathrm{SG}}^r}
	\arrow["\Xi", from=1-1, to=1-3]
	\arrow["{\pi^H}"', from=1-1, to=3-2]
	\arrow["{\tilde{\pi}^H}", from=1-3, to=3-2]
    \end{tikzcd}\]
    where $\tilde{\mathcal{F}}^H_{\mathrm{SG}}\coloneqq \mathcal{F}^r_{\mathrm{SG}}\times F_f$ is a product of -1--symplectic manifolds. The symplectomorphism $\Xi$ is found by noticing that at any spacetime point $x\in M$, we can find a unique orthogonal transformation $\Lambda_x$ such that $\tilde{e}_0=\Lambda_x \tilde{e}$, hence providing an isomorphism  $$\ker{W_{\tilde{e}_0}^{(1,2)}}=\{ \tilde{v}\in\Omega_\partial^{(1,2)}\hspace{1mm}|\hspace{1mm} \forall x\in M, \hspace{1mm} \Lambda_x \tilde{e}_x \tilde{v}_x= 0  \}\simeq \ker{W_{\tilde{e}}^{(1,2)}}.$$ This extends as a symplectomorphism $\mathcal{F}_f=T^*[1]\ker{W_{\tilde{e}_0}^{(1,2)}}\simeq T^*[1]\ker{W_{\tilde{e}}^{(1,2)}}$. Therefore, considering an open $\mathcal{U}\subset\mathcal{F}_{\mathrm{SG}}^r$, with the local trivialization of $\mathcal{F}_{\mathrm{SG}}^H\rightarrow\mathcal{F}_{\mathrm{SG}}^r$ is given by $$\mathcal{F}_{\mathrm{SG}}^H\bigg|_{\mathcal{U}}\simeq \mathcal{U}\times T^*[1]\ker{W_{\tilde{e}}^{(1,2)}},$$ we see that the above symplectomorphism induces  $\Xi:\mathcal{F}^H_{\mathrm{SG}}\rightarrow\tilde{\mathcal{F}}^H_{\mathrm{SG}}$. 
\end{remark}

\begin{theorem}
        \begin{equation*}
            \tilde{\mathcal{P}}_{\mathcal{L}_f}:\mathrm{Dens}^{\frac{1}{2}}(\mathcal{F}_{\mathrm{SG}})\rightarrow\mathrm{Dens}^{\frac{1}{2}}(\mathcal{F}^r_{\mathrm{SG}})
        \end{equation*}
    defined by the composition 
        \begin{equation*}
            \mathrm{Dens}^{\frac{1}{2}}(\mathcal{F}_{\mathrm{SG}})\overset{\Phi^*}{\longrightarrow}\mathrm{Dens}^{\frac{1}{2}}(\mathcal{F}^H_{\mathrm{SG}})\overset{\mathcal{P}_{\mathcal{L}_f}}{\longrightarrow}\mathrm{Dens}^{\frac{1}{2}}(\mathcal{F}_{\mathrm{SG}}^r),
        \end{equation*}
    wehere $\Phi=\phi_2\circ\phi_1$ as defined in proposition \ref{prop: split sg action} and $\mathcal{P}_{\mathcal{L}}$ is the BV pushforward of the BV bundle $\mathcal{F}^H_{\mathrm{SG}}\rightarrow\mathcal{F}^r_{\mathrm{SG}}$ along the Lagrangian submanifold $\mathcal{L}_f\coloneqq\{ (\tilde{v},\tilde{v}^\Finv)\in\mathcal{F}_f\hspace{1mm}|\hspace{1mm}\tilde{v}^\Finv=0 \}$. 

    Then 
        \begin{equation*}
            \mu_r^{\frac{1}{2}}\exp\left( \frac{i}{\hbar}\mathcal{S}^r_{\mathrm{SG}} \right)=\tilde{P}_{\mathcal{L}}\mu^{\frac{1}{2}}\exp\left( \frac{i}{\hbar}\mathcal{S}_{\mathrm{SG}} \right)
        \end{equation*}
\end{theorem}
\begin{proof}
    The proof amounts to showing that $\mu_r^{\frac{1}{2}}\exp\left( \frac{i}{\hbar}\mathcal{S}^r_{\mathrm{SG}} \right)={P}_{\mathcal{L}}\mu^{\frac{1}{2}}\exp\left( \frac{i}{\hbar}\mathcal{S}^H_{\mathrm{SG}} \right)$. This is a consequence of the fact that on $\mathcal{L}_f$ we have
        \begin{equation*}
            \mathcal{S}^H_{\mathrm{SG}}\bigg\vert_{\mathcal{L}_f}=\mathcal{S}^r_{\mathrm{SG}}+\int_{I\times\Sigma}\frac{1}{2}\tilde{e}_n \tilde{e}[\tilde{v}-y^\Finv,\tilde{v}-y^\Finv]. 
        \end{equation*}
    and that the quadratic form $\frac{1}{2}\tilde{e}_n\tilde{e}[-,-]:\Omega^{(1,2)}_\partial\times\Omega^{(1,2)}_\partial\rightarrow\mathcal{C^1}(M)$ is non--degenerate \cite{C25,canepa20254dpalatinicartangravityhamiltonian}, hence producing a well-defined Gaussian integral
        \begin{equation*}
            \int_{\mathcal{L}_f}\mu_{f}^{\frac{1}{2}}\exp\left( \frac{i}{\hbar} \int_{I\times\Sigma}\frac{1}{2}\tilde{e}_n \tilde{e}[\tilde{v}-y^\Finv,\tilde{v}-y^\Finv]\right),
        \end{equation*}
    which contributes to ${P}_{\mathcal{L}}\mu^{\frac{1}{2}}\exp\left( \frac{i}{\hbar}\mathcal{S}^H_{\mathrm{SG}} \right)$ just with a factor. 
\end{proof}

\section{The BFV description of $N=1,D=4$ supergravity}\label{sec: BFV SG}

So far, we worked on a cylinder to obtain the reduced BV action of $N=1,D=4$ supergravity. Such theory is a suitable candidate to obtain a BFV structure on the boundary, since the induced boundary symplectic form is now non-degenerate, hence solving the original issue outlined at the beginning of the previous section. 

Traditionally, one would need to compute the boundary potential 1--form $\vartheta^\Sigma_{\mathrm{SG}}$ arising as a boundary term from 
    \begin{equation*}
        \delta S_{\mathrm{SG}}^r=\iota_{Q^r_{\mathrm{SG}}} \varpi^r_{\mathrm{SG}} +({\pi}^{r,\Sigma}_{\mathrm{SG}})^*(\vartheta^{r,\Sigma}_{\mathrm{SG}}),
    \end{equation*}
where ${\pi}^{r,\Sigma}_{\mathrm{SG}}:\mathcal{F}^r_{\mathrm{SG}}\rightarrow \mathcal{F}^{r,\Sigma}_{\mathrm{SG}} $ is the surjective submersion to the space of boundary fields. 
In particular, we notice that, from the variation of $S^H_{\mathrm{SG}}$, we have
    \begin{equation*}
        \delta S_{\mathrm{SG}}^H=\iota_{Q^H_{\mathrm{SG}}} \varpi^H_{\mathrm{SG}}+({\pi}^\Sigma_{\mathrm{SG}})^*\left(\vartheta^{r,\Sigma}_{\mathrm{SG}} +\int_{\Sigma} \tilde{\xi}^n\tilde{v}^\Finv\delta\tilde{v}\right).
    \end{equation*}

The induced action on the boundary is obtained as the boundary term given by the (failure of) the CME in the bulk
    \begin{equation*}
        \frac{1}{2}\iota_{Q_{\mathrm{SG}}}\iota_{Q_{\mathrm{SG}}}\varpi^M_{\mathrm{SG}}=\mathcal{S}^{r,\Sigma}_{\mathrm{SG}} + \mathcal{S}^\Sigma_f,
    \end{equation*}
where $\mathcal{S}^\Sigma_{\mathrm{SG}}$ is exactly the boundary BFV action we are seeking, given as $\mathcal{S}^\Sigma_{\mathrm{SG}}=\frac{1}{2}\iota_{Q^r_{\mathrm{SG}}}\iota_{Q^r_{\mathrm{SG}}}\varpi^r_{\mathrm{SG}}$, while $\mathcal{S}^\Sigma_f$ can be seen as the Hamiltonian generating the gauge transformations of $v$, defining a coisotropic submanifold.\footnote{Indeed by construction one has $\{\mathcal{S}^\Sigma_f,\mathcal{S}^\Sigma_f\}=0$, where $\{-,-\}$ is the Poisson bracket induced by $\varpi^\Sigma_f=\int_\Sigma\delta(\tilde{\xi}^n\tilde{v}^\Finv)\delta\tilde{v} $.} 
In particular, one sees that
    \begin{equation}\label{eq: Hamiltonian v}
        \mathcal{S}^\Sigma_f=\int_\Sigma \left(\mathrm{L}_{\tilde{\xi}}^{\hat{\omega}}\tilde{v} +\iota_z\tilde{v}d\tilde{\xi}^n -[\tilde{c},\tilde{v}] + \delta_\chi v + \mathbb{q}\tilde{v}  \right)\tilde{v}^\Finv\tilde{\xi}^n +\frac{1}{2}\epsilon_n \tilde{\xi}^n \tilde{e} [\tilde{v},\tilde{v}]
    \end{equation}

This is not significantly relevant at the moment, as we are are interested in $\mathcal{S}^{r,\Sigma}_{\mathrm{SG}}$. In particular, we know from Section \ref{subsec: BFV attempt} that a BFV structure $\mathfrak{F}^\Sigma\coloneq(\mathcal{F}^\Sigma_{\mathrm{SG}}, \mathcal{S}^\Sigma_{\mathrm{SG}}, \varpi^\Sigma_{\mathrm{SG}})$ must exist such that $H^0_{Q^\Sigma_{\mathrm{SG}}}(\mathcal{F}^\Sigma_{\mathrm{SG}})\simeq\mathcal{C^1}(\underline{C}^\Sigma_{\mathrm{SG}}). $ The goal is then to find a symplectomorphsim
    \begin{equation*}
        (\mathcal{F}^\Sigma_{\mathrm{SG}}, \varpi^\Sigma_{\mathrm{SG}})\overset{\Psi_r}{\longrightarrow}(\mathcal{F}^{r,\Sigma}_{\mathrm{SG}}, \varpi^{r,\Sigma}_{\mathrm{SG}})
    \end{equation*}
and then define $\mathcal{S}^\Sigma_{\mathrm{SG}}\coloneq\Psi_r^*\left( \mathcal{S}^{\Sigma,r}_{\mathrm{SG}}\right)$, as we know that it satisfies the CME by construction.

\subsection{Symplectomorphism via 1--D AKSZ}
In this subsection, we employ the methods from the 1--D AKSZ construction described in Appendix \ref{sec: AKSZ theory} to obtain the following -1--symplectic supermanifold
    \begin{align*}
        &\mathcal{F}^{AKSZ}_{\mathrm{SG}}\coloneqq \mathrm{Map}(T[1]I,\mathcal{F}^\Sigma_{\mathrm{SG}})\\
        &\varpi^{AKSZ}_{\mathrm{SG}}\coloneq \mathfrak{T}^{(2)}_I(\varpi^\Sigma_{\mathrm{SG}}).
    \end{align*}
and look for a symplectomorphism $\Phi_r\colon (\mathcal{F}^{r}_{\mathrm{SG}},\varpi^{r}_{\mathrm{SG}})\rightarrow(\mathcal{F}^{AKSZ}_{\mathrm{SG}},\varpi^{AKSZ}_{\mathrm{SG}})$. 

For starters we define the AKSZ fields 
    \begin{align}\label{eq: AKSZ SG variable def}
        &\nonumber\mathfrak{e}=e+f^\Finv & \mathfrak{w}=\omega+ u^\Finv\\
        &\nonumber \mathfrak{p} = \psi + \varsigma^\Finv & \bar{\mathfrak{p}}=\bar{\psi}+\bar{\varsigma}^\Finv\\
        &\nonumber\mathfrak{x}= \chi + \epsilon &\bar{\mathfrak{x}}=\bar{\chi}+ \bar{\epsilon}\\
        &\mathfrak{c}= c + w & \mathfrak{z}= \xi + z\\
        &\nonumber\mathfrak{l}= \lambda + \mu & \mathfrak{c}^\Finv= k^\Finv + c^\Finv\\
        &\nonumber\mathfrak{x}^\Finv= \theta^\Finv + \chi^\Finv &\bar{\mathfrak{x}}^\Finv=\bar{\theta}^\Finv + \bar{\chi}^\Finv\\
        &\nonumber\mathfrak{y}^\Finv= e^\Finv + y^\Finv,
    \end{align}
having used the splitting $\mathcal{F}^{AKSZ}_{\mathrm{SG}}\coloneqq \Omega^\bullet(I)\times \mathcal{F}^\Sigma_{\mathrm{SG}}$. Recalling \eqref{eq: split AKSZ fields}, we have
    \begin{align*}
        & e\in \mathcal{C^1}(I)\otimes\Omega^{(1,1)}_\partial & f^\Finv \in \Omega^1[-1](I) \otimes\Omega^{(1,1)}_\partial  \\
        & \omega \in \mathcal{C^1}(I) \otimes \Omega^{(1,2)}_\partial &u^\Finv \in \Omega^1[-1](I) \otimes \Omega^{(1,2)}_\partial \\
        & \psi \in \mathcal{C^1}(I) \otimes \Omega^{(1,0)}_\partial(\Pi\mathbb{S}_M) & \varsigma
        ^\Finv\in \Omega^1[-1](I) \otimes \Omega^{(1,0)}_\partial(\Pi\mathbb{S}_M)\\
        &\chi \in  \mathcal{C^1}(I)\otimes \Omega^{(0,0)}_\partial[1](\Pi \mathbb{S}_M) &\epsilon \in  \Omega^1[-1](I) \otimes  \Omega^{(0,0)}_\partial[1](\Pi \mathbb{S}_M) \\
        &\xi \in  \mathcal{C^1}(I)\otimes \mathfrak{X}[1](\Sigma) &z \in  \Omega^1[-1](I) \otimes \mathfrak{X}[1](\Sigma)  \\
        &c\in  \mathcal{C^1}(I)\otimes \Omega^{(0,2)}_\partial[1] & w\in \Omega^1[-1](I) \otimes  \Omega^{(0,2)}_\partial[1]\\
        &\lambda \in \mathcal{C^1}(I)\otimes \mathcal{C^1}[1](\Sigma) &\mu \in \Omega^1[-1](I) \otimes \mathcal{C^1}[1](\Sigma)   \\
        & k^\Finv \in \mathcal{C^1}(I)\otimes \Omega^{(3,2)}_\partial[-1] & c^\Finv \in \Omega^1[-1](I) \otimes \Omega^{(3,2)}_\partial[-1] \\
        & \theta^\Finv\in \mathcal{C^1}(I)\otimes \Omega^{(3,4)}_\partial[-1](\Pi\mathbb{S}_M) & \chi^\Finv \in \Omega^1[-1](I) \otimes  \Omega^{(3,4)}_\partial[-1](\Pi\mathbb{S}_M)\\ 
        & \eta^\Finv\in \mathcal{C^1}(I)\otimes\Omega^{(3,3)}_\partial[-1] & e^\Finv \in \Omega^1[-1](I) \otimes \Omega^{(3,3)}_\partial[-1]
    \end{align*}
where we have omitted the $\sim$ symbol for boundary fields, as in this section there is not ambiguity of meaning.%\footnote{Note that, starting from \ref{sec: lengthy proofs}, we will reintroduce the $\sim$ notation, and the "untilded" fields will refer to the bulk.}

The AKSZ symplectic form is then given by
    \begin{equation*}
    \begin{split}
        \varpi_{\mathrm{SG}}^{AKSZ}=\varpi_{PC}^{AKSZ}+\int_{I\times \Sigma}&\frac{1}{3!}\left( {\underline{\bar{\varsigma}}^\Finv\gamma^3 \delta\psi \delta e } +  {\bar{\psi}\gamma^3 \delta\underline{\varsigma}^\Finv \delta e} +  {\bar{\psi}\gamma^3\delta\psi \delta\underline{f}^\Finv} +  {\underline{f}^\Finv \delta\bar{\psi}\gamma^3 \delta\psi } \right)\\
        & + \frac{1}{3} { e \delta\bar{\psi}\gamma^3 \delta\underline{\varsigma}^\Finv} + i  {\delta\underline{\bar{\epsilon}}\delta\theta^\Finv }+  {i \delta\bar{\chi}\delta\underline{\chi}^\Finv} + i\delta\underline{\bar{\varsigma}}( {\iota_{\delta\xi} \theta^\Finv} + { \iota_\xi \delta\theta^\Finv} ) \\
        & + i \delta\bar{\psi}\left( { \iota_{\delta\underline{z}}\theta^\Finv }+  {\iota_{\underline{z}}\delta\theta^\Finv} + {\iota_{\delta\xi} \underline{\chi}^\Finv} + { \iota_\xi \delta\underline{\chi}^\Finv}  \right) 
    \end{split}
    \end{equation*}
\begin{proposition}\label{prop: AKSZ sympl}
    There exists a symplectomorphism $\Phi_r\colon\mathcal{F}^r_{\mathrm{SG}}\rightarrow\mathcal{F}^{AKSZ}_{\mathrm{SG}}$. In particular, it acts on the fields as
    \begin{align*}
        &\Phi_r^*( {e} ) = e + \lambda\mu^{-1} f^\Finv &\Phi_r^*(\un{ {e}}_n )= \mu \epsilon_n + \iota_{\un{z}} e + \lambda \epsilon_n^i \un{f}^\Finv_i\\
        &\Phi_r^*( {\omega}) = \omega - \lambda\mu^{-1} u^\Finv &\Phi_r^*(\un{ {\omega}}_n)=\un{w} - \iota_\xi \un{u}^\Finv -\lambda \epsilon_n^i \un{u}_i\\
        &\Phi_r^*( {\psi}) = \psi + \lambda\mu^{-1} \varsigma^\Finv &\Phi_r^*(\un{\psi}_n)=\un{\epsilon} -\iota_\xi \un{\varsigma}_dag + \lambda\mu^{-1}\iota_{\un{z}}\varsigma^\Finv  \\
        &\Phi_r^*(\psi^\Finv)=\theta^\Finv  & \Phi_r^*(\un{ {\psi}}^\Finv_n)=\iota_{\un{z}}\theta^\Finv + \iota_\xi \un{\chi}_{\Finv} -\frac{e}{3}\gamma^3\un{\varsigma}^\Finv - \frac{1}{3}\un{f}^\Finv\gamma^3\psi + \frac{1}{3}\lambda\mu^{-1} \un{f}\gamma^3\varsigma^\Finv  \\
        & \Phi_r^*(\chi)=\chi + \lambda\mu^{-1}\iota_\xi \varsigma^\Finv & \Phi_r^*(\un{\chi}^\Finv_n)=\un{\chi}_{\Finv}\\
        &\Phi_r^*( c)=c-\lambda\mu^{-1} \iota_\xi( u^\Finv) &\Phi_r^*(\un{ {c}}^\Finv_n)=\un{c}^\Finv\\
        & \Phi_r^*( {\omega}^\Finv)= k^\Finv & \Phi_r^*(\un{ {\omega}}^\Finv_n)= e\un{f}^\Finv + \iota_{\un{z}}k^\Finv + \iota_\xi \un{c}^\Finv\\
        &\Phi_r^*( {\xi}^i)=\xi^i - \lambda\mu^{-1} z^i &\Phi_r^*(\un{ {\xi}}^\Finv)= e \un{y}^\Finv + \un{f}^\Finv e^\Finv - \un{u}^\Finv k^\Finv + \un{c}^\Finv \lambda\mu^{-1} u^\Finv+i\un{\bar{\varsigma}^\Finv}\theta^\Finv  -  i \lambda\mu^{-1}\bar{\varsigma}^\Finv \un{\chi}^\Finv  \\
        &\Phi_r^*( {e}^\Finv)=e^\Finv - \lambda \mu^{-1}y^\Finv & \Phi_r^*( {\xi}^n)=\lambda\mu^{-1}
    \end{align*}
    \begin{align*}
        \Phi_r^*(\un{ {e}}^\Finv_n)=&e\un{u}^\Finv + \iota_{\un{z}}e^\Finv - \lambda\epsilon_n^i y^\Finv_i + \lambda\mu^{-1}f^\Finv \un{u}^\Finv -\frac{1}{3!}\un{\bar{\varsigma}}^\Finv\gamma^3\psi  - \frac{1}{3!}\lambda\mu^{-1} \un{\bar{\varsigma}}^\Finv\gamma^3\varsigma^\Finv   \\
        \Phi_r^*(\un{ {\xi}}^\Finv_n)=& e_n\un{y}^\Finv + e f^\Finv \un{u}\Finv + f^\Finv \iota_{\un{z}}e^\Finv + u^\Finv \iota_{\un{z}}k^\Finv + c^\Finv \lambda\epsilon_n^i \un{u}^\Finv_i -\frac{1}{3!}\un{f}^\Finv\bar{\psi}\gamma^3\un{\varsigma}^\Finv \\
        &- \frac{1}{3!}e\bar{\un{\varsigma}}^\Finv\gamma^3\un{\varsigma}^\Finv+ i \iota_{\un{z}}\bar{\varsigma}^\Finv \theta^\Finv - i \lambda\mu^{-1}\iota_{\un{z}}\bar{\varsigma}^\Finv{\chi}^\Finv  
    \end{align*}
    
\end{proposition}
\begin{proof}
    We leave this long calculation for Appendix \ref{proof: AKSZ SYMPL}.    
\end{proof}

With the above isomorphism, we can define $\Psi_r$ by restricting it to the fields on the boundary and define the boundary BFV action as $\mathcal{S}_{\mathrm{SG}}^\Sigma\coloneq (\Psi_r)^*\mathcal{S}^{\Sigma,r}_{\mathrm{SG}} $ and 
    \begin{equation*}
        S^{AKSZ}_{\mathrm{SG}}\coloneqq\mathfrak{T}_I^{(0)}(\mathcal{S}^\Sigma_{\mathrm{SG}}) + \iota_{d_I}\mathfrak{T}_I^{(1)}(\alpha^\Sigma_{\mathrm{SG}}).
    \end{equation*}
By construction one has
    \begin{align*}
        &\mathcal{S}^\Sigma_{\mathrm{SG}}=\frac{1}{2}\iota_{Q^{AKSZ}_{\mathrm{SG}}}\iota_{Q^{AKSZ}_{\mathrm{SG}}}\varpi^{AKSZ}_{\mathrm{SG}}\\
        &\mathcal{S}^{\Sigma,r}_{\mathrm{SG}}=\frac{1}{2}\iota_{Q^r_{\mathrm{SG}}}\iota_{Q^r_{\mathrm{SG}}}\varpi^{r}_{\mathrm{SG}},
    \end{align*}
but since $\varpi^{AKSZ}_{\mathrm{SG}}=\Phi^*_r(\varpi^{r}_{\mathrm{SG}})$ and $\mathcal{S}^\Sigma_{\mathrm{SG}}=\Phi^*_r\left(\mathcal{S}^{\Sigma,r}_{\mathrm{SG}}\right)$, then one must have $Q^{AKSZ}_{\mathrm{SG}}=(\Phi_r)_*Q^r_{\mathrm{SG}}$.

At the same time
    \begin{align*}
        \delta S^{AKSZ}_{\mathrm{SG}}&=\iota_{Q^{AKSZ}_{\mathrm{SG}}}\varpi^{AKSZ}_{\mathrm{SG}} + \pi^*\vartheta_\Sigma\\
       &=\Phi_r^*\left(\iota_{Q^r_{\mathrm{SG}}}\varpi^r_{\mathrm{SG}}  \right)+ \pi^*\vartheta_\Sigma\\
       &= \Phi_r^*\left( \delta S^r_{\mathrm{SG}} - \pi^*(\vartheta^r_{\mathrm{SG}})  \right)+ \pi^*\vartheta_\Sigma.
    \end{align*}
Now, defining $\Psi_r$ such that $\Phi_r^*\circ \pi^* = \pi^* \circ \Psi_r $, and noticing that $\delta$ commutes with the pullback, we have 
    \[
    \delta S^{AKSZ}_{\mathrm{SG}}= \delta(\Phi_r^* S^r_{\mathrm{SG}}) + \pi^*\left( \vartheta_\Sigma - \Psi^*_r \vartheta^r_{\mathrm{SG}} \right)
    \]
Taking the variation of the above expression gives $\delta\vartheta_\Sigma - \delta\Psi^*_r(\vartheta^r_{\mathrm{SG}})=0$, implying
\begin{equation*}
    \Psi_r^*(\varpi^{r,\Sigma}_{\mathrm{SG}})=\varpi^\Sigma_{\mathrm{SG}},
\end{equation*}
which is exactly the symplectomorphism we were seeking on the space of BFV boundary fields.
\begin{remark}
    We notice that, in Section \ref{subsec: BFV attempt} we defined the BFV space of fields $\mathcal{F}^\Sigma_{\mathrm{SG}}$ as the space of boundary fields subject to the structural constraint 
        \begin{equation*}
            \epsilon_n\left( d_\omega e - \frac{1}{2}\bar{\psi}\gamma\psi \right) + \cdots = e\sigma,
        \end{equation*}
    where the terms $(\cdots)$ were left undefined to account for contribution that would render the structural constraint invariant with respect to the cohomological vector field $Q^\Sigma_{\mathrm{SG}}$, which was yet to be obtained in Section \ref{subsec: BFV attempt}. In principle, when defining $\mathcal{F}^{AKSZ}_{\mathrm{SG}}$, the structural constraints splits into a tangent constraint to $\Sigma$ and a part containing the transversal component along the interval $I$. In order to overcome the ambiguity in the definition of $\mathcal{F}^{\mathrm{SG}}_{AKSZ}$, we simply define the structural constraints to be
        \begin{align}
            &\Phi_r^*(\epsilon_n \check{k}-e \check{a})=0\label{AKSZ constr1}\\
            &\Phi^*_r\left[ \epsilon_n\left( d_\omega e - \frac{1}{2}\bar{\psi}\gamma\psi  -\tau^\Finv d\xi^n  \right) + \iota_X k^\Finv + \mathfrak{Q} - e\sigma \right]=0\label{AKSZ constr2}.
        \end{align}
    In particular, we see that \eqref{AKSZ constr1} is automatically satisfied by construction. Indeed, \eqref{AKSZ constr1} is equivalent to $\mathfrak{W}^\Finv\in\mathrm{Im}(W_{e}^{\partial(1,1)})$, therefore, recalling the definition \eqref{eq: BV PC constr omega dag} of $\mathfrak{W}^\Finv$, we have    
    \begin{align*}
        &\Phi_r^*\left({ {\omega}}_n^\Finv - \iota_{{z}} {\omega}^\Finv - \iota_{ {\xi}}{ {c}}_n^\Finv + \iota_{{z}}{ {c}}_n^\Finv{\xi}^n\right)=\\
        &=ef^\Finv +\iota_z k^\Finv + \iota_\xi c^\Finv - z^ik^\Finv_i -\iota_\xi c^\Finv - z^i \lambda\mu^{-1} c^\Finv +  z^i \lambda\mu^{-1} c^\Finv \\
        &=ef^\Finv.
    \end{align*}
    This tells us that the only relevant constraint in the space of AKSZ fields is given by \eqref{AKSZ constr2}, which in turn splits into the structural constraint of $\mathcal{F}^\Sigma_{\mathrm{SG}}$, simply given as
        \begin{equation}
            \epsilon_n\left( d_\omega e - \frac{1}{2}\bar{\psi}\gamma\psi    \right) + \iota_X k^\Finv + \mathfrak{Q}' = e\sigma
        \end{equation}
    and the part proportional to $\lambda\mu^{-1}$, which is given by
        \begin{equation*}
            \epsilon_n\left( [u^\Finv,e] + d_\omega f^\Finv - \bar{\psi}\gamma\varsigma^\Finv \right) + \left( \mathrm{L}_z^\omega\epsilon_n + [\iota_\xi u^\Finv-w,\epsilon_n]  \right)^i k^\Finv_i + \iota_X c^\Finv + (X^i f^\Finv_i)^j k^\Finv_j + \mathfrak{U}=^\Finv\sigma +  e B,
        \end{equation*}
    having defined $\Phi^*_r(\mathfrak{Q})=\mathfrak{Q}'+\lambda\mu^{-1}\mathfrak{U}$. The above two equations then define respectively the tangential and transversal AKSZ constraints. 
\end{remark}

\section{A short summary}

In this paper we investigated the boundary and phase space structure of
$N=1$, $D=4$ supergravity in the Palatini--Cartan formalism within the
BV--BFV framework. The main goal was to describe the reduced phase space
of the theory and to relate it consistently to the BV formulation in the
bulk.

We first reviewed the BV description of $N=1$, $D=4$ supergravity,
recalling that the bulk BV action correctly encodes diffeomorphism
invariance, local Lorentz symmetry, and local supersymmetry. We then
analyzed the classical boundary structure using the
Kijowski--Tulczijew construction. This allowed us to identify the space
of pre-boundary fields, the induced presymplectic form, and the
geometric phase space. Under suitable regularity assumptions on the
boundary metric, we derived the boundary constraints by restricting the
Einstein, torsion, and Rarita--Schwinger equations to the boundary.

A detailed  analysis showed that these constraints form a
first-class system. Their Hamiltonian vector fields generate,
respectively, diffeomorphisms, internal Lorentz transformations, and
local supersymmetry. As a consequence, the space of Cauchy data is a
coisotropic submanifold, and its coisotropic reduction defines the
reduced phase space of the theory. Since this reduced space proves challenging to study with traditional methods, our aim was to employ the BFV formalism to obtain a
cohomological resolution of the algebra of observables.

A key difficulty addressed in this work is that the BV theory of
supergravity does not directly induce a compatible BFV theory on the
boundary, due to obstructions analogous to those appearing in
Palatini--Cartan gravity. To overcome this issue, we used the BV
pushforward. Working on a cylindrical spacetime $M = I \times \Sigma$,
we showed that the original BV theory is equivalent, in the BV sense, to
a reduced BV theory in which the obstructing degrees of freedom are
eliminated. This reduced BV theory is BV-BFV compatible and admits a
well-defined BFV extension on the boundary.

The diagram below summarizes these relations. 

% https://q.uiver.app/#q=WzAsNyxbMCwxLCJcXG1hdGhjYWx7Rn1fe1xcbWF0aHJte1NHfX0iXSxbNCwxLCJcXG1hdGhjYWx7Rn1eXFxTaWdtYV97XFxtYXRocm17U0d9fSJdLFsyLDIsIlxcbWF0aGNhbHtGfV57XFxtYXRocm17QUtTWn19X3tcXG1hdGhybXtTR319Il0sWzEsMCwiRl97XFxtYXRocm17U0d9fSJdLFszLDAsIkZeXFxTaWdtYV97XFxtYXRocm17U0d9fSJdLFsxLDMsIlxcbWF0aGNhbHtGfV5yX3tcXG1hdGhybXtTR319Il0sWzMsMywiXFxtYXRoY2Fse0Z9XntcXFNpZ21hLHJ9X3tcXG1hdGhybXtTR319Il0sWzAsMSwiXFx0ZXh0e0JWLUJGVn0iLDEseyJzdHlsZSI6eyJib2R5Ijp7Im5hbWUiOiJkYXNoZWQifX19XSxbMSwyLCJcXHRleHR7MUQgQUtTWn0iLDEseyJjdXJ2ZSI6MX1dLFsyLDEsIlxcdGV4dHtCVi1CRlZ9IiwxLHsiY3VydmUiOjF9XSxbNCwxLCJcXHRleHR7QkZWfSIsMSx7ImN1cnZlIjotMSwic3R5bGUiOnsiYm9keSI6eyJuYW1lIjoiZGFzaGVkIn19fV0sWzMsNCwiXFx0ZXh0e0tUfSIsMSx7ImN1cnZlIjotMX1dLFsyLDUsIlxcUGhpX3IiLDEseyJzdHlsZSI6eyJ0YWlsIjp7Im5hbWUiOiJhcnJvd2hlYWQifX19XSxbNSw2LCJcXHRleHR7QlYtQkZWfSIsMSx7ImN1cnZlIjoxfV0sWzEsNiwiXFxQc2lfciIsMSx7ImN1cnZlIjotMiwic3R5bGUiOnsidGFpbCI6eyJuYW1lIjoiYXJyb3doZWFkIn0sImJvZHkiOnsibmFtZSI6ImRvdWJsZSBiYXJyZWQifX19XSxbMywwLCJcXHRleHR7QlZ9IiwxLHsiY3VydmUiOjF9XSxbMCw1LCJcXHRleHR7QlYgcGZ9IiwxLHsiY3VydmUiOjIsInN0eWxlIjp7ImhlYWQiOnsibmFtZSI6ImVwaSJ9fX1dXQ==
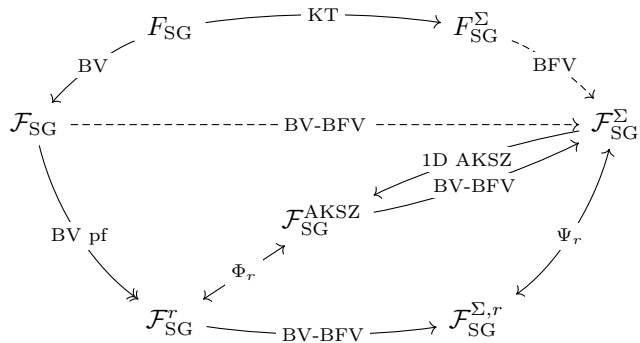
\[\begin{tikzcd}
	& {F_{\mathrm{SG}}} && {F^\Sigma_{\mathrm{SG}}} \\
	{\mathcal{F}_{\mathrm{SG}}} &&&& {\mathcal{F}^\Sigma_{\mathrm{SG}}} \\
	&& {\mathcal{F}^{\mathrm{AKSZ}}_{\mathrm{SG}}} \\
	& {\mathcal{F}^r_{\mathrm{SG}}} && {\mathcal{F}^{\Sigma,r}_{\mathrm{SG}}}
	\arrow["{\text{KT}}"{description}, curve={height=-6pt}, from=1-2, to=1-4]
	\arrow["{\text{BV}}"{description}, curve={height=6pt}, from=1-2, to=2-1]
	\arrow["{\text{BFV}}"{description}, curve={height=-6pt}, dashed, from=1-4, to=2-5]
	\arrow["{\text{BV-BFV}}"{description}, dashed, from=2-1, to=2-5]
	\arrow["{\text{BV pf}}"{description}, curve={height=12pt}, two heads, from=2-1, to=4-2]
	\arrow["{\text{1D AKSZ}}"{description}, curve={height=6pt}, from=2-5, to=3-3]
	\arrow["{\Psi_r}"{description},  curve={height=-12pt}, tail reversed, from=2-5, to=4-4]
	\arrow["{\text{BV-BFV}}"{description}, curve={height=6pt}, from=3-3, to=2-5]
	\arrow["{\Phi_r}"{description}, tail reversed, from=3-3, to=4-2]
	\arrow["{\text{BV-BFV}}"{description}, curve={height=6pt}, from=4-2, to=4-4]
\end{tikzcd}\]
The vertical arrows represent the passage from classical theories to their
BV or BFV counterparts, while the horizontal arrows describe the
restriction to the boundary. The dashed arrows indicate the failure of
a direct BV-BFV construction for the unreduced theory. The BV
pushforward provides the missing link, yielding a reduced BV theory that
fits into a consistent BV-BFV pair. The resulting BFV theory is shown
to be equivalent, via a symplectomorphism constructed using the
one-dimensional AKSZ formalism, to the BFV theory resolving the reduced
phase space.

\appendix
\section{Quick review of the AKSZ construction}\label{subsec: AKSZ theory}

A useful realization of the BV-BFV formalism can be obtained through a construction \cite{AKSZ} due to Aleksandrov, Kontsevich, Schwartz and Zaboronsky (AKSZ). It is particularly useful when tackling the problem of inducing a compatible BV structure in the bulk from a well defined BFV one on the boundary, in the case of cylindrical spacetimes. 

We first consider the finite-dimensional setting. Let $\mathcal{M}$ be a $\mathbb{Z}$--graded supermanifold, and $N$ be a regular manifold. The parity of the supermanifold $\mathcal{M}$ is usually set to be the grading modulo 2. When spinors are present, the parity is the sum of the grading and the intrinsic parity of the object at hand modulo 2. We assume there exist a degree n function $S$ on $\mathcal{M}$ and a non degenerate exact 2-form (hence symplectic) $\omega=d\alpha$, where $\alpha$ is of degree $n-1$, assuming $n>0$. With these assumptions, $\mathcal{M}$ is called a dg-Hamiltonian manifold if $\{S,S\}=0$, i.e. if and only if $Q\coloneqq\{S,\cdot\}$ is cohomological. 

\begin{definition}
Given the following diagram 
\[\begin{tikzcd}
	{\mathrm{Map}(T[1]N,\mathcal{M})\times T[1]N} && \mathcal{M} \\
	{\mathrm{Map}(T[1]N,\mathcal{M})}
	\arrow["{\mathrm{ev}}", from=1-1, to=1-3]
	\arrow["p"', from=1-1, to=2-1]
\end{tikzcd}\]
we define the transgression map
    \begin{equation*}
        \mathfrak{T}^{(\bullet)}_N\colon \Omega^\bullet(\mathcal{M})\longrightarrow \Omega^\bullet(\mathrm{Map}(T[1]N,\mathcal{M}))
    \end{equation*}
as $\mathfrak{T}^{(\bullet)}_N\coloneqq p_*\mathrm{ev}^* $, setting $p_*=\int_N \mu_N$ where $\mu_N$ is the canonical Berezinian on $T[1]N.$     
\end{definition}
\begin{remark}
    Notice that the differential on $N$ can be reinterpreted as a vector field on $\mathrm{Map}(T[1]N,\mathcal{M})$. Indeed, considering coordinates $(u^i,\xi^i)$ on $T[1]N$, we map $\xi^i\longmapsto du^i$ provides provides the isomorphism $\mathcal{C^1}(T[1]N)\simeq \Omega^\bullet(N)$, and $d=du^i \partial_i$ is the image of $\xi^i \partial_i \in \mathfrak{X}(T[1]N)$.
\end{remark}
\begin{theorem}[\cite{AKSZ}]\label{thm: AKSZ}
     Letting $\mathcal{F}^{AKSZ}\coloneqq\mathrm{Map}(T[1]N,\mathcal{M})$, $\varpi^{AKSZ}\coloneqq\mathfrak{T}^2_N(\omega)$ and $\mathcal{S}_{AKSZ}:=\mathfrak{T}^0_N(S) + \iota_{d_N}\mathfrak{T}^1_N(\alpha) $, the data
        \begin{equation*}
            \mathfrak{F}_{AKSZ}\coloneqq(\mathcal{F}^{AKSZ},\varpi^{AKSZ},\mathcal{S}_{AKSZ},Q_{AKSZ})
        \end{equation*}
    define a BV theory. 
\end{theorem}
To better understand the content of the above theorem, let us consider $f\in\mathcal{C^1}(\mathcal{M})$ and a homogeneous degree map $X\in\mathcal{F}_{AKSZ} $. Then we can define the composition $X_f =f\circ X \in \mathcal{C^1}(T[1]N)\simeq \Omega^\bullet(N)$. In particular, if $(u^i,\xi^i)$ are coordinates over $T[1]N$, we can see $X_f$ can be ralized as a differential form on $N$ as
    \begin{equation}
        X_f(u)=X_f(u)_{i_1\cdots i_k} du^{i_1}\wedge\cdots\wedge du^{i_k}.
    \end{equation}
If we let $(x^a)$ be coordinates on $\mathcal{M}$, we can compose the coordinate fucntions $x^a$ with $X$ and see $X^a(u)=X^a(u)_{i_1\cdots i_k} du^{i_1}\wedge\cdots\wedge du^{i_k}$. Then the AKSZ action and symplectic form take the form
    \begin{align*}
        &\mathcal{S}_{AKSZ}= \int_N S[X^a(u)] + \alpha_a(X(u))\wedge dX^a(u)\\
        & \varpi_{AKSZ}=\int_N \omega_{ab}(X(u))\delta X^a(u)\wedge \delta X^b(u).
    \end{align*}
\subsection{The 1--dimensional case and its relation to the boundary}\label{sec: AKSZ theory}
In this section we investigate how a BFV theory on $\Sigma$ can induce a BV theory on $I\times \Sigma,$ where $I$ is an interval.

Letting $\mathfrak{F}_\Sigma$ be the data of an exact BFV theory, as it appears after the KT construction, we notice that it automatically satisfies the definition of Hamiltonian dg manifold with $n=1$. We can therefore apply Theorem \ref{thm: AKSZ} to see that the data $\mathfrak{F}_{AKSZ}$ given by 
\begin{align*}
    &\mathcal{F}^{AKSZ}\coloneqq\mathrm{Map}(T[1]I,\mathcal{F}_\Sigma)\\
    &\varpi^{AKSZ}\coloneqq\mathfrak{T}^2_I(\varpi_\Sigma)\\
    &\mathcal{S}_{AKSZ}\coloneqq\mathfrak{T}^0_I(\mathcal{S}_\Sigma) + \iota_{d_I}\mathfrak{T}^1_I(\alpha_\Sigma) \\
    &Q_{AKSZ} \quad \mathrm{s.t.}\quad \iota_{Q_{AKSZ}}\varpi^{AKSZ}=\delta\mathcal{S}_{AKSZ}
\end{align*} yield a BV theory on $\Sigma\times I$. Observe that the target is infinite-dimensional, but the construction still holds.

The setting gets simplified when we notice that $\mathcal{F}^{AKSZ}\simeq \Omega^\bullet(I)\times\mathcal{F}_\Sigma$, i.e. we see that AKSZ fields are just boundary fields times sections of a graded vector space, where in particular
    \begin{equation}\label{eq: split AKSZ fields}
    \Omega^\bullet(I)=\mathcal{C^1}(I)\oplus\Omega^1(I)[-1],     
    \end{equation}
we obtain two fields for every boundary field in $\mathcal{F}_\Sigma$. In particular, if $\phi^I$ are fields on $\mathcal{F}_\Sigma$, we obtain that the AKSZ fields are $\Phi^I\coloneqq\phi^I(t)+\psi^I(t)dt$, where we point out that $\phi^I(t)\in\mathcal{C^1}(I)\times\mathcal{F}_\Sigma$. With this definition, the AKSZ action and symplectic form become
    \begin{align}
        &\varpi_{AKSZ}=\int_{I\times\Sigma } (\varpi_\Sigma[\Phi])_{IJ}\delta\phi^I(t)\wedge\delta\psi^J(t) dt\\
        &\mathcal{S}_{AKSZ}=\int_{I\times\Sigma} (\alpha_\Sigma[\Phi])_I\delta\psi^i(t) dt + (\mathcal{S}_\Sigma[\Phi])^\mathrm{top}
    \end{align}
Furthermore, one can easily see that the BV theory obtained in this way automatically gives rise to a a BV-BFV extendible theory, where the BFV one is given by the data on the target of the AKSZ.

\subsection{The AKSZ PC theory and the relation with the reduced BV theory}
As an application of the above construction, we review the results in \cite{Graksz21}. We denote by $\mathfrak{F}^\Sigma_{PC}=( \mathcal{F}^\Sigma_{PC}, \mathcal{S}^\Sigma_{PC}, \varpi^\Sigma_{PC} )$ the BFV theory of Palatini--Cartan gravity, which can be obtained by "turning off" the gravitino interaction in the supegravity case, e.g. from equations \eqref{eq: BFV linear action new variables} and \eqref{eq: BFV SG symplectic form}.

 We promote the fields in $\mathcal{F}^\Sigma_{PC}$ to fields in $\mathcal{F}^{AKSZ}_{PC}$ by considering \cite{Graksz21}
    \begin{align}\label{eq: AKSZ PC variable def}
        \nonumber&\mathfrak{e}=e+f^\Finv & \mathfrak{w}=\omega+ u^\Finv\\
        &\mathfrak{c}= c + w & \mathfrak{z}= \xi + z\\
        \nonumber&\mathfrak{l}= \lambda + \mu & \mathfrak{c}^\Finv= k^\Finv + c^\Finv\\
        \nonumber&\mathfrak{y}^\Finv= e^\Finv + y^\Finv
    \end{align}
where we used the same letters for the boundary fields which are now promoted to fields in $\mathcal{C^1}(I)\otimes\mathcal{F}^\Sigma_{PC}$. In particular, if $\phi\in\mathcal{F}^\Sigma_{PC}$, the corresponding AKSZ field becomes
    \begin{equation*}
        \mathfrak{P}=\phi + \psi^\Finv, \qquad \mathrm{where} \qquad \phi\in\mathcal{C^1}(I)\otimes\mathcal{F}^\Sigma_{PC}, \quad\text{ and }\quad  \psi^\Finv\in\Omega^1[-1](I)\otimes\mathcal{F}^\Sigma_{PC}
    \end{equation*}
\begin{theorem}[\cite{Graksz21}]
    The AKSZ data $\mathfrak{F}^{AKSZ}_{PC}$ on $M=I\times\Sigma$ are given by 
    \begin{align*}
        &\mathcal{F}^{AKSZ}_{PC}=T^*[-1](\mathrm{Map}(I,\mathcal{F}^\Sigma_{PC}),
    \end{align*}
    \begin{equation*}
        \varpi^{AKSZ}_{PC}=\int_{I\times\Sigma}\mathfrak{e}\delta\mathfrak{e}\delta\mathfrak{w} + \delta\mathfrak{c}\delta\mathfrak{c}^\Finv + \delta\mathfrak{w}\delta(\iota_{\mathfrak{z}}\mathfrak{c}^\Finv ) - \delta\mathfrak{l}\epsilon_n \delta\mathfrak{y}^\Finv + \iota_{\delta\mathfrak{z}}\delta(\mathfrak{e}\mathfrak{y}^\Finv),
    \end{equation*}
    \begin{align*}
        \mathcal{S}^{AKSZ}_{PC}=\int_{I\times \Sigma}&\mathfrak{e}^2d_i\mathfrak{w} + \mathfrak{c} d_I \mathfrak{c}^\Finv + d_I\mathfrak{w}\iota_{\mathfrak{z}}\mathfrak{c}^\Finv -\iota_{d_I\mathfrak{z}}\mathfrak{e}\mathfrak{y}^\Finv + d_I\mathfrak{l}\epsilon_n\mathfrak{y}^\Finv\\
        &+\mathfrak{c}\mathfrak{e}d_{\mathfrak{w}}\mathfrak{e}+\iota_{\mathfrak{z}}\mathfrak{e}\mathfrak{e}F_{\mathfrak{w}}+\epsilon_n\mathfrak{l}\mathfrak{e}F_{\mathfrak{w}} +\frac{1}{2}[\mathfrak{c},\mathfrak{c}]\mathfrak{c}^\Finv -\mathrm{L}^{\mathfrak{w}}_{\mathfrak{z}}\mathfrak{c}\mathfrak{c}^\Finv + \frac{1}{2}\iota_{\mathfrak{z}}\iota_{\mathfrak{z}}F_{\mathfrak{w}}\mathfrak{c}^\Finv\\
        & -[\mathfrak{c},\epsilon_n \mathfrak{l}]\mathfrak{y}^\Finv + \mathrm{L}^{\mathfrak{w}}_{\mathfrak{z}}(\epsilon_n \mathfrak{l})\mathfrak{y} +\frac{1}{2}\iota_{[\mathfrak{z},\mathfrak{z}]}(\mathfrak{e}\mathfrak{y}^\Finv),
    \end{align*}
    where it is understood that only the terms containing fields in $\Omega^1[-1](I)$ should be selected in the above expressions, to obtain a top form on $I\times\Sigma$.
\end{theorem}
The above theorem provides a compatible BV-BFV theory of gravity in the first order formalism. 

The question arises on how to relate the AKSZ PC theory $\mathfrak{F}_{PC}^{AKSZ}$ with the full BV PC theory in the bulk $\mathfrak{F}_{PC}$. The main obvious difference is that the former is BV-BFV extendible, while the latter is not, and the reason for that is that in $\mathfrak{F}_{PC}^{AKSZ}$ the connection is constrained by the structural constraint on the connection $\omega$ of the BFV PC theory $\mathfrak{F}_{PC}^\partial$, as found in \cite{CCS2020}. Such constraint restricts the space AKSZ of fields in the bulk in such a way that the induced boundary symplectic form is non-degenerate. This motivates the following theorem
\begin{theorem}[\cite{Graksz21}]\label{thm: AKSZ PC}
    There exists a map $\Phi \colon \mathfrak{F}^{AKSZ}_{PC}\rightarrow \mathfrak{F}_{PC}$ such that $\Phi\colon\mathcal{F}^{AKSZ}_{PC}\rightarrow \mathcal{F}_{PC}$ is an embedding and $\Phi^*(\varpi^M_{PC})=\varpi^{AKSZ}_{PC}$. Such map is called a BV embedding.
\end{theorem}

\begin{remark}
     The map $\Phi$ actually splits as the composition of two maps. In particular, the authors showed that there exists a BV equivalence $\varphi\colon\mathfrak{F}^{AKSZ}_{PC}\rightarrow\mathfrak{F}^{r}_{PC}$ between the reduced BV theory of  \ref{def: reduced PC theory} and the AKSZ PC theory, which is defined to be a symplectomorphism $\varphi\colon (\mathcal{F}^{AKSZ}_{PC},\varpi^{AKSZ}_{PC})\rightarrow (\mathcal{F}^{r}_{PC},\varpi^{r}_{PC})$ such that $\varphi^*\left(\mathcal{S}^{r}_{PC}\right)=\mathcal{S}^{AKSZ}_{PC}$. Then the map $\Phi$ of the previous theorem splits as follows
        \[\begin{tikzcd}
	&& {\mathfrak{F}_{PC}} \\
	{\mathfrak{F}^{AKSZ}_{PC}} && {\mathfrak{F}^{r}_{PC}}
	\arrow["\Phi", from=2-1, to=1-3]
	\arrow["\varphi", from=2-1, to=2-3]
	\arrow["{\iota_r}", from=2-3, to=1-3]
\end{tikzcd}\]
 where $\iota_r$ is the BV inclusion $\iota_{r}\colon\mathcal{F}^r_{PC}\hookrightarrow\mathcal{F}_{PC}$.
\end{remark}

\section{Tools, lemmata and identities }\label{app: tools}
    This section provides some technical lemmas, useful throughout the paper. In the following we will denote by $e$ both the veilbein $\in \Omega^{1}(M,\mathcal{V}) $ and its restriction to the boundary $e\in \Omega^{1}(\Sigma,\mathcal{V}_\Sigma)$ and in the same way there will be no distinction between $\mathcal{V}$ and its restriction to $\Sigma$, as their meaning will be cleared by the context in which they are used.
    We start by recalling the definitions of the following spaces
        \begin{equation*}
            \Omega^{(k,l)}:=\Omega^k(M,\wedge^l \mathcal{V}) \qquad \Omega_\partial^{(k,l)}:=\Omega^k(\Sigma,\wedge^l \mathcal{V}),
        \end{equation*}
    and maps 
        \begin{eqnarray}
            W_{k}^{ (i,j)}: \Omega^{i,j}  \longrightarrow &\Omega^{i+k,j+k}  \label{map: W_e-bulk}\\
            X  \longmapsto  &  \frac{1}{k!}\underbrace{e \wedge \dots \wedge e}_{k-times}\wedge X  \nonumber\\
            W_{k}^{ \partial, (i,j)}: \Omega_{\partial}^{i,j}  \longrightarrow & \Omega_{\partial}^{i+k,j+k} \label{map: W_e-boundary} \\
            X  \longmapsto &  \frac{1}{k!} \underbrace{e \wedge \dots \wedge e}_{k-times}\wedge X\nonumber
        \end{eqnarray}
        \begin{eqnarray}    
            \varrho^{i,j}: \Omega^{i,j}  \longrightarrow &\Omega^{i+1,j-1}  \label{map: [e,]-bulk}\\
            X \longmapsto & [e,X]\nonumber\\
            \varrho_\partial^{i,j}: \Omega_\partial^{i,j}  \longrightarrow &\Omega_\partial^{i+1,j-1}  \label{map: [e,]-boundary}\\
            X \longmapsto & [e,X]\nonumber.
        \end{eqnarray}

            Such maps have been studied in previous papers (notably in \cite{CCS2020}, \cite{CS2019} and \cite{C25}). We quickly recall their main results.

        The following diagrams \cite{C25} indicate the properties of the maps $W_1^{(i,j)}$ and $W_1^{\partial,(i,j)} $, in particular a hooked arrow indicates injectivity while a two--headed arrow indicates surjectivity. In the bulk we have    
    \begin{equation}\label{diag: prop e bulk}
    \begin{tikzcd}
    	{\Omega^{(0,0)}} & {\Omega^{(0,1)}} & {\Omega^{(0,2)}} & {\Omega^{(0,3)}} & {\Omega^{(0,4)}} \\
    	{\Omega^{(1,0)}} & {\Omega^{(1,1)}} & {\Omega^{(1,2)}} & {\Omega^{(1,3)}} & {\Omega^{(1,4)}} \\
    	{\Omega^{(2,0)}} & {\Omega^{(2,1)}} & {\Omega^{(2,2)}} & {\Omega^{(2,3)}} & {\Omega^{(2,4)}} \\
    	{\Omega^{(3,0)}} & {\Omega^{(3,1)}} & {\Omega^{(3,2)}} & {\Omega^{(3,3)}} & {\Omega^{(3,4)}} \\
    	{\Omega^{(4,0)}} & {\Omega^{(4,1)}} & {\Omega^{(4,2)}} & {\Omega^{(4,3)}} & {\Omega^{(4,4)}}
    	\arrow[hook, from=2-1, to=3-2]
    	\arrow[hook, two heads, from=3-2, to=4-3]
    	\arrow[two heads, from=4-3, to=5-4]
    	\arrow[hook, from=1-1, to=2-2]
    	\arrow[hook, from=2-2, to=3-3]
    	\arrow[two heads, from=3-3, to=4-4]
    	\arrow[two heads, from=4-4, to=5-5]
    	\arrow[hook, from=1-2, to=2-3]
    	\arrow[hook, two heads, from=2-3, to=3-4]
    	\arrow[two heads, from=3-4, to=4-5]
    	\arrow[hook, from=1-3, to=2-4]
    	\arrow[two heads, from=2-4, to=3-5]
    	\arrow[hook, two heads, from=1-4, to=2-5]
    	\arrow[hook, from=3-1, to=4-2]
    	\arrow[two heads, from=4-2, to=5-3]
    	\arrow[hook, two heads, from=4-1, to=5-2]
    \end{tikzcd}        
    \end{equation}
    whereas on the boundary one obtains
    \begin{equation}\label{diag: prop e bdry}
    \begin{tikzcd}
    	{\Omega_\partial^{(0,0)}} & {\Omega_\partial^{(0,1)}} & {\Omega_\partial^{(0,2)}} & {\Omega_\partial^{(0,3)}} & {\Omega_\partial^{(0,4)}} \\
    	{\Omega_\partial^{(1,0)}} & {\Omega_\partial^{(1,1)}} & {\Omega_\partial^{(1,2)}} & {\Omega_\partial^{(1,3)}} & {\Omega_\partial^{(1,4)}} \\
    	{\Omega_\partial^{(2,0)}} & {\Omega_\partial^{(2,1)}} & {\Omega_\partial^{(2,2)}} & {\Omega_\partial^{(2,3)}} & {\Omega_\partial^{(2,4)}} \\
    	{\Omega_\partial^{(3,0)}} & {\Omega_\partial^{(3,1)}} & {\Omega_\partial^{(3,2)}} & {\Omega_\partial^{(3,3)}} & {\Omega_\partial^{(3,4)}}
    	\arrow[two heads, from=3-4, to=4-5]
    	\arrow[two heads, from=3-3, to=4-4]
    	\arrow[two heads, from=3-2, to=4-3]
    	\arrow[hook, from=3-1, to=4-2]
    	\arrow[hook, from=2-1, to=3-2]
    	\arrow[hook, from=2-2, to=3-3]
    	\arrow[two heads, from=2-3, to=3-4]
    	\arrow[two heads, from=2-4, to=3-5]
    	\arrow[two heads, from=1-4, to=2-5]
    	\arrow[hook, from=1-3, to=2-4]
    	\arrow[hook, from=1-2, to=2-3]
    	\arrow[hook, from=1-1, to=2-2]
    \end{tikzcd}
    \end{equation}
    \begin{lemma}\cite{CCS2020,C25,FR25}\label{lem: useful isos}
        The following maps are isomorphisms:
        \begin{enumerate}
            \item \label{lem: iso W2 (0,2)}$W_2^{(0,2)}\colon\Omega^{(0,2)}\rightarrow \Omega^{(2,4)} $,
            \item \label{lem: iso W2 (2,0)}$W_2^{(2,0) } \colon\Omega^{(2,0)}\rightarrow \Omega^{(4,2)} $,
            \item \label{lem: iso W2 (1,1)} $W_2^{(1,1) } \colon\Omega^{(2,0)}\rightarrow \Omega^{(3,3)} $,
            \item \label{lem: iso W4 (0,0)}$W_4^{(0,0) }\colon\Omega^{(0,0)}\rightarrow \Omega^{(4,4)} $,
            \item \label{lem: iso rho (0,1)}$\varrho^{(0,1)}\colon\Omega^{(0,1)}\rightarrow\Omega^{(1,0)} $.
            \item \label{lem: iso rho (3,4)}$\varrho^{(3,4)}\colon\Omega^{(3,4)}\rightarrow\Omega^{(4,3)} $.
        \end{enumerate}
    \end{lemma}

    \begin{lemma}[\cite{CCS2020}]\label{lem: splitting (2,1)}
    Let $\alpha \in \Omega^{2,1}_\partial$. Then 
    \begin{align}\label{ConditionforOmega21_d4}
    \alpha=0 \qquad \Longleftrightarrow  \qquad \begin{cases}
    e\alpha =0 \\
    \epsilon_n \alpha \in \Ima W_{1}^{\partial, (1,1)}
    \end{cases}.
    \end{align}
    \end{lemma}
    \begin{lemma}[\cite{CCS2020}]\label{lem: splitting (2,2)}
    Let $\beta \in \Omega^{2,2}_\partial$. If $g^\partial$ is nondegenerate, there exist a unique $v \in \mathrm{Ker} W_{1}^{\partial, (1,2)}$ and a unique $\rho \in \Omega_{\partial}^{1,1}$ such that 
    \begin{align*}
    \beta = e \rho + \epsilon_n [e,v].
    \end{align*}
    \end{lemma}
    \begin{lemma}\label{lem: splitting (1,2)}
        Let $a\in\Omega^{(1,2)}_\partial$. Then 
        \begin{equation*}
            a=0 \quad \Longleftrightarrow \quad \begin{cases}
                ea=0\\
                \epsilon_na\in\mathrm{Im}(W_e^{\partial(0,2)})
            \end{cases}.
        \end{equation*}
    \end{lemma}
    \begin{proof}
        Let $I\subset \mathbb{R}$ be an interval with $\{x^n\}$ coordinate on it and let $\tilde{M}:=\Sigma\times I$. Then $E:=e+\epsilon_ndx^n \in \Gamma(\tilde{M},\mathcal{V})$ defines a non-degenerate vielbein on $M$. Let $A:=a + b dx^n\in \Omega^1(\tilde{M},\wedge^2 \mathcal{V})$, with $b\in \Omega_\partial^{(0,2)}$.

        Then the above system is equivalent to the the equation $E\wedge A= e a + (eb - \epsilon_n a) dx^n=0$. By diagram \eqref{diag: prop e bulk}, $E\wedge\cdot$ is injective, hence $E\wedge A=0$ iff $A=0$, implying $ a=0$. 
    \end{proof}
    \begin{lemma}\label{lem:split cdag}
        For all $\tilde{k}\in\Omega^{(2,1)}_\partial$ there exists a unique decomposition $\tilde{k}=\check{k}+r$ such that 
        \begin{equation*}
            er=0, \qquad \epsilon_n \check{k} \in \text{Im}(W_e^{\partial,(1,1)}).
        \end{equation*}
    \end{lemma}
    \begin{proof} 
        From Lemma \ref{lem: splitting (2,2)}, we know there exists a unique decomposition
            \begin{equation*}
                \epsilon_n \tilde{k}=e\check{a} +  \epsilon_n [e,\check{b}],
            \end{equation*}
        with $\check{b}\in\Ker{W_e^{\partial,(1,2)} }$. Define $r:=[e,\check{b}]$ and $\check{k}:=\tilde{k} - r$. This implies
            \begin{equation*}
                \epsilon_n\check{k}=e\check{a}\in\text{Im}(W_e^{\partial,(1,1)}) \qquad er=e[e,\check{b}]=[e,e\check{b}]-[e,e]\check{b}=0.
            \end{equation*}
        For uniqueness, assume there exist $\tilde{k}=\check{k}_1+r_1=\check{k}_2+r_2  $ such that $r_1,r_2\in\Ker{W_e^{\partial(1,2)}}$ and $\epsilon_n\check{k}_1,\epsilon_n\check{k}_2\in \text{Im}(W_e^{\partial,(1,1)})$. Then we obtain the following system
            \begin{equation*}
            \begin{cases}
                \epsilon_n(\check{k}_1 -\check{k}_2)=\epsilon_n(r_2-r_1)\in\text{Im}(W_e^{\partial,(1,1)})\\
                e(r_2-r_1)=0.
            \end{cases}
            \end{equation*}
        By Lemma \ref{lem: splitting (1,2)}, we have that $r_2=r_1$, implying $\check{k}_1 -\check{k}_2=r_2-r_1=0$.
    \end{proof}
    \begin{lemma}\label{lem: decomposition (1,3)}
        Let $\Theta\in\Omega_\partial^{(1,3)}$. Then there exist unique $\alpha\in \Omega^{(0,2)}_\partial$ and $\beta\in \Ker{W_e^{\partial(1,2)}}$ such that
            \begin{equation*}
                \Theta=e\alpha + \epsilon_n \beta.
            \end{equation*}
    \end{lemma}
    \begin{proof}
        Consider the map 
        \begin{align*}
            \rho\colon\Ker{W_e^{\partial(1,2)}} &\rightarrow \Omega^{(1,3)}_\partial\\
            \beta&\mapsto \epsilon_n \beta.
        \end{align*}
        Then assume $\exists \beta \in \Ker{W_e^{\partial(1,2)}} $ such that $\rho(\beta)=\epsilon_n\beta=0$. Lemma \ref{lem: splitting (1,2)} implies that $\beta=0$, hence $\rho$ is injective.

        We can then deduce that dim(Im$\rho$)=dim( $\Ker{W_e^{\partial(1,2)}}$ )=6. In the same way, since $W_e^{\partial(0,2)}$ is injective, dim(Im$(W_e^{\partial(0,2)})$)=6=dim$(\Omega^{(0,2)}_\partial)$.
        Hence dim(Im$(W_e^{\partial(0,2)}))+$ dim(Im$\rho$)=12=dim($\Omega_\partial^{(1,3)}$).

        Now we just need to prove that Im$(W_e^{\partial(0,2)})\cap \mathrm{Im}\rho=\{0\}$. 
        
        Assume $\exists 0\neq \beta\in\Ker{W_e^{\partial(1,2)}}$ such that for some $\alpha \in \Omega^{(0,2)}_\partial$
            \begin{equation*}
                \epsilon_n \beta=e\alpha.
            \end{equation*}
        Then, by Lemma \ref{lem: splitting (1,2)}, setting $a=v$, we automatically obtain $\beta=0$, contradicting the hypothesis. Hence Im$(W_e^{\partial(0,2)})\cap \mathrm{Im}\rho=\{0\}$, implying $\Omega_\partial^{(1,3)}\simeq \text{Im}W_e^{\partial(0,2)} \oplus \text{Im}\rho$.
        Uniqueness follow from the injectivity of $\rho$ and $W_e^{(0,2)}$.
    \end{proof}

    \begin{definition}\label{def: alpha e beta}
        The previous Lemma allows to define maps
            \begin{align*}
                \alpha_\partial \colon& \Omega^{(1,3)}_\partial \rightarrow  \Omega^{(0,2)}_\partial &\beta_\partial\colon\Omega^{(1,3)}_\partial \rightarrow \Omega^{(1,2)}_\partial\\
                &\Theta\mapsto \alpha_\partial(\Theta) &\Theta \mapsto \beta_\partial(\Theta);
            \end{align*}
        such that $\Theta= e\alpha_\partial(\Theta) + \epsilon_n\beta_\partial(\Theta)$.    
    \end{definition}    

    \begin{lemma}\label{lem: iso (3,4)=(0,4)}
        The map $\frac{1}{3!}\epsilon_n e^3\colon\Omega_\partial^{(0,0)}\rightarrow \Omega_\partial^{(3,4)} $  is an isomorphism.
    \end{lemma}
    \begin{proof}
        It is immediate to see that $\Omega^{(3,0)}_\partial\simeq \Omega^{(3,4)}_\partial $ as $\Gamma(\Sigma,\wedge^4 \mathcal{V})\simeq \mathcal{C}^\infty(\Sigma)  $ upon choice of an orientation. The same is true for $\Omega^{(3,0)}_\partial\simeq\Omega^{(0,0)}_\partial$, hence we have $\Omega^{(0,0)}_\partial\simeq \Omega^{(3,4)}_\partial$. Therefore showing that the above maps are isomorphisms is equivalent to showing that they are nowhere vanishing, which is obvious from their defininition.
    \end{proof}
    \begin{remark}
        If one does the computation directly, letting $\tilde{e}$ be the coframe in the bulk and denoting the transversal (to the boundary) index by $n$,  it is possible to see that $\text{Vol}_M=\frac{1}{4!}\tilde{e}^4=\frac{1}{3!}\tilde{e}^3 \tilde{e}_n dx^n $. Restricting to the boundary, we have $\tilde{e}_n=\upsilon \epsilon_n + \iota_\zeta e$, where $\upsilon\in\mathcal{C}^\infty(\Sigma) $ and $\zeta \in \mathfrak{X}(\Sigma)$. In particular, $\upsilon$ is a nowhere vanishing function, hence, upon restriction to the boundary, one finds
            \begin{equation*}
                \text{Vol}_\Sigma=\frac{1}{3!}\left( \upsilon \epsilon_n e^3 + \iota_\zeta(e) e^3 \right)=\frac{1}{3!} \upsilon \epsilon_n e^3 \quad \Rightarrow \quad \frac{1}{3!}\epsilon_n e^3 = \frac{1}{\upsilon}\text{Vol}_\Sigma,
            \end{equation*}
    \end{remark}
    \begin{corollary}
        The map 
        \begin{equation*}
            \frac{1}{3!}e^3\gamma:\Omega_\partial^{(0,0)}(\Pi\mathbb{S}_M)\rightarrow\Omega_\partial^{(3,4)}(\Pi\mathbb{S}_M)
        \end{equation*}
        is an isomorphism.
    \end{corollary}
    \begin{proof}
        By direct inspection, $\frac{1}{3!}e^3\gamma=\frac{1}{3!}\epsilon_n e^3 \gamma^n$. Since $\gamma^n$ is invertible,\footnote{$\{\gamma_n,\gamma_n\}=2(\gamma_n)^2=-2g_{(nn)}\neq 0$ implies $(\gamma_n)^{-1}=-\frac{1}{g_{nn}}\gamma_n$} Lemma  \ref{lem: iso (3,4)=(0,4)} implies the desired result.
    \end{proof}
    
 \subsection{Results for Gamma matrices and Majorana spinors in $D=4$}
    
    The following results are either well known from literature, or have been elucidated in \cite{FR25}. We start by reviewing some known identities for gamma matrices

    \begin{align}
        &\label{app_A_gamma2} \gamma^a\gamma_a=-D;\\
        & \gamma^a\gamma^b\gamma_a=(D-2)\gamma^b;\\
        & \gamma^a\gamma^b\gamma^c\gamma_a=(4-D)\gamma^b\gamma^c + 4 \eta^{bc}\mathbb{1};\\ &\gamma^a\gamma^b\gamma^c\gamma^d\gamma_a=(D-6)\gamma^b\gamma^c\gamma^d - 12\gamma^{[b}\eta^{cd]}.\\
    \end{align}
    The following hold in $d=4$, having set $\gamma^5:=i\gamma^0\gamma^1\gamma^2\gamma^3$ and $\gamma_{a_1,\cdots,a_n}:= \gamma_{[a_1} \gamma_{a_2}\cdots \gamma_{a_n]}  $
    \begin{align}
        &\gamma^a\gamma^b\gamma^c\gamma^d\gamma_a=2\gamma^d\gamma^c\gamma^b;\\
        & \gamma^a\gamma^b\gamma^c=-\eta^{ab}\gamma^c-\eta^{bc}\gamma^a+\eta^{ac}\gamma^b + i \epsilon^{dabc}\gamma_d\gamma^5;\\
        &\gamma^5\gamma^{[c}\gamma^{d]}=-\frac{i}{2}\epsilon^{abcd}\gamma_{ab};\\
        &\label{id: gamma5gammac} \gamma^5\gamma^c=\frac{i}{6}\epsilon^{abcd}\gamma_{abc}.
    \end{align}
    Considering $\{v_a\}$ a basis for $V$, setting $\gamma:=\gamma^a v_a$ we have
        \begin{align*}
            [v_a,\cdot]\colon &\wedge^k V \longrightarrow  \wedge^{k-1}V\\
            &\alpha=\frac{1}{k!}\alpha^{a_1\cdots a_k}v_{a_1}\cdots v_{a_k} \longmapsto\frac{1}{(k-1)!}\eta_{a a_1} \alpha^{a_1\cdots a_k} v_{a_2}\cdots v_{a_k}
        \end{align*}
    obtaining
        \begin{align}
            & \label{id:v_a,gamma^N}[v_a,\Gamma^N]=N[v_a,\Gamma]\Gamma^{N-1} + N(N-1)v_a\Gamma^{N-2} ,\quad N\geq 2;\\%=(-1)^{N-1}[ N\Gamma^{N-1}[v_a,\Gamma] + N(N-1)\Gamma^{N-2}v_a ] \\
            &\label{id:v_a,gamma^N2}\hspace{11.5mm}=(-1)^{N-1}(N\Gamma^{N-1}\Gamma_a + N(N-1)\Gamma^{N-2}v_a).
        \end{align}
        
            \subsubsection{Majorana flip relations}
            Given any two Majorana spinors $\psi$ and $\chi$ of arbitrary parity, denoting $\gamma=\gamma^a v_a\in \mathrm{End}(S)\otimes V$, we have the following 
                \begin{align}
                    &  \label{flip:0}\bar{\chi}\psi= -(-1)^{|\chi||\psi|}\bar{\psi}\chi;\\
                    &  \label{flip:1}\bar{\chi}\gamma\psi=(-1)^{|\psi|+|\chi|+|\psi||\chi|}\bar{\psi}\gamma\chi; \\
                    & \label{flip:2} \bar{\chi}\gamma^2\psi=(-1)^{|\psi||\chi|}\bar{\psi}\gamma^2\chi  ;\\
                    &   \label{flip:3}\bar{\chi}\gamma^3\psi=-(-1)^{|\psi|+|\chi|+|\psi||\chi|}\bar{\psi}\gamma^3\chi.
                \end{align}
    
            In general, one finds 
                \begin{equation}
                    \label{flip:N}\bar{\chi}\gamma^N\psi=-t_N(-1)^{N(|\psi|+|\chi|)+|\psi||\chi|}\bar{\psi}\gamma^N\chi,
                \end{equation}
            where $t_N$ is defined from $(C\gamma^N)^t=-t_NC \gamma^N$ and is such that $t_{N+4}=t_N$. The first 4 parameters read
                \begin{equation*}
                    t_0=1,\qquad t_1=-1, \qquad t_2=-1, \qquad t_3=1.
                \end{equation*}

            Furthermore, another important identity derived from the ones above, is the following
                \begin{equation}\label{id: action of omega}
                     \bar{\chi}\gamma[\alpha,\psi]=3\bar{\chi}\gamma\psi \alpha + \frac{1}{2}\bar{\chi}[\alpha,\gamma^3]_V\psi,
                \end{equation}
            which is true for all $\chi,\psi\in\mathbb{S}_M$  and $\alpha\in\wedge^2 V$.   
            \subsubsection{Fierz identities}
            Using the fact that the gamma matrices generate the whole Clifford algebra, which in the gamma representation is obtained as the algebra of matrices acting on $\mathbb{C}^4$, from the completeness relation one obtains the following identity
                \begin{equation}\label{id: Fierz identiy}
                    (\gamma^a)^\cdot_{\alpha(\delta}(\gamma^a)^\cdot_{\rho\beta)}=0.
                \end{equation}
            Contracting with 4 Majorana spinors $\lambda_i$'s ($i=1, \cdots, 4$) of arbitary parity, we obtain the following Fierz identities
                \begin{align}
                    & \label{Fierz:1}\bar{\lambda}_1 \gamma^3 \lambda_2 \bar{\lambda}_3 \gamma \lambda_4=(-1)^{|\lambda_2||\lambda_3|}\bar{\lambda}_1\gamma\lambda_3 \bar{\lambda}_2 \gamma^3 \lambda_4 + (-1)^{|\lambda_4|(|\lambda_2|+|\lambda_3|+1)+|\lambda_3|} \bar{\lambda}_1\gamma\lambda_4 \bar{\lambda}_2 \gamma^3 \lambda_3;\\
                    & \label{Fierz:2} \bar{\lambda}_1 \gamma^3 \lambda_2 \bar{\lambda}_3 \gamma \lambda_4= - (-1)^{|\lambda_2||\lambda_3|}\bar{\lambda}_1\gamma^3\lambda_3 \bar{\lambda}_2 \gamma \lambda_4 - (-1)^{|\lambda_4|(|\lambda_2|+|\lambda_3|+1)+|\lambda_3|} \bar{\lambda}_1\gamma^3\lambda_4 \bar{\lambda}_2 \gamma \lambda_3 .
                \end{align}

    \subsubsection{Lemmata}
    In the following we provide a series of lemmata holding for Majorana (and, when unspecified, Dirac) spinors. They appear verbatim, with proofs, in \cite{FR25}.
    \begin{lemma}\label{lem: injectivity egamma^3}
        The map 
        \begin{align*}
            \Theta^{(1,0)}\colon  \Omega^{(1,0)}(\Pi \mathbb{S}_D)& \longrightarrow \Omega^{(2,4)}(\Pi \mathbb{S}_D)\\
            \psi& \longmapsto \frac{1}{3!}e\gamma^3 \psi
        \end{align*}
        is injective.
    \end{lemma}

     \begin{lemma}\label{lem: iso (1,0) (3,4)}
        The map 
        \begin{align*}
            \Theta_\gamma^{(1,0)}\colon  \Omega^{(1,0)}(\Pi \mathbb{S}_D)& \longrightarrow \Omega^{(3,4)}(\Pi \mathbb{S}_D)\\
            \psi& \longmapsto \frac{1}{3!}e\gamma^3 \underline{\gamma}\psi
        \end{align*}
        is an isomorphism, where $\underline{\gamma}:= [e,\gamma]=\gamma_\mu dx^\mu$
    \end{lemma}

    \begin{remark}
        By the same reasoning (or just by taking the Dirac conjugate of the above expression), one finds that also the map
            \begin{equation*}
                 \psi \longmapsto \frac{1}{3!}e\ugam\gamma^3 \psi
            \end{equation*}
        is an isomorphism.    
    \end{remark}
    \begin{lemma}\label{lem: splitting (3,1)}
        For all $\theta \in \Omega^{(3,1)}(\Pi \mathbb{S}_M) $ there exist unique $\alpha\in \Omega^{(1,0)}(\Pi \mathbb{S}_M) $ and $\beta \in \Omega^{(3,1)}(\Pi \mathbb{S}_M) $ such that
            \begin{equation}
                \theta = i e \underline{\gamma} \alpha + \beta \qquad \mathrm{and}\qquad \gamma^3 \beta = 0.
            \end{equation}
    \end{lemma}
    
    \begin{lemma}\label{lem: splitting (2,1) spin}
        Let $\ugam\gamma^n_{(i,j)}$ be the map
            \begin{equation*}
                \ugam\gamma^n_{(i,j)}\colon\Omega^{(i,j)}\rightarrow \Omega^{(i,j+n)}\colon\beta\mapsto \gamma^n \beta,
            \end{equation*}
        then, for all $\theta \in \Omega^{(2,1)}(\Pi \mathbb{S}_M)(\Pi \mathbb{S}_M)$ there exist unique $\alpha\in\Omega^{(1,0)}(\Pi \mathbb{S}_M)$ and $\beta \in \ker{\ugam\gamma^3_{(2,1)}}$ such that 
            \begin{equation*}
                \theta= e \alpha + \beta. 
            \end{equation*}
    \end{lemma}
 
    \begin{definition}
        Thanks to the previous lemmata, we can define maps, 
        \begin{align*}
            &\alpha\colon\Omega^{(3,1)}(\Pi \mathbb{S}_M)\rightarrow \Omega^{(1,0)}(\Pi \mathbb{S}_M) &\beta\colon\Omega^{(3,1)}(\Pi \mathbb{S}_M)\rightarrow\ker{(\gamma^3_{(3,1)})}\\
            &\kappa\colon\Omega^{(2,1)}(\Pi \mathbb{S}_M)\rightarrow\Omega^{(1,0)}(\Pi \mathbb{S}_M) & \varkappa\colon \Omega^{(2,1)}(\Pi \mathbb{S}_M)\rightarrow\ker{(\ugam\gamma^3_{(2,1)})}
        \end{align*}
        such that for all $\theta\in\Omega^{(3,1)} $ and $\omega\in\Omega^{(2,1)}$, we have
            \begin{align*}
               & \theta= i e \ugam\alpha(\theta) + \beta(\theta), \qquad\omega = e \kappa(\omega) + \varkappa(\omega).
            \end{align*}
    \end{definition}
    
    \begin{lemma}\label{lem: Fierz}
        For all $\lambda,\psi,\chi\in \mathbb{S}_M$  such that $|\chi|=0$ and $|\psi|=1$, the following identities hold
            \begin{equation*}
                \bar{\lambda}\gamma^3 \chi \bar{\chi}\gamma\psi =0, \qquad  \bar{\chi}\gamma\chi \bar{\lambda}\gamma^3 \psi=0 \qquad \mathrm{and} \qquad \bar{\lambda}\gamma\chi \bar{\chi}\gamma^3\psi=0.
            \end{equation*}
    \end{lemma}

\section{Poisson brackets of the constraints}\label{app: PB constraints}

\begin{proof}[Proof of \ref{thm: PB of constraints}]
        Having obtained the Hamiltonian vector fields of the constraints, we can compute their Poisson brackets.
        The pure gravity sector has been computed in \cite{CCS2020}, we refer to it for the details, and concentrate on the Rarita--Schwinger sector. 
        \begin{remark}
            In the following, instead of the definition via the symplectic form $\{F,G\}=\iota_{\mathbb{X}_F}\iota_{\mathbb{X}_G}\varpi$, we use the equivalent formulation
            $\{F,G\}= \mathbb{X}_F(G)(=\iota_{\mathbb{X}_F}\delta G=\iota_{\mathbb{X}_F}\iota_{\mathbb{X}_G}\varpi)$.
        \end{remark}
            \begin{align*}
                \{L_c,L_c\}&=\int_\Sigma -[c,c]ed_\omega e + \frac{1}{3!}[c,e]\bar{\psi}\gamma^3[c,\psi] + \frac{1}{3!}e \left([c,\bar{\psi}]\gamma^3[c,\psi] + \bar{\psi}\gamma^3[c,[c,\psi]]\right)\\
                &=\int_\Sigma -[c,c]ed_\omega e +\frac{1}{3!}e \left( - \bar{\psi}[c,\gamma^3]_v [c,\psi] + \bar{\psi}[c,\gamma^3]_V [c,\psi] + 2 \bar{\psi}\gamma^3[c,[c,\psi]]  \right)\\
                &=\int_\Sigma -[c,c]ed_\omega e +\frac{1}{3!}e  \bar{\psi}\gamma^3[[c,c],\psi] \\
                &=-\int_\Sigma [c,c] e\left( d_\omega e - \frac{1}{2}\bar{\psi}\gamma\psi \right)=-L_{[c,c]};
            \end{align*}
            \begin{align*}
                \{L_c,M_\chi\}&=\frac{1}{3!}  \int_\Sigma [c,e] (d_\omega \bar{\chi}\gamma^3\psi + \bar{\chi}\gamma^3 d_\omega \psi )+ e ([d_\omega c ,\bar{\chi}]\gamma^3 \psi + d_\omega \bar{\chi}\gamma^3 [c,\psi] -\bar{\chi}\gamma^3 [d_\omega c, \psi] + \bar{\chi}\gamma^3 d_\omega[c,\psi] )\\
                &=\frac{1}{3!}\int_\Sigma e ( d_\omega \bar{\chi}[c,\gamma^3]_V \psi - \bar{\chi}[c,\gamma^3]_V d_\omega\psi +[d_\omega c,\bar{\chi}]\gamma^3 \psi - d_\omega \bar{\chi}[c,\gamma^3]_V \psi - [c,d_\omega \bar{\chi}]\gamma^3 \psi - \bar{\chi} \gamma^3 [c,d_\omega \psi] ) \\
                &=\frac{1}{3!}\int_\Sigma e ( d_\omega[c,\bar{\chi}]\gamma^3 \psi - [c,\bar{\chi}]\gamma^3 d_\omega \psi ) = M_{[c,\chi]},
            \end{align*}
            having used Leibniz rule and an expression analogous to \eqref{eq: Leibniz spinors}.
            \begin{align*}
                \{L_c,P_\xi\}&=\int_\Sigma \mathrm{L}_{\xi}^{\omega_0} c e d_\omega e - \frac{1}{3!}[c,e] (\bar{\psi}\gamma^3 \mathrm{L}_{\xi}^{\omega_0}\psi) - \frac{1}{3!}e \left( [c,\bar{\psi}]\gamma^3\mathrm{L}_{\xi}^{\omega_0}\psi + \bar{\psi}\gamma^3 \mathrm{L}_{\xi}^{\omega_0}[c,\psi] \right)\\
                &=\int_\Sigma \mathrm{L}_{\xi}^{\omega_0} c e d_\omega e - \frac{1}{3!}e \left(  \bar{\psi}\gamma^3[c,\mathrm{L}_{\xi}^{\omega_0}\psi] + \bar{\psi}\gamma^3[\mathrm{L}_{\xi}^{\omega_0}c, {\psi}] - \bar{\psi} \gamma^3 [c,\mathrm{L}_{\xi}^{\omega_0}\psi]\right)\\
                 &=\int_\Sigma \mathrm{L}_{\xi}^{\omega_0} c e d_\omega e - \frac{1}{3!}e \bar{\psi}\gamma^3[\mathrm{L}_{\xi}^{\omega_0}c, {\psi}]=\int_\Sigma  \mathrm{L}_{\xi}^{\omega_0} c  e \left(d_\omega e - \frac{1}{2}\bar{\psi}\gamma\psi\right)\\
                 &=L_{\mathrm{L}_{\xi}^{\omega_0} c};
            \end{align*}
            \begin{align*}
                \{L_c,H_\lambda\}&=\int_\Sigma -[c,\lambda \epsilon_n]e F_\omega + \frac{1}{3!}\lambda \epsilon_n \left( [c,\bar{\psi}]\gamma^3 d_\omega \psi - \bar{\psi}\gamma^3 [d_\omega c, \psi] + \bar{\psi}\gamma^3 d_\omega [c,\psi] \right)\\
                &=\int_\Sigma -[c,\lambda \epsilon_n]e F_\omega + \frac{1}{3!}\lambda \epsilon_n  \left( [c,\bar{\psi}]\gamma^3 d_\omega \psi - \bar{\psi}\gamma^3  [c,d_\omega\psi] \right)\\
                &=\int_\Sigma -[c,\lambda \epsilon_n]e F_\omega + \frac{1}{3!}\lambda \epsilon_n  \bar{\psi}[c,\gamma^3]_V d_\omega \psi\\
                &=\int_\Sigma -[c,\lambda \epsilon_n]\left(e F_\omega + \frac{1}{3!} \bar{\psi}\gamma^3 d_\omega\psi\right)\\
                &=-P_{X}+L_{\iota_X(\omega-\omega_0)} + M_{\iota_X\psi}-H_{X^n} ,
            \end{align*}
                where, letting $\{x^i\}$ be coordinates on $\Sigma$, we have $X=e^i_a[c,\lambda \epsilon_n]^{(i)}\partial_i$ and $X^{(n)}=[c,\lambda \epsilon_n]^{(n)}$, having set $e^i_a e^a_j=\delta^i_j$.
            \begin{align*}
                \{P_\xi,M_\chi\}&=\int_\Sigma -\frac{1}{3!}\mathrm{L}_{\xi}^{\omega_0}e (d_\omega \bar{\chi}\gamma^3 \psi + \bar{\chi}\gamma^3 d_\omega\psi)  - i e \bar{\chi}\gamma^3\psi(\iota_\xi F_{\omega_0}+ \mathrm{L}_{\xi}^{\omega_0}(\omega-\omega_0))\\
                &\qquad -\frac{1}{3!}e ( d_\omega \bar{\chi} \gamma^3 \mathrm{L}_{\xi}^{\omega_0}\psi - \bar{\chi}\gamma^3 d_\omega \mathrm{L}_{\xi}^{\omega_0}\psi)\\
                &\overset{\star}{=}\int_\Sigma \frac{1}{3!}e \left( \mathrm{L}_{\xi}^{\omega_0}\bar{\chi}\gamma^3 d_\omega \psi - d_\omega \mathrm{L}_{\xi}^{\omega_0}\bar{\psi}\gamma^3 \psi \right)=-M_{\mathrm{L}_{\xi}^{\omega_0}\chi},
            \end{align*}
            where we have used integration by parts and
            \begin{align*}
                 &\qquad [\mathrm{L}_{\xi}^{\omega_0},d_\omega]\psi=[\iota_\xi F_{\omega_0} + \mathrm{L}_{\xi}^{\omega_0}(\omega-\omega_0),\psi]  &&(\star)
            \end{align*}
            \begin{align*}
                \{P_\xi,H_\lambda\}&=\int_\Sigma \mathrm{L}_{\xi}^{\omega_0}(\lambda \epsilon_n) e F_\omega - \frac{1}{3!}\lambda \epsilon_n \left( \mathrm{L}_{\xi}^{\omega_0}\bar{\psi}\gamma^3d_\omega \psi - \bar{\psi}\gamma^3[\iota_\xi F_{\omega_0} + \mathrm{L}_{\xi}^{\omega_0}(\omega-\omega_0),\psi] + \bar{\psi}\gamma^3 d_\omega \mathrm{L}_{\xi}^{\omega_0}\psi \right)\\
                &\overset{\star}{=}\int_\Sigma \mathrm{L}_{\xi}^{\omega_0}(\lambda \epsilon_n) e F_\omega + \frac{1}{3!}\mathrm{L}_{\xi}^{\omega_0}(\lambda \epsilon_n)\bar{\psi} \gamma^3 d_\omega \psi\\
                &=P_Y -L_{\iota_Y(\omega-\omega_0)} -M_{\iota_Y \psi} + H_{Y^{(n)}},
            \end{align*}
        where $Y=e^i_a \mathrm{L}_{\xi}^{\omega_0}(\lambda \epsilon_n)^{(i)}\partial_i$, $Y^{(n)}= \mathrm{L}_{\xi}^{\omega_0}(\lambda \epsilon_n)^{(n)} $.
            \begin{align*}
                \{M_\chi,M_\chi\}&=\frac{1}{2} \int_\Sigma -\frac{1}{3!}\bar{\chi}\gamma\psi(d_\omega\bar{\chi}\gamma^3\psi + \bar{\chi}\gamma^3d_\omega\psi) - e\mathbb{M}_\omega\bar{\chi}\gamma\psi + \frac{e}{3!}d_\omega\bar{\chi}\gamma d_\omega\chi \\
                &\qquad\quad  - \frac{1}{2\cdot 3!}d_\omega\bar{\chi}d_\omega e \gamma^3 \chi + \frac{1}{3!} e \bar{\chi}\gamma^3[F_\omega,\chi] + \frac{1}{3!}e\bar{\chi}\gamma^3 d_\omega\mathbb{M_\psi^e} \\
                &\overset{\ref{lem: Fierz}}{=}\frac{1}{2}\int_\Sigma -\frac{1}{ 3!} d_\omega e \bar{\chi}\gamma^3d_\omega\chi + \frac{1}{3}\bar{\chi}\gamma^3[F_\omega,\chi]+ \frac{1}{3!}d_\omega e d_\omega\bar{\chi}\gamma^3\chi + \frac{1}{3!}d_\omega e \bar{\chi}\gamma^3 \mathbb{M}^e_\psi \\
                &\overset{\eqref{id: action of omega}}{=}\int_\Sigma\frac{1}{2}\bar{\chi}\gamma\chi e F_\omega \\
                &=\int_\Sigma\frac{1}{2}\bar{\chi}\gamma\chi e F_\omega - e\alpha^\partial(\chi,d_\omega\chi) \left( d_\omega e - \frac{1}{2}\bar{\psi}\gamma\psi \right) - \epsilon_n \beta^\partial(\chi,d_\omega\chi)\left( d_\omega e - \frac{1}{2}\bar{\psi}\gamma\psi \right)\\
                &=\frac{1}{2}P_\varphi - \frac{1}{2}L_{\vf(\omega-\omega_0)} - \frac{1}{2}M_{\vf\psi} + H_{\varphi^n} 
            \end{align*}
        where $\varphi^i:= \bar{\chi}\gamma^a\chi e_a^i$ and $\varphi^n:= \bar{\chi}\gamma^a\chi e_a^n$, having used that $\mathbb{M}_\psi^e\propto \chi$, hence $\bar{\chi}\gamma^3 \mathbb{M}_\psi^e \propto \bar{\chi}\gamma^3\chi=0$.
        %We remark that in a few steps we used a combination of flip relations and Fierz identities to show that some terms vabish. One example is $d_\omega \bar{\chi} \gamma^3 \psi \bar{\chi}\gamma \psi $. By applying separately \ref{Fierz:1} and a combination of \ref{flip:3} and \ref{Fierz:2} we find
        %    \begin{align*}
        %        d_\omega \bar{\chi} \gamma^3 \psi \bar{\chi}\gamma \psi &\overset{\ref{Fierz:1}}{=}d_\omega \bar{\chi}\gamma \psi \bar{\psi}\gamma^3 \chi\\
        %        &\overset{\ref{flip:3}}{=}\bar{\psi}\gamma^3 d_\omega \chi \bar{\chi}\gamma \psi  \overset{\ref{Fierz:2}}{=} - d_\omega \bar{\chi}\gamma \psi \bar{\psi}\gamma^3 \chi,
        %    \end{align*}
        %showing the expression is 0. An analogous process has been employed for $\bar{\chi}\gamma^3d_\omega \psi \bar{\psi}\gamma\psi$ and $\bar{\chi}\gamma\chi \bar{\psi}\gamma^3 d_\omega \psi$.

        %Furthemore, defining 
        %\begin{itemize}
          %  \item $\hat{(\cdot)}:\Gamma(\mathcal{V}_\Sigma)\rightarrow \mathfrak{X}(\Sigma)\colon A=A^i e_i + A^n \epsilon_n \mapsto A^i \partial_i,$;
           % \item $A_\partial= A^i e_i = A - A^n \epsilon_n$ for all $A\in\mathcal{V}$;
            %\item $<e,\cdot>:=g^{ij}_\partial e_i \iota_{\partial_i}( \cdot)$ on $\Omega^\bullet(\Sigma)$,
        %\end{itemize} 
        %then one can find\footnote{We leave the proof of this statement for the appendix.} 
        %\begin{equation}
         %   \label{def:alpha}
         %  \alpha=-\frac{1}{3}\bar{\chi}\tilde{\gamma}^2 <\hat{\gamma},d_\omega \chi>.
        %\end{equation}

            \begin{align*}
                \{P_\xi,P_\xi\}&=\int_\Sigma e d_\omega e \iota_{[\xi,\xi]}(\omega-\omega_0) + \frac{1}{2}\iota_{[\xi,\xi]}(e^2) F_\omega - e d_\omega e F_{\omega_0} + \frac{1}{3!}\mathrm{L}_\xi^{\omega_0}e \bar{\psi}\gamma^3\mathrm{L}_\xi^{\omega_0}\psi \\
                &\qquad + \frac{1}{3!}e\left( \mathrm{L}_\xi^{\omega_0}\bar{\psi}\gamma^3 \psi + \bar{\psi}\gamma^3 \mathrm{L}_\xi^{\omega_0}\mathrm{L}_\xi^{\omega_0}\psi \right)\\
                &\overset{\ref{id:L^2}}{=} \int_\Sigma e d_\omega e \iota_{[\xi,\xi]}(\omega-\omega_0) + \frac{1}{2}\iota_{[\xi,\xi]}(e^2) F_\omega +\frac{1}{3!}e\bar{\psi}\gamma^3 \mathrm{L}_{[\xi,\xi]}^{\omega_0}\psi \\
                &\qquad - e d_\omega e F_{\omega_0}  + \frac{1}{3!}e\bar{\psi}\gamma^3[\iota_\xi\iota_\xi F_{\omega_0},\psi]\\
                &=P_{[\xi,\xi]} - L_{\iota_\xi\iota_\xi F_{\omega_0}};
            \end{align*}
        having used 
            \begin{equation}\label{id:L^2}
                \mathrm{L}_\xi^{\omega_0}\mathrm{L}_\xi^{\omega_0} A = \frac{1}{2}\mathrm{L}_{[\xi,\xi]}^{\omega_0} A + \frac{1}{2}[\iota_\xi\iota_\xi F_{\omega_0},A].
            \end{equation}

        \begin{align*}
            &=\int_\Sigma d_\omega(\lambda \epsilon_n)\left( \lambda \epsilon_n F_\omega -\frac{1}{3!}\bar{\psi} \gamma^3 \mathbb{H}_\psi \right) + \lambda \epsilon_n \left( d_\omega e - \frac{1}{2}\bar{\psi}\gamma\psi \right)\mathbb{H}_\omega\\
            &\qquad + \frac{1}{3!}\lambda \epsilon_n \left( \mathbb{H}_{\bar{\psi}}\gamma^3 d_\omega\psi + \bar{\psi}\gamma^3 d_\omega\mathbb{H}_\psi \right)\\
            &=\int_\Sigma \lambda e \mathbb{H}_\omega \sigma + \frac{1}{3}\lambda\epsilon_n \bar{\psi}\gamma^3\mathbb{H}_\psi= \int_\Sigma\lambda^2 (\cdots)= 0,
        \end{align*}
        having used the fact that $\lambda^2=\epsilon_n^2=0$ and that $d_\omega(\lambda \epsilon_n)\lambda \epsilon_n=0$;
        \begin{align*}
            \{M_\chi,H_\lambda\}&=\mathbb{M}_\chi(H_\lambda)\\
            &=\int_\Sigma -\lambda\epsilon_n \bar{\chi}\gamma\psi F_\omega + d_\omega(\lambda\epsilon_n e )\mathbb{M}_\omega + \frac{1}{3!}\lambda\epsilon_n d_\omega\bar{\chi}\gamma^3d_\omega\psi -\frac{1}{2}\lambda\epsilon_n\bar{\psi}\gamma\psi\mathbb{M}_\omega+\frac{1}{3!}\lambda\epsilon_n\bar{\psi}\gamma^3[F_\omega,\chi]\\
            &=\int_\Sigma -\lambda\epsilon_n \bar{\chi}\gamma\psi F_\omega  +\lambda\epsilon_n\left(d_\omega e -\frac{1}{2}\bar{\psi}\gamma\psi\right)\mathbb{M}_\omega +\frac{1}{3!}\lambda\epsilon_n\left(\bar{\psi}\gamma^3[F_\omega,\chi]+\bar{\chi}\gamma^3[F_\omega,\psi]\right)\\
            &\overset{\eqref{id: action of omega}}{=}\int_\Sigma \lambda\epsilon_n\mathbb{M}_\omega \left(d_\omega e -\frac{1}{2}\bar{\psi}\gamma\psi\right)=\int_\Sigma\big(e\alpha^\partial(\mathbb{M}_\omega ) +\epsilon_n \beta^\partial(\mathbb{M}_\omega)\big)\left(d_\omega e -\frac{1}{2}\bar{\psi}\gamma\psi\right)\\
            &=L_{\alpha^\partial(\epsilon_n\mathbb{M}_\omega)}
        \end{align*}
        having used the structural constraint and that $e\beta^{\partial}(\chi,\psi) =0$.
 
\end{proof}

\section{Proofs of section \ref{sec: BFV extension} and \ref{sec: BFV SG}}\label{app: pullback restricted action}

\subsection{Proposition \ref{prop: equivalence of BV PC r constraints}}
\begin{proof}[Proof of \ref{prop: equivalence of BV PC r constraints}]\label{proof: equiv BVPCr}
    We begin by unpacking the terms inside $\underline{\mathfrak{W}^\Finv}=\underline{\tilde{\omega}}_n^\Finv - \iota_{\underline{z}}\tilde{\omega}^\Finv - \iota_{\tilde{\xi}}\underline{\tilde{c}}_n^\Finv + \iota_{\underline{z}}{\tilde{c}}_n^\Finv\tilde{\xi}^n$, starting from the definition $\bm{k}^\Finv:=\bm{\omega}^\Finv -\iota_{\bm{\xi}}\bm{c}^\Finv = \bm{e}\bm{\check{k}}=\tilde{e}\tilde{\check{k}} + \underline{\tilde{e}}_n\tilde{\check{k}} +\tilde{e}\underline{\tilde{\check{k}}}_n $. In particular, one sees    
        \begin{align*}
            &\underline{\tilde{\omega}}_n^\Finv-\iota_{\tilde{\xi}}\underline{\tilde{c}}^\Finv_n=\underline{\tilde{k}}^\Finv_n= \tilde{e}\underline{\tilde{\check{k}}}_n +\underline{\mu}\epsilon_n \tilde{\check{k}} +\iota_{\underline{z}}\tilde{e}\tilde{\check{k}}\\
            &\iota_{\underline{z}}\big( \tilde{\omega}^\Finv - \tilde{c}^\Finv_n\tilde{\xi}^n \big)=\iota_{\underline{z}}\tilde{k}^\Finv=\iota_{\underline{z}}\tilde{e}\tilde{\check{k}} +\tilde{e}\iota_{\underline{z}}\tilde{\check{k}},
        \end{align*}
    hence, noticing $\underline{\mathfrak{W}}^\Finv=\underline{\tilde{k}}^\Finv_n -\iota_{\underline{z}}\tilde{k}^\Finv$, we have
        \begin{align*}
            \underline{\mathfrak{W}}^\Finv=&\tilde{e}\underline{\tilde{\check{k}}}_n +\underline{\mu}\epsilon_n \tilde{\check{k}} +\iota_{\underline{z}}\tilde{e}\tilde{\check{k}} -\iota_{\underline{z}}\tilde{e}\tilde{\check{k}} -\tilde{e}\iota_{\underline{z}}\tilde{\check{k}}\\
            =&\tilde{e}\underline{\tilde{\check{k}}}_n +\underline{\mu}\epsilon_n \tilde{\check{k}}  -\tilde{e}\iota_{\underline{z}}\tilde{\check{k}}\in\mathrm{Im}(W_{\tilde{e}}^{(1,1)})\\
            \Leftrightarrow &\quad\epsilon_n\tilde{\check{k}}\in\mathrm{Im}(W_{\tilde{e}}^{(1,1)})
        \end{align*}
    We can see that if $\underline{\mathfrak{W}}^\Finv=\tilde{e}\underline{\tilde{\tau}}^\Finv$ for some $\underline{\tilde{\tau}}^\Finv\in\Omega^1(I)\otimes\Omega_\partial^{(1,1)}[-1]$, and $\epsilon_n\tilde{\check{k}}=\tilde{e}\tilde{\check{a}}$ for some $\tilde{\check{a}}\in\mathcal{C^1}(I)\otimes\Omega_\partial^{(1,1)}[-1] $, then
        \begin{equation}\label{eq: def tau dag a check}
            \underline{\tilde{\tau}}^\Finv=\underline{\tilde{\check{k}}}_n+ \underline{\mu}\tilde{\check{a}} - \iota_{\underline{z}}\check{k}
        \end{equation}
    For the second part of the proposition, we first apply $Q_{PC}$ to  $\check{\bm{k}}=\check{\bm{\omega}}-\iota_{\bm{\xi}}e\check{\bm{c}}-\frac{\bm{e}}{2}\iota_{\bm{\xi}}\check{\bm{c}}$, obtaining,
        \begin{align*}
            Q_{PC}\left( \check{\bm{k}} \right)&=d_{\bm{\omega}}\bm{e} + \mathrm{L}_{\bm{\xi}}^{\bm{\omega}}(\check{\bm{k}} ) - \left[\bm{c},\check{\bm{k}} \right].
        \end{align*}
     From $Q_{PC}(\bm{\check{k}})$ we can extract $Q_{PC}(\tilde{\check{k}})$, and since \eqref{eq: new BV PC constr omegadag}  is equivalent to $\epsilon_n \tilde{\check{k}}- \tilde{e}\tilde{\check{a}}=0$, for some $\tilde{\check{a}}\in \mathcal{C^1}(I)\otimes\Omega_\partial^{(1,1)}[-1] $, yielding
        \begin{align*}
            Q_{PC}(\epsilon_n \tilde{\check{k}}- \tilde{e}\tilde{\check{a}})=&-\epsilon_n\left(d_{\tilde{\omega}}\tilde{e} + \mathrm{L}_{\tilde{\xi}}^{\tilde{\omega}}(\tilde{\check{k}}) + d\tilde{\xi}^n \tilde{\check{k}}_n  - \tilde{\xi}^n d_{\tilde{\omega}_n}\tilde{\check{k}} - [\tilde{c},\tilde{\check{k}}]\right)\\
            &-( \mathrm{L}_{\tilde{\xi}}^{\tilde{\omega}}\tilde{e} - d\tilde{\xi}^n \tilde{e}_n  - d_{\tilde{\omega}_n}\tilde{e}\tilde{\xi}^n -[\tilde{c},\tilde{e}])\tilde{\check{a}}  -\tilde{e}Q_{PC}\tilde{\check{a}}=0\\
            \Leftrightarrow\hspace{2mm} \epsilon_n d_{\tilde{\omega}}\tilde{e}+ d\tilde{\xi}^n(\epsilon_n &\tilde{\check{k}}_n - \tilde{e}_n \tilde{\check{a}}) + (\mathrm{L}_{\tilde{\xi}}^{\tilde{\omega}}\epsilon_n - d_{\tilde{\omega}_n}(\epsilon_n)\tilde{\xi}^n  - [\tilde{c},\epsilon_n])\tilde{\check{k}}\in \mathrm{Im}(W_{\tilde{e}}^{(1,1)}).
        \end{align*}
    Using \eqref{eq: def tau dag a check} we see 
        \begin{align*}
            \tilde{\check{k}}_n - \tilde{e}_n \tilde{\check{a}}&=\epsilon_n\tilde{\tau}^\Finv + \mu \epsilon_n \tilde{\check{a}}+\iota_z(\epsilon_n \tilde{\check{k}}) - (\mu\epsilon_n + \iota_z\tilde{e})\tilde{\check{a}}\\
            &=\epsilon_n\tilde{\tau}^\Finv + \iota_z(\tilde{e}\tilde{\check{a}}) -\iota_z\tilde{e}\tilde{\check{a}}\\
            &=\epsilon_n\tilde{\tau}^\Finv + \tilde{e}\iota_z\tilde{a},
        \end{align*} 
    obtaining 
        \begin{equation*}
             \epsilon_n d_{\tilde{\omega}}\tilde{e}+ \epsilon_n d\tilde{\xi}^n\tilde{\tau}^\Finv + (\mathrm{L}_{\tilde{\xi}}^{\tilde{\omega}}\epsilon_n - d_{\tilde{\omega}_n}(\epsilon_n)\tilde{\xi}^n  - [\tilde{c},\epsilon_n])\tilde{\check{k}}\in\mathrm{Im}(W_{\tilde{e}}^{(1,1)}).
        \end{equation*}
\end{proof}
\subsection{BV pushforward computations}\label{proofs of BVPFSG}
\begin{proof}[Proof of \ref{lem: phi_2 Sg}]
    Adapting the proof from \cite{C25}, one can easily see that, considering the extra terms inside the supergravity structural constraint, we have 
        \begin{equation}\label{eq: phi_2 srpc}
        \begin{split}
            &\left(\phi_2^{PC}\right)^*\left( \mathcal{S}^r_{PC} +\int_{I\times\Sigma} \frac{1}{2}\underline{\tilde{e}}_n \tilde{e}[\tilde{v}-y^\Finv,\tilde{v}-y^\Finv] + g(\tilde{v}^\Finv)\right)=\mathcal{S}^r_{PC} +\int_{I\times\Sigma} \frac{1}{2}\underline{\tilde{e}}_n \tilde{e}[\tilde{v},\tilde{v}] + h_{PC}(\tilde{v}^\Finv)\\
            &+\int_{I\times\Sigma}\unl{ \mathrm{L}_{\tilde{\xi}}^{\hat{\omega}}\tilde{e} x^\Finv \tilde{\mu}^\Finv}{x1} -\unl{ \epsilon_n[\mathrm{L}_{\tilde{\xi}}^{\hat{\omega}}e, y^\Finv]\tilde{\mu}^\Finv }{x3}\\
            &\qquad -\unl{ \epsilon_n [e,\mathrm{L}_{\tilde{\xi}}^{\hat{\omega}} y^\Finv]\tilde{\mu}^\Finv}{x4}   + \unl{\epsilon_n \mathrm{L}_{\tilde{\xi}}^{\hat{\omega}}(\tilde{\bar{\psi}})\gamma\tilde{\psi}\tilde{\mu}^\Finv  }{x5}+\unl{ \frac{1}{2}\mathrm{L}_{\tilde{\xi}}^{\hat{\omega}}(\epsilon_n)^i \tilde{\mu}^\Finv_i e \tilde{\bar{\psi}}\gamma\tilde{\psi}}{x6}\\
            &\qquad + \unl{\frac{1}{2}(d_{\tilde{\omega}_n}\epsilon_n)^i \tilde{\mu}^\Finv_i\tilde{\xi}^n\tilde{\bar{\psi}}\gamma\tilde{\psi}}{x7} + \unl{(d_{\tilde{\omega}_n}e)\tilde{\xi}^n x^\Finv \tilde{\mu}^\Finv}{x8} + \unl{\epsilon_n [d_{\tilde{\omega}_n}e, y^\Finv]\tilde{\xi}^n \tilde{\mu}^\Finv}{x9} \\
            &\qquad +\unl{ \epsilon_n[e,d_{\tilde{\omega}_n}y^\Finv]\tilde{\xi}^n\tilde{\mu}^\Finv}{x10} -\unl{\epsilon_nd_{\tilde{\omega}_n}\tilde{\bar{\psi}}\gamma\tilde{\psi}\tilde{\xi}^n\tilde{\mu}^\Finv }{x11}+ \unl{\frac{1}{2}\tilde{e}_n \nu \tilde{\bar{\psi}}\gamma\tilde{\psi}}{x12} + \unl{\frac{1}{2}\tilde{e}\iota_z \nu \tilde{\bar{\psi}}\gamma\tilde{\psi} }{x13}\\
            &\qquad - \unl{\tilde{e}_n x^\Finv d\tilde{\xi}^n \tilde{\mu}^\Finv}{x14} + \unl{\epsilon_n [\tilde{e}_n,y^\Finv]d\tilde{\xi}^n \tilde{\mu}^\Finv}{x15}+\unl{ \iota_z y^\Finv d\tilde{\xi}^n \tilde{v}^\Finv }{x16}-\unl{ \epsilon_n [\tilde{c},\tilde{\bar{\psi}}]\gamma\tilde{\psi}\tilde{\mu}^\Finv}{x17} \\
            &\qquad  + \unl{\epsilon_n [[\tilde{c},\tilde{e}],y^\Finv]\tilde{\mu}^\Finv}{x18} -\unl{ [\tilde{c},y^\Finv]\tilde{v}^\Finv}{x19} - \unl{[c,e]x^\Finv\tilde{\mu}^\Finv}{x20} -\unl{\frac{1}{2} [\tilde{c},\epsilon_n]^i\tilde{\mu}^\Finv_i \tilde{e}\tilde{\bar{\psi}}\gamma\tilde{\psi}}{x21}+\unl{ \left(\delta_\chi \nu + \mathbb{q} \nu\right)\underline{\tilde{v}}^\Finv}{x22}
        \end{split}
        \end{equation}
    where $h_{PC}(\tilde{v}^\Finv)=f(v^\Finv) +  (\iota_{\tilde{\xi}}F_{\hat{\omega}} + F_{\tilde{\omega}_n}\tilde{\xi}^n + d_{\hat{\omega}}\tilde{c})\tilde{v}^\Finv$.
    We also easily see that
        \begin{equation}\label{eq: phi3 srpc}
        \begin{split}
            &\phi_3^*\left( \mathcal{S}^r_{PC} +\int_{I\times\Sigma} \frac{1}{2}\underline{\tilde{e}}_n \tilde{e}[\tilde{v}-y^\Finv,\tilde{v}-y^\Finv] + g(\tilde{v}^\Finv)\right)=\\
            &=\int_{I\times\Sigma} \unl{(\delta_\chi y^\Finv + \mathbb{q}y^\Finv)\tilde{v}^\Finv }{d1} + \mathrm{L}_{\tilde{\xi}}^{\hat{\omega}}\tilde{e}\left( -\unl{ x^\Finv \tilde{\mu}^\Finv }{d2}+  \unl{ [\epsilon_n\tilde{\mu}^\Finv,y^\Finv]}{d3} \right)+ d_{\tilde{\omega}\tilde{e}}\tilde{\xi}^n \left( -\unl{ x^\Finv \tilde{\mu}^\Finv }{d4}+  \unl{ [\epsilon_n\tilde{\mu}^\Finv,y^\Finv]}{d5} \right)\\
            &\qquad  + \tilde{e}_n d\tilde{\xi}^n \left( -\unl{ x^\Finv \tilde{\mu}^\Finv }{d6}+  \unl{ [\epsilon_n\tilde{\mu}^\Finv,y^\Finv]}{d7} \right) - [\tilde{c},\tilde{e}]\left( -\unl{ x^\Finv \tilde{\mu}^\Finv }{d8}+  \unl{ [\epsilon_n\tilde{\mu}^\Finv,y^\Finv]}{d9} \right) + \unl{\mathrm{L}_{\tilde{\xi}}^{\hat{\omega}}y^\Finv \tilde{v}^\Finv}{d10}\\
            &\qquad + \unl{d_{\tilde{\omega}_n}y^\Finv \tilde{\xi}^n \tilde{v}^\Finv}{d11} -\unl{ [c,y^\Finv]\tilde{v}^\Finv}{d12}  .
        \end{split}
        \end{equation}
    We immediately see
    \begin{itemize}
        \item $\reft{eq: phi3 srpc}{d2} + \reft{eq: phi_2 srpc}{x1}=0$.
        \item $\reft{eq: phi3 srpc}{d3} + \reft{eq: phi_2 srpc}{x3}=0$.
        \item $\reft{eq: phi3 srpc}{d4} + \reft{eq: phi_2 srpc}{x8}=0$.
        \item $\reft{eq: phi3 srpc}{d5} + \reft{eq: phi_2 srpc}{x9}=0$.
        \item $\reft{eq: phi3 srpc}{d6} + \reft{eq: phi_2 srpc}{x14}=0$.
        \item $\reft{eq: phi3 srpc}{d7} + \reft{eq: phi_2 srpc}{x15}=0$.
        \item $\reft{eq: phi3 srpc}{d8} + \reft{eq: phi_2 srpc}{x20}=0$.
        \item $\reft{eq: phi3 srpc}{d9} + \reft{eq: phi_2 srpc}{x19}=0$.
        \item $\reft{eq: phi3 srpc}{d10} + \reft{eq: phi_2 srpc}{x3}=0$.
        \item $\reft{eq: phi3 srpc}{d11} + \reft{eq: phi_2 srpc}{x9}=0$.
        \item $\reft{eq: phi3 srpc}{d12} + \reft{eq: phi_2 srpc}{x18}=0$.
    \end{itemize}
    We are left with computing $\phi_2^*(s^r_\psi+s^r_2)$, where $s_2$ is given by \eqref{eq: quadratic BV action} and
    \begin{align*}
        s^r_\psi\coloneqq\int_{I\times M}& \frac{1}{3!}\left(\underline{\tilde{e}}_n\tilde{\bar{\psi}}\gamma^3 d_{\hat{\omega}}\tilde{\psi}+ \tilde{e}\underline{\tilde{\bar{\psi}}}_n\gamma^3 d_{\hat{\omega}}\tilde{\psi} + \tilde{e}\tilde{\bar{\psi}}\gamma^3 d_{\underline{\tilde{\omega}}_n}\tilde{\psi} + \tilde{e}\tilde{\bar{\psi}}\gamma^3d_{\hat{\omega}}\underline{\tilde{\psi}}_n   \right)\\
        & - \tilde{\bar{\chi}}\gamma\underline{\tilde{\psi}}_n \tilde{e}^\Finv -\tilde{\bar{\chi}}\gamma\tilde{\psi}\underline{\tilde{e}}^\Finv_n - \frac{1}{3!}\tilde{\bar{\chi}}\gamma^3\left(d_{\underline{\tilde{\omega}}_n}\tilde{\psi} +d_{\hat{\omega}}\underline{\tilde{\psi}}_n  \right)\tilde{\check{k}} - \frac{1}{3!}\tilde{\bar{\chi}}\gamma^3d_{\hat{\omega}}\tilde{\psi}\underline{\tilde{\check{k}}}_n \\
        &- i\left( \mathrm{L}_{\tilde{\xi}}^{\hat{\omega}}\tilde{\bar{\psi}} + d_{\tilde{\omega}_n}\tilde{\bar{\psi}}\tilde{\xi}^n  + \tilde{\bar{\psi}}_nd\tilde{\xi}^n -[\tilde{c},\tilde{\bar{\psi}}] - d_{\hat{\omega}}\tilde{\bar{\chi}} \right)\underline{\tilde{\psi}^\Finv}_{n}\\
        &- i\left(\mathrm{L}_{\tilde{\xi}}^{\hat{\omega}}\underline{\tilde{\bar{\psi}}}_n +\iota_{\underline{\partial_n}\tilde{\xi}}\tilde{\bar{\psi}}  -d_{\underline{\tilde{\omega}}_n}(\tilde{\bar{\psi}}_n\tilde{\xi}_n) -[\tilde{c},\tilde{\bar{\psi}}_n] - d_{\underline{\tilde{\omega}}_n}\tilde{\bar{\chi}}  \right)\tilde{\psi}^\Finv\\
        &- i\left( \mathrm{L}_{\tilde{\xi}}^{\hat{\omega}}\tilde{\bar{\chi}} + d_{\tilde{\omega}_n}\tilde{\bar{\chi}}\tilde{\xi}^n   -[\tilde{c},\tilde{\bar{\chi}}] - \frac{1}{2}\iota_{\tilde{\varphi}}\tilde{\bar{\psi}} -\frac{1}{2}\tilde{\bar{\psi}}_n\tilde{\varphi}^n  \right)\underline{\tilde{\chi}^\Finv}_{n}\\
        &+ \frac{1}{2}\iota_{\tilde{\varphi}}\underline{\tilde{\xi}}^\Finv + \frac{1}{2}\underline{\tilde{\xi}}^\Finv_n \tilde{\varphi}^n.
    \end{align*}
    Notice that $\phi_2^*(\tilde{k}^\Finv)=0$, hence we obtain
    \begin{equation}\label{eq: phi2 srsg}
    \begin{split}
        &\phi_2^*\left( s^r_\psi+s^r_2\right)=s^r_\psi+s^r_2\\
        =\int_{I\times\Sigma}&\frac{1}{3!}\left( \unl{\tilde{e}_n\tilde{\bar{\psi}}\gamma^3[\nu,\tilde{\psi}]}{g1} + \unl{\tilde{e}\tilde{\bar{\psi}}_n\gamma^3[\nu,\tilde{\psi}]}{g2} + \unl{ \tilde{e}\tilde{\bar{\psi}}\gamma^3[\iota_z \nu}{g3} + \unl{\iota_{\tilde{X}}\tilde{\mu}^\Finv,\tilde{\psi}]}{g4} + \unl{e \tilde{\bar{\psi}}\gamma^3[\nu,\tilde{\psi}_n]}{g5} \right)\\
        &+\unl{\epsilon_n\mathbb{q}_{\tilde{\bar{\psi}}}\gamma\tilde{\psi}\tilde{\mu}^\Finv}{g6}+ \tilde{\bar{\chi}}\gamma\tilde{\psi}\left(\unl{ d_{\hat{\omega}}(\epsilon_n \tilde{\mu}^{\Finv})}{g7} + \unl{\sigma\tilde{\mu}^\Finv}{g8} + \unl{\tilde{\check{k}}_n \nu}{g9} + \unl{\iota_z \nu \tilde{\check{k}}}{g10} + \unl{\iota_{\tilde{X}}(\tilde{\check{k}}\tilde{\mu}^\Finv)}{g11} \right)\\
        &+\tilde{\bar{\chi}}\gamma\tilde{\psi}\left( \unl{x^\Finv\tilde{\mu}^\Finv}{g12} + \unl{\epsilon_n[\tilde{e},y^\Finv]}{g13} \right) + \unl{\tilde{\bar{\chi}}\gamma\tilde{\psi}_n\tilde{\check{k}}}{g14} - \frac{1}{3!}\tilde{\bar{\psi}}\gamma^3\left([\unl{\iota_z \nu}{g15} +\unl{ \iota_{\tilde{X}}\tilde{\mu}^\Finv}{g16},\psi] + \unl{[\nu,\tilde{\psi}_n]}{g17}  \right)\tilde{\check{k}} \\
        &- \unl{\frac{1}{3!}\tilde{\bar{\chi}}\gamma^3[\nu,\tilde{\psi}]\tilde{\check{k}}_n}{g18} - \unl{i [\nu,\tilde{\bar{\chi}}]\tilde{\psi}^\Finv_n}{g19} + \epsilon_n \left( \unl{\mathrm{L}_{\tilde{\xi}}^{\hat{\omega}}\tilde{\bar{\psi}}}{g20} +\unl{ d_{\tilde{\omega}_n}\tilde{\bar{\psi}}\tilde{\xi}^n}{g21} + \unl{\tilde{\bar{\psi}}_n d\tilde{\xi}^n}{g22} - \unl{[\tilde{c},\tilde{\bar{\psi}}]}{g23}  \right)\gamma\tilde{\psi}\tilde{\mu}^\Finv\\
        &- \epsilon_n \left(\unl{ d_{\hat{\omega}}\tilde{\bar{\chi}}}{g24} + \unl{[\nu,\tilde{\bar{\chi}}] }{g25}\right)\gamma\tilde{\psi}\tilde{\mu}^\Finv - i[\unl{\iota_z \nu}{g26} + \unl{\iota_{\tilde{X}}\tilde{\mu}^\Finv}{g27},\tilde{\bar{\chi}}]\tilde{\psi}^\Finv + \frac{1}{2}\iota_{\tilde{\varphi}}\nu\left(\unl{ \tilde{c}^\Finv_n}{g28} + \unl{[\epsilon_n , \tilde{\check{k}}\tilde{\mu}^\Finv] }{g29}\right) \\
        &+ \unl{\frac{1}{2}\mathrm{L}_{\tilde{\varphi}}^{\hat{\omega}}(\epsilon_n)\tilde{\check{k}\tilde{\mu}^\Finv}}{g30} + \frac{1}{2}\tilde{\varphi}^n\left(\unl{ \iota_{\tilde{X}}\tilde{c}^\Finv_n \tilde{\mu}^\Finv + d(\epsilon_n {\tau}^\Finv \tilde{\mu}^\Finv)}{g31} +\unl{ \iota_z\tilde{c}^\Finv_n \nu }{g32} +\unl{ \iota_z\nu[\epsilon_n,\tilde{\check{k}}\tilde{\mu}^\Finv]}{g33}\right)\\
        &+ \tilde{\varphi}^n\left( \unl{(d_{\tilde{\omega}_n}\epsilon_n)\tilde{\check{k}}\tilde{\mu}^\Finv }{g34}+ \unl{\iota_{\tilde{X}} \tilde{\mu}^\Finv[\epsilon_n,\tilde{\check{k}}\tilde{\mu}^\Finv]}{g35} \right) + \frac{1}{2}\left(\unl{ d_{\hat{\omega}}(\epsilon_n\tilde{\mu}^\Finv)}{g36} + \unl{\sigma\tilde{\mu}^\Finv}{g37} + \unl{\tilde{\check{k}}_n \nu}{g38} + \unl{\iota_z\nu \tilde{\check{k}}}{g39}  \right)\iota_{\tilde{\varphi}}\tilde{\check{k}}\\
        &+ \frac{1}{2}\left(\unl{\iota_{\tilde{X}}(\tilde{\check{k}}\tilde{\mu}^\Finv)}{g40}+\unl{x^\Finv\tilde{\mu}^\Finv}{g41} + \unl{\epsilon_n[\tilde{e},y^\Finv] }{g42}\right)\iota_{\tilde{\varphi}}\tilde{\check{k}} +\frac{1}{2}\left( \unl{d_{\hat{\omega}}(\epsilon_n\tilde{\mu}^\Finv) }{g43}+ \unl{\sigma\tilde{\mu}^\Finv}{g44} + \unl{\tilde{\check{k}}_n \nu}{g45} + \unl{\iota_z\nu \tilde{\check{k}} }{g46}   \right)\tilde{\check{k}}_n\tilde{\varphi}^n\\
        &+ \frac{1}{2}\left(\unl{\iota_{\tilde{X}}(\tilde{\check{k}}\tilde{\mu}^\Finv)}{g47}+\unl{x^\Finv\tilde{\mu}^\Finv}{g48} + \unl{\epsilon_n[\tilde{e},y^\Finv] }{g49}\right)\tilde{\check{k}}_n\tilde{\varphi}^n + \unl{\frac{1}{2}\nu\tilde{\check{k}}\iota_{\tilde{\varphi}}\tilde{\check{k}}_n}{g50}
    \end{split}
    \end{equation}
    We can then see
    \begin{itemize}
        \item $\reft{eq: phi2 srsg}{g1} + \reft{eq: phi_2 srpc}{x12}=0$.
        \item $\reft{eq: phi2 srsg}{g2} +\reft{eq: phi2 srsg}{g5}+\reft{eq: phi2 srsg}{g21} =0$.
        \item $\reft{eq: phi2 srsg}{g3} + \reft{eq: phi_2 srpc}{x13}=0$.
        \item $\reft{eq: phi2 srsg}{g4} + \reft{eq: phi_2 srpc}{x6} + \reft{eq: phi_2 srpc}{x7}+ \reft{eq: phi_2 srpc}{x21}=0$.
        \item $\reft{eq: phi2 srsg}{g20} + \reft{eq: phi_2 srpc}{x5}=0$.
        \item $\reft{eq: phi2 srsg}{g21} + \reft{eq: phi_2 srpc}{x11}=0$.
        \item $\reft{eq: phi2 srsg}{g23} + \reft{eq: phi_2 srpc}{x17}=0$.
    \end{itemize}
    Therefore, noticing that $\mathcal{S}^r_{\mathrm{SG}}=\mathcal{S}^r_{PC}+s_\psi^r + s_2^r$, we have 
        \begin{equation}\label{eq: phi3 SG}
        \begin{split}
            &\phi_2^*\left( \mathcal{S}^r_{\mathrm{SG}}+\int_{I\times\Sigma} \frac{1}{2}\underline{\tilde{e}}_n \tilde{e}[\tilde{v}-y^\Finv,\tilde{v}-y^\Finv] + g(\tilde{v}^\Finv)\right)=\\
            &=\mathcal{S}^r_{\mathrm{SG}}+\int_{I\times\Sigma}\frac{1}{2}\underline{\tilde{e}}_n \tilde{e}[\tilde{v},\tilde{v}] + f(\tilde{v}^\Finv)+ \left((\delta_{\tilde{\chi}} + \mathbb{q})\tilde{v} + \unl{(\delta_{\tilde{\chi}} + \mathbb{q})y^\Finv  }{f1}+ 
            \right)\tilde{v}^\Finv\\
            &\qquad +\unl{\epsilon_n\mathbb{q}_{\tilde{\bar{\psi}}}\gamma\tilde{\psi}\tilde{\mu}^\Finv}{f3}+ \tilde{\bar{\chi}}\gamma\tilde{\psi}\left(\unl{ d_{\hat{\omega}}(\epsilon_n \tilde{\mu}^{\Finv})}{f4} + \unl{\sigma\tilde{\mu}^\Finv}{f5} + \unl{\tilde{\check{k}}_n \nu}{f6} + \unl{\iota_z \nu \tilde{\check{k}}}{f7} + \unl{\iota_{\tilde{X}}(\tilde{\check{k}}\tilde{\mu}^\Finv)}{f8} \right)\\
            &\qquad +\tilde{\bar{\chi}}\gamma\tilde{\psi}\left( \unl{x^\Finv\tilde{\mu}^\Finv}{f9} + \unl{\epsilon_n[\tilde{e},y^\Finv]}{f10} \right) + \unl{\tilde{\bar{\chi}}\gamma\tilde{\psi}_n\tilde{\check{k}}\nu}{f11} - \frac{1}{3!}\tilde{\bar{\chi}}\gamma^3\left([\unl{\iota_z \nu}{f12} +\unl{ \iota_{\tilde{X}}\tilde{\mu}^\Finv}{f13},\psi] + \unl{[\nu,\tilde{\psi}_n]}{f14}  \right)\tilde{\check{k}} \\
            &\qquad - \unl{\frac{1}{3!}\tilde{\bar{\chi}}\gamma^3[\nu,\tilde{\psi}]\tilde{\check{k}}_n}{f15} - \unl{i [\nu,\tilde{\bar{\chi}}]\tilde{\psi}^\Finv_n}{f16}- \epsilon_n \left(\unl{ d_{\hat{\omega}}\tilde{\bar{\chi}}}{f17} + \unl{[\nu,\tilde{\bar{\chi}}] }{f18}\right)\gamma\tilde{\psi}\tilde{\mu}^\Finv \\
            &\qquad - i[\unl{\iota_z \nu}{f19} + \unl{\iota_{\tilde{X}}\tilde{\mu}^\Finv}{f20},\tilde{\bar{\chi}}]\tilde{\psi}^\Finv   + \frac{1}{2}\iota_{\tilde{\varphi}}\nu\left(\unl{ \tilde{c}^\Finv_n}{f21} + \unl{[\epsilon_n , \tilde{\check{k}}\tilde{\mu}^\Finv] }{f22}\right) \\
            &\qquad + \unl{\frac{1}{2}\mathrm{L}_{\tilde{\varphi}}^{\hat{\omega}}(\epsilon_n)\tilde{\check{k}\tilde{\mu}^\Finv}}{f23} + \frac{1}{2}\tilde{\varphi}^n\left(\unl{ \iota_{\tilde{X}}\tilde{c}^\Finv_n \tilde{\mu}^\Finv + d(\epsilon_n {\tau}^\Finv \tilde{\mu}^\Finv)}{f24} +\unl{ \iota_z\tilde{c}^\Finv_n \nu }{f25} +\unl{ \iota_z\nu[\epsilon_n,\tilde{\check{k}}\tilde{\mu}^\Finv]}{f26}\right)\\
            &\qquad + \tilde{\varphi}^n\left( \unl{(d_{\tilde{\omega}_n}\epsilon_n)\tilde{\check{k}}\tilde{\mu}^\Finv }{f27}+ \unl{\iota_{\tilde{X}} \tilde{\mu}^\Finv[\epsilon_n,\tilde{\check{k}}\tilde{\mu}^\Finv]}{f28} \right) + \frac{1}{2}\left(\unl{ d_{\hat{\omega}}(\epsilon_n\tilde{\mu}^\Finv)}{f29} + \unl{\sigma\tilde{\mu}^\Finv}{f30} + \unl{\tilde{\check{k}}_n \nu}{f31} + \unl{\iota_z\nu \tilde{\check{k}}}{f32}  \right)\iota_{\tilde{\varphi}}\tilde{\check{k}}\\
            &\qquad + \frac{1}{2}\left(\unl{\iota_{\tilde{X}}(\tilde{\check{k}}\tilde{\mu}^\Finv)}{f33}+\unl{x^\Finv\tilde{\mu}^\Finv}{f34} + \unl{\epsilon_n[\tilde{e},y^\Finv] }{f35}\right)\iota_{\tilde{\varphi}}\tilde{\check{k}} +\frac{1}{2}\left( \unl{d_{\hat{\omega}}(\epsilon_n\tilde{\mu}^\Finv) }{f36}+ \unl{\sigma\tilde{\mu}^\Finv}{f37} + \unl{\tilde{\check{k}}_n \nu}{f38} + \unl{\iota_z\nu \tilde{\check{k}} }{f39}   \right)\tilde{\check{k}}_n\tilde{\varphi}^n\\
            &\qquad + \frac{1}{2}\left(\unl{\iota_{\tilde{X}}(\tilde{\check{k}}\tilde{\mu}^\Finv)}{f40}+\unl{x^\Finv\tilde{\mu}^\Finv}{f41} + \unl{\epsilon_n[\tilde{e},y^\Finv] }{f42}\right)\tilde{\check{k}}_n\tilde{\varphi}^n + \unl{\frac{1}{2}\nu\tilde{\check{k}}\iota_{\tilde{\varphi}}\tilde{\check{k}}_n}{f43}.
        \end{split}
        \end{equation}
    We are therefore left with showing that all the underlined terms above exacltly correspond to $(\delta_{\tilde{\chi}}+\mathbb{q} )\hat{\omega}\tilde{v}^\Finv$. We already remarked that we only know $e(\delta_{\tilde{\chi}}+\mathbb{q} )$ and not the full expression. However, we notice
        \begin{equation*}
            (\delta_{\tilde{\chi}}+\mathbb{q} )\hat{\omega}\tilde{v}^\Finv=\epsilon_n[(\delta_{\tilde{\chi}}+\mathbb{q} )\hat{\omega},\tilde{e}]\tilde{\mu}^\Finv.
        \end{equation*}
    It is then enough to apply the operator $(\delta_{\tilde{\chi}}+\mathbb{q} )$ to the constraint \eqref{constr: BV SG red omega}, and isolate the term we need exactly. In particular, we see that 
        \begin{equation*}
            (\delta_{\tilde{\chi}}+\mathbb{q} )\left[ \epsilon_n \left( d_{\hat{\omega}}\tilde{e} - \frac{1}{2}\tilde{\bar{\psi}}\gamma\tilde{\psi}  + (\tilde{\check{k}}_n - \iota_z \tilde{\check{k}})d\tilde{\xi}^n\right) + \iota_{\tilde{X}}\tilde{k}^\Finv + \tilde{e}x^\Finv + \epsilon_n[\tilde{e},y^\Finv] - \tilde{e}\sigma \right]\tilde{\mu^{\Finv}}=0
        \end{equation*}
     yields the relevant term $\epsilon_n[(\delta_{\tilde{\chi}}+\mathbb{q} )\hat{\omega},\tilde{e}]\tilde{\mu}^\Finv$.\footnote{We have used that $\tau^\Finv= \tilde{\check{k}}_n - \iota_z \tilde{\check{k}} + \check{a} $ and that $\epsilon_n \check{a}=0$.}   We apply the operator to each addend one by one, obtaining
        \begin{equation}\label{eq: delta q omega}
            \begin{split}
                \epsilon_n[(\delta_{\tilde{\chi}}+\mathbb{q} )&(\hat{\omega}+\nu),\tilde{e}]\tilde{\mu}^\Finv=\\
                =&\epsilon_n\left( - \unl{\tilde{\bar{\chi}}\gamma d_{\hat{\omega}}\tilde{\psi}}{j1}  + \unl{d_{\hat{\omega}}(\iota_{\tilde{\varphi}}\tilde{\check{k}}}{j2} +\unl{\tilde{\check{k}}_n\tilde{\varphi}^n )}{j3} + \unl{\mathbb{q}_{\tilde{\bar{\psi}}}\gamma\tilde{\psi}}{j4}  \right)\mu^\Finv +\unl{\frac{1}{3!}[\nu,\tilde{\bar{\chi}}]\gamma^3\tilde{\psi}_n\tilde{\check{k}}}{j5} \\
                &+\unl{ \frac{1}{3!}[\nu,\tilde{\bar{\chi}}]\gamma^3 \tilde{\psi}\tilde{\check{k}}_n}{j6} -\unl{i[\nu,\tilde{\bar{\chi}}]\tilde{\psi}^\Finv_n}{j7} -\unl{\frac{1}{2}\nu\iota_{\tilde{\varphi}}\tilde{c}^\Finv_n }{j8}+\unl{\frac{1}{3!}[\iota_z \nu,\tilde{\bar{\chi}}]   \gamma^3\tilde{\psi}\tilde{\check{k}}}{j9}\\
                &-\unl{i[\iota_z\nu,\tilde{\bar{\chi}}]\tilde{\psi}^\Finv}{j10} -\unl{\frac{1}{2}\iota_z\nu \tilde{c}^\Finv_n \tilde{\varphi}^n}{j11}-\unl{\frac{1}{2}\nu\iota_{\tilde{\varphi}}(\tilde{\check{k}}_n \tilde{\check{k}})}{j12} -\unl{\frac{1}{2}\iota_z \nu \tilde{\check{k}}(\iota_{\tilde{\varphi}}\tilde{\check{k}}+\tilde{\check{k}}\tilde{\varphi}^n}{j13}\\
                &+ \unl{( \tilde{\check{k}}_n -\iota_z \tilde{\check{k}} )d\tilde{\varphi}^n \tilde{\mu}^\Finv}{j14} +\frac{1}{2}\tilde{\mu}^\Finv\left( \unl{\mathrm{L}_{\tilde{\varphi}}^{\hat{\omega}}\epsilon_n}{j15} -\unl{d_{\tilde{\omega}_n}\epsilon_n \tilde{\varphi} ^n}{j16}\right)\tilde{\check{k}} + \unl{\frac{1}{3!}\tilde{\bar{\chi}}\gamma^3[\iota_{\tilde{X}}\tilde{\mu}^\Finv,\tilde{\psi}]}{j17}\\
                &-\unl{\frac{1}{2}\tilde{c}^\Finv_n \tilde{\varphi}^n \iota_{\tilde{X}}\tilde{\mu}^\Finv}{j18} - \unl{i[\iota_{\tilde{X}}\tilde{\mu}^\Finv,\tilde{\bar{\chi}}]\tilde{\psi}^\Finv }{j19}+ \unl{\tilde{\bar{\chi}}\gamma\tilde{\psi}x^\Finv\tilde{\mu}^\Finv }{j20}\\
                &+\unl{\tilde{\bar{\chi}}\gamma\tilde{\psi}\iota_{\tilde{X}}(\tilde{\check{k}}\tilde{\mu})}{j31} + \unl{\tilde{\mu}^\Finv (\mathbb{q}\tilde{X})\tilde{\check{k}}}{j32} -\frac{1}{2}\left(\unl{\iota_{\tilde{\varphi}}\tilde{\check{k}}}{j33} +\unl{\tilde{\check{k}}_n\tilde{\varphi}^n}{j34}\right)\iota_{\tilde{X}}(\tilde{\check{k}}\tilde{\mu}^\Finv)\\
                &+ \frac{1}{2}\left( \unl{\iota_{\tilde{\varphi}}\tilde{\check{k}}+\tilde{\check{k}}_n\tilde{\varphi}^n }{j21}\right)x^\Finv\tilde{\mu}^\Finv + \epsilon_n\left[ \unl{\tilde{\bar{\chi}}\gamma\tilde{\psi}}{j22} + \frac{1}{2}\left( \unl{\iota_{\tilde{\varphi}}\tilde{\check{k}} +\tilde{\check{k}}_n\tilde{\varphi}^n}{j23} \right), y^\Finv\right]\tilde{\mu}^\Finv\\
                &+\unl{\epsilon_n[\tilde{e},(\delta_\chi + \mathbb{q})y^\Finv]\tilde{\mu}^\Finv}{j24} + \unl{\tilde{\bar{\chi}}\gamma\tilde{\psi}\sigma\tilde{\mu}^\Finv }{j25}+ \frac{1}{2}\left(\unl{ \iota_{\tilde{\varphi}}\tilde{\check{k}} +\tilde{\check{k}}_n\tilde{\varphi}^n}{j26} \right)\sigma\tilde{\mu}^\Finv \\
                &-\unl{[\nu,\tilde{\bar{\chi}}]\gamma\tilde{\psi}\epsilon_n\tilde{\mu}^\Finv}{j27} -\unl{\frac{1}{2}\iota_{\tilde{\varphi}} \nu [\epsilon_n, \tilde{\check{k}}\tilde{\mu}^\Finv]}{j28} -\unl{\frac{1}{2}\iota_z \nu [\epsilon_n, \tilde{\check{k}}\tilde{\mu}^\Finv]\tilde{\varphi}^n }{j29}-\unl{\frac{1}{2}[\epsilon_n, \tilde{\check{k}}\tilde{\mu}^\Finv]\tilde{\varphi}^n \iota_{\tilde{X}}\tilde{\mu}^\Finv}{j30}  
           \end{split}
        \end{equation}
    where we have used the definition of $\delta_\chi $ and $\mathbb{q}$ from \eqref{eq: q fields} and \eqref{eq: Q_0 fields}.\footnote{For completeness, we compute explicitly the terms $\epsilon_n(\delta_\chi \tilde{\check{k}}_n + \iota_z \delta\tilde{\check{k}} )d\tilde{\xi}^n\tilde{\mu}^\Finv$. The computation works for $\mathbb{q}$ equivalently. We know from \eqref{eq: Q_0 fields} that in the bulk
        \begin{equation}\label{eq: e delta k}
            \bm{e}\delta_\chi \bm{\check{k}}= -\frac{1}{2}\bm{\bar{\chi}}\gamma\bm{\psi}\bm{\check{k}} - \frac{1}{2\cdot 3!}\bm{\bar{\chi}}[\bm{\check{k}},\gamma^3]\bm{\psi} + i[\bm{\bar{\psi}}^\Finv,\bm{\chi}] -\frac{1}{2}\bm{\vf}\bm{c}^\Finv.
        \end{equation}
    Selecting the component along $dx^n$, we obtain     
        \begin{equation*}
            \tilde{e}_n \delta_\chi \tilde{\check{k}} + \tilde{e}\delta_{\chi}\tilde{\check{k}}_n= \iota_z(\tilde{e}\delta_\chi\tilde{\check{k}}) + \mu\epsilon_n \delta_\chi \tilde{\check{k}} + \tilde{e}(\delta_\chi \tilde{\check{k}}_n -\iota_z \delta_\chi \tilde{\check{k}}),
        \end{equation*}
    hence, using $\epsilon_n d\tilde{\xi}^n\mu^\Finv= \tilde{e}\nu$, with $\epsilon_n \nu = 0$, we have
        \begin{align*}
            \epsilon_n(\delta_\chi \tilde{\check{k}}_n + \iota_z \delta\tilde{\check{k}} )d\tilde{\xi}^n\tilde{\mu}^\Finv&=\nu \tilde{e}(\delta_\chi \tilde{\check{k}}_n -\iota_z \delta_\chi \tilde{\check{k}})\\
            &=\nu\left( \tilde{e}_n \delta_\chi \tilde{\check{k}} + \tilde{e}\delta_{\chi}\tilde{\check{k}}_n - \iota_z(\tilde{e}\delta_\chi\tilde{\check{k}})  \right),
        \end{align*}
    which gives the desired result, as  $\tilde{e}_n \delta_\chi \tilde{\check{k}} + \tilde{e}\delta_{\chi}\tilde{\check{k}}_n$ and $\tilde{e}\delta_\chi\tilde{\check{k}}$ can be found respectively as the transversal and tangential components of \eqref{eq: e delta k}. 
    } Furthermore, we have used the fact that $ \hat{\omega}+\nu=\phi_2(\hat{\omega})$ to compute $\epsilon_n\mu^\Finv[\tilde{e},(\delta_\chi+\mathbb{q})\nu]=(\delta_\chi+\mathbb{q})\nu\tilde{v}^\Finv$.
    
    We are just left with checking that all the terms in \eqref{eq: delta q omega} appear in \eqref{eq: phi3 SG}. We notice
    \begin{itemize}
        \item $\reft{eq: phi3 SG}{f1}=\reft{eq: delta q omega}{j24}$;
        \item $\reft{eq: phi3 SG}{f3}=\reft{eq: delta q omega}{j4}$;
        \item $\reft{eq: phi3 SG}{f4}+\reft{eq: phi3 SG}{f17}=\reft{eq: delta q omega}{j1}$ having used integration by parts;
        \item $\reft{eq: phi3 SG}{f5}=\reft{eq: delta q omega}{j25}$;
        \item $\reft{eq: phi3 SG}{f6}+\reft{eq: phi3 SG}{f15}=\reft{eq: delta q omega}{j6}$;
        \item $\reft{eq: phi3 SG}{f7}+\reft{eq: phi3 SG}{f12}=\reft{eq: delta q omega}{j9}$;
        \item $\reft{eq: phi3 SG}{f8}=\reft{eq: delta q omega}{j31}$;
        \item $\reft{eq: phi3 SG}{f10}=\reft{eq: delta q omega}{j22}$;
        \item $\reft{eq: phi3 SG}{f11}+\reft{eq: phi3 SG}{f14}=\reft{eq: delta q omega}{j5}$;
        \item $\reft{eq: phi3 SG}{f13}=\reft{eq: delta q omega}{j17}$;
        \item $\reft{eq: phi3 SG}{f16}=\reft{eq: delta q omega}{j7}$;
        \item $\reft{eq: phi3 SG}{f18}=\reft{eq: delta q omega}{j27}$;
        \item $\reft{eq: phi3 SG}{f19}=\reft{eq: delta q omega}{j10}$;
        \item $\reft{eq: phi3 SG}{f20}=\reft{eq: delta q omega}{j19}$;
        \item $\reft{eq: phi3 SG}{f21}=\reft{eq: delta q omega}{j8}$;
        \item $\reft{eq: phi3 SG}{f22}=\reft{eq: delta q omega}{j28}$;
        \item $\reft{eq: phi3 SG}{f23}=\reft{eq: delta q omega}{j15}$;
        \item $\reft{eq: phi3 SG}{f24}=\reft{eq: delta q omega}{j18}+\reft{eq: delta q omega}{j14}$;
        \item $\reft{eq: phi3 SG}{f25}=\reft{eq: delta q omega}{j11}$;
        \item $\reft{eq: phi3 SG}{f26}=\reft{eq: delta q omega}{j29}$;
        \item $\reft{eq: phi3 SG}{f27}=\reft{eq: delta q omega}{j16}$;
        \item $\reft{eq: phi3 SG}{f28}=\reft{eq: delta q omega}{j30}$;
        \item $\reft{eq: phi3 SG}{f29}+\reft{eq: phi3 SG}{f36}=\reft{eq: delta q omega}{j2}+\reft{eq: delta q omega}{j3}$;
        \item $\reft{eq: phi3 SG}{f30}+\reft{eq: phi3 SG}{f37}=\reft{eq: delta q omega}{j26}$;
        \item $\reft{eq: phi3 SG}{f31}+\reft{eq: phi3 SG}{f43}=\reft{eq: delta q omega}{j12}$;
        \item $\reft{eq: phi3 SG}{f32}+\reft{eq: phi3 SG}{f39}=\reft{eq: delta q omega}{j13}$;
        \item $\reft{eq: phi3 SG}{f33}+\reft{eq: phi3 SG}{f40}=\reft{eq: delta q omega}{j33}+\reft{eq: delta q omega}{j34}$;
        \item $\reft{eq: phi3 SG}{f34}+\reft{eq: phi3 SG}{f41}=\reft{eq: delta q omega}{j21}$;
        \item $\reft{eq: phi3 SG}{f35}+\reft{eq: phi3 SG}{f42}=\reft{eq: delta q omega}{j23}$;
        \item $\reft{eq: phi3 SG}{f38}=0$, because $\tilde{\check{k}}_n^2=0$, since it is an odd quantity,
    \end{itemize}
    which concludes the proof.
\end{proof}
\begin{proof}[Proof of \ref{lem: phi1 SG}]
    We start by noticing that all the terms of the quadratic part $s_2$ of $\mathcal{S}_{\mathrm{SG}}$ are left unchanged by $\phi_1$, except for the term $\frac{1}{2}\check{\bm{k}}\iota_{\bm{\varphi}}\bm{e}^\Finv$.
    Therefore, letting
        \begin{align*}
            \mathcal{A}_{\mathrm{SG}}:&=\int_{I\times \Sigma} \frac{\bm{e}^2}{2} F_{\bm{\omega}} + \frac{1}{3!} \bm{e}\bar{\bm{\psi}}\gamma^3d_{\bm{\omega}}\bm{\psi} - (\mathrm{L}_{\bm{\xi}}^{\bm{\omega}} \bm{e} - [{\bm{c}},\bm{e}] + \bar{\bm{\chi}}\gamma\bm{\psi}) \bm{e}^\Finv\\
            &\qquad + (\iota_{\bm{\xi}} F_{\bm{\omega}} - d_{\bm{\omega}} {\bm{c}} + \delta_{\bm{\chi}}\bm{\omega})\bm{\omega}^\Finv -i(\mathrm{L}_{\bm{\xi}}^{\bm{\omega}} \bar{\bm{\psi}} - [{\bm{c}},\bar{\bm{\psi}}] - d_{\bm{\omega}}\bar{\bm{\chi}} )\bm{\psi}^\Finv\\
            &\qquad + \left(\frac{1}{2} \iota_{\bm{\xi}} \iota_{\bm{\xi}} F_{\bm{\omega}} -\frac{1}{2}  [{\bm{c}},{\bm{c}}] + \iota_{\bm{\xi}}\delta_{\bar{\chi}}\bm{\omega} \right) {\bm{c}}^\Finv+ \frac{1}{2} \iota_{[{\bm{\xi}},{\bm{\xi}}]} {\bm{\xi}}^\Finv + \frac{1}{2} \iota_{\bm{\varphi}} {\bm{\xi}}^\Finv\\
            &\qquad  -i\left(\mathrm{L}_{\bm{\xi}}^{\bm{\omega}} \bar{\bm{\chi}} - [{\bm{c}},\bar{\bm{\chi}}] - \frac{1}{2}\iota_{\bm{\varphi}}\bar{\bm{\psi}} \right)\bm{\chi}^\Finv + \frac{1}{2}\left(\check{\bm{\omega}} - \iota_{\bm{\xi}}\bm{e}\check{\bm{c}} -\frac{\bm{e}}{2}\iota_{\bm{\xi}}\check{\bm{c}} \right)\iota_{\bm{\varphi}}\bm{e}^\Finv,
        \end{align*}
    we just have to show that 
        \begin{equation*}
            \phi_1^*\left(\mathcal{A}^r_{\mathrm{SG}}+\int_{I\times\Sigma}\underline{\tilde{e}}_n \tilde{e}^2[\tilde{v},\tilde{v}] + h(\tilde{v}^\Finv) \right)=\mathcal{A}_{\mathrm{SG}}.
        \end{equation*}
    In order to do so, we start by looking at the proof of the corresponding Lemma for the pure PC theory in \cite{C25}, we see that, with the new definition of structural constraints, we have
        \begin{equation*}
            \phi_1^*\left( \mathcal{S}_{{PC}}^r +\int_{I\times\Sigma}\frac{1}{2}\tilde{e}_n\tilde{e}[\tilde{v},\tilde{v}]+h(\tilde{v}^\Finv)\right)=\mathcal{S}_{PC} + \int_{I\times\Sigma}\frac{1}{2}\tilde{v}\underline{\mu}\epsilon_n\bar{\tilde{\psi}}\gamma\tilde{\psi} - \tilde{v}\underline{\mu}\mathfrak{Q} + \left(\delta_\chi(\hat{\omega} + \tilde{v}) + \mathbb{q}(\hat{\omega} + \tilde{v})\right)\tilde{v}^\Finv.
        \end{equation*}
    At this point, we see
    \begin{align*}
        \mathcal{A}_{\mathrm{SG}}=\mathcal{S}_{PC}+ \int_{I\times\Sigma}& \frac{1}{3!} \bm{e}\bar{\bm{\psi}}\gamma^3d_{\bm{\omega}}\bm{\psi}-  \bar{\bm{\chi}}\gamma\bm{\psi} \bm{e}^\Finv + \delta_{\bm{\chi}}\bm{\omega}\bm{\omega}^\Finv  \\
        &-i(\mathrm{L}_{\bm{\xi}}^{\bm{\omega}} \bar{\bm{\psi}} - [{\bm{c}},\bar{\bm{\psi}}] - d_{\bm{\omega}}\bar{\bm{\chi}} )\bm{\psi}^\Finv +  \iota_{\bm{\xi}}\delta_{\bar{\chi}}\bm{\omega} {\bm{c}}^\Finv \\
        &+ \frac{1}{2} \iota_{\bm{\varphi}} {\bm{\xi}}^\Finv -i\left(\mathrm{L}_{\bm{\xi}}^{\bm{\omega}} \bar{\bm{\chi}} - [{\bm{c}},\bar{\bm{\chi}}] - \frac{1}{2}\iota_{\bm{\varphi}}\bar{\bm{\psi}} \right)\bm{\chi}^\Finv \\
        &+ \frac{1}{2}\left(\check{\bm{\omega}} - \iota_{\bm{\xi}}\bm{e}\check{\bm{c}} -\frac{\bm{e}}{2}\iota_{\bm{\xi}}\check{\bm{c}} \right)\iota_{\bm{\varphi}}\bm{e}^\Finv
    \end{align*}
    obtaining
    \begin{align*}
        \mathcal{A}_{\mathrm{SG}}^r=\mathcal{S}^r_{PC}+\int_{I\times M}& \frac{1}{3!}\left(\underline{\tilde{e}}_n\tilde{\bar{\psi}}\gamma^3 d_{\hat{\omega}}\tilde{\psi}+ \tilde{e}\underline{\tilde{\bar{\psi}}}_n\gamma^3 d_{\hat{\omega}}\tilde{\psi} + \tilde{e}\tilde{\bar{\psi}}\gamma^3 d_{\underline{\tilde{\omega}}_n}\tilde{\psi} + \tilde{e}\tilde{\bar{\psi}}\gamma^3d_{\hat{\omega}}\underline{\tilde{\psi}}_n   \right)\\
        & - \tilde{\bar{\chi}}\gamma\underline{\tilde{\psi}}_n \tilde{e}^\Finv -\tilde{\bar{\chi}}\gamma\tilde{\psi}\underline{\tilde{e}}^\Finv_n - \frac{1}{3!}\tilde{\bar{\chi}}\gamma^3\left(d_{\underline{\tilde{\omega}}_n}\tilde{\psi} +d_{\hat{\omega}}\underline{\tilde{\psi}}_n  \right)\tilde{\check{k}} - \frac{1}{3!}\tilde{\bar{\chi}}\gamma^3d_{\hat{\omega}}\tilde{\psi}\underline{\tilde{\check{k}}}_n \\
        &- i\left( \mathrm{L}_{\tilde{\xi}}^{\hat{\omega}}\tilde{\bar{\psi}} + d_{\tilde{\omega}_n}\tilde{\bar{\psi}}\tilde{\xi}^n  + \tilde{\bar{\psi}}_nd\tilde{\xi}^n -[\tilde{c},\tilde{\bar{\psi}}] - d_{\hat{\omega}}\tilde{\bar{\chi}} \right)\underline{\tilde{\psi}^\Finv}_{n}\\
        &- i\left(\mathrm{L}_{\tilde{\xi}}^{\hat{\omega}}\underline{\tilde{\bar{\psi}}}_n +\iota_{\underline{\partial_n}\tilde{\xi}}\tilde{\bar{\psi}}  -d_{\underline{\tilde{\omega}}_n}(\tilde{\bar{\psi}}_n\tilde{\xi}_n) -[\tilde{c},\tilde{\bar{\psi}}_n] - d_{\underline{\tilde{\omega}}_n}\tilde{\bar{\chi}}  \right)\tilde{\psi}^\Finv\\
        &- i\left( \mathrm{L}_{\tilde{\xi}}^{\hat{\omega}}\tilde{\bar{\chi}} + d_{\tilde{\omega}_n}\tilde{\bar{\chi}}\tilde{\xi}^n   -[\tilde{c},\tilde{\bar{\chi}}] - \frac{1}{2}\iota_{\tilde{\varphi}}\tilde{\bar{\psi}} -\frac{1}{2}\tilde{\bar{\psi}}_n\tilde{\varphi}^n  \right)\underline{\tilde{\chi}^\Finv}_{n}\\
        &+ \frac{1}{2}\iota_{\tilde{\varphi}}\underline{\tilde{\xi}}^\Finv + \frac{1}{2}\underline{\tilde{\xi}}^\Finv_n \tilde{\varphi}^n+\frac{1}{2}\tilde{\check{k}}\iota_{\tilde{\varphi}}\underline{\tilde{e}}^\Finv_n +\frac{1}{2}\underline{\tilde{\check{k}}}_n\left(\iota_{\tilde{\varphi}}\tilde{e}^\Finv + \tilde{e}^\Finv_n\tilde{\varphi}^n  \right),
    \end{align*}
    and    
    \begin{equation}\label{eq: pullbacl A_L}
    \begin{split}
        \mathcal{A}_{\mathrm{SG}}-\mathcal{S}_{PC}=&\mathcal{A}_{\mathrm{SG}}^r-\mathcal{S}_{PC}^r+\int_{I\times\Sigma} \frac{1}{3!}\left(\unl{\underline{\tilde{e}_n}\tilde{\bar{\psi}}\gamma^3[\tilde{v},\tilde{\psi}]}{m.1}+ \unl{\tilde{e}\underline{\tilde{\bar{\psi}}}_n\gamma^3 [\tilde{v},\tilde{\psi}]}{m.2} + \unl{\tilde{e}\tilde{\bar{\psi}}\gamma^3 [\tilde{v},\underline{\tilde{{\psi}}}_n] }{m.3}\right)\\
        &+\int_{I\times\Sigma}-\frac{1}{3!}\tilde{\bar{\chi}}\gamma^3\left(\unl{[\tilde{v},\underline{\tilde{{\psi}}}_n]\tilde{\check{k}}}{m.4}+\unl{ [\tilde{v},\tilde{\psi}]\underline{\tilde{\check{k}}}_n}{m.5}\right) - i\left(\unl{ [\iota_{\tilde{\xi}}\tilde{v},\tilde{\bar{\psi}}]}{m.6}  - \unl{[\tilde{v},\tilde{\bar{\chi}}]}{m.7}\right)\underline{\tilde{\psi}^\Finv}_{n}\\
        &\qquad \qquad -i\left(   \unl{[\iota_{\tilde{\xi}}\tilde{v},\underline{\tilde{\bar{\psi}}}_n]}{m.8}\right)\tilde{\psi}^\Finv- i\left( \unl{[\iota_{\tilde{\xi}}\tilde{v},\tilde{\bar{\chi}}] }{m.9}\right)\underline{\tilde{\chi}^\Finv}_{n} + \left(\delta_\chi(\hat{\omega} + \tilde{v}) + \mathbb{q}(\hat{\omega} + \tilde{v})\right)\tilde{v}^\Finv,
    \end{split}
    \end{equation}
Using $\eqref{id: action of omega}$, we see that $\reft{eq: pullbacl A_L}{m.1}=\frac{1}{2}\underline{\tilde{e_n}}\tilde{v}\bar{\psi}\gamma\psi$, and that $\reft{eq: pullbacl A_L}{m.2}+\reft{eq: pullbacl A_L}{m.3}$ give $$\frac{1}{3!}\left(\underline{\tilde{\bar{\psi}}}_n\gamma^3 [\tilde{v},\tilde{\psi}]+\tilde{e}\tilde{\bar{\psi}}\gamma^3 [\tilde{v},\underline{\tilde{{\psi}}}_n]\right)\propto \tilde{e}\tilde{v}\underline{\tilde{\bar{\psi}}}_n\gamma\tilde{\psi}=0.$$
 A straightforward computation gives
    \begin{equation}\label{eq: phi_1^*(A_SG)}
    \begin{split}
         &\hspace{-12mm}\phi_1^*\left(\mathcal{A}^r_{\mathrm{SG}} -\mathcal{S}^r_{PC}\right)+ \int_{I\times\Sigma}\frac{1}{2}\tilde{v}\underline{\mu}\epsilon_n\bar{\tilde{\psi}}\gamma\tilde{\psi} - \tilde{v}\underline{\mu}\mathfrak{Q}-\left(\delta_\chi(\hat{\omega} + \tilde{v}) + \mathbb{q}(\hat{\omega} + \tilde{v})\right)\tilde{v}^\Finv\\
         &=\mathcal{A}^r_{\mathrm{SG}}-\mathcal{S}^r_{PC}
        +\int_{I\times\Sigma} \unl{\frac{1}{3!}\tilde{e}\tilde{\bar{\psi}}\gamma^3[\iota_{\underline{z}}\tilde{v},\tilde{\psi}]}{as.1}- \unl{\tilde{\bar{\chi}}\gamma\underline{\tilde{\psi}}_n \tilde{v}\tilde{\check{k}}  }{as.2} + \unl{\tilde{\bar{\chi}}\gamma\tilde{\psi}\tilde{v}\underline{\tilde{\check{k}}}_n  }{as.3}  +\unl{\tilde{\bar{\chi}}\gamma\tilde{\psi}\iota_{\underline{z}}\tilde{v}\tilde{\check{k}}  }{as.5}\\
        &\qquad -\unl{\frac{1}{3!} \tilde{\bar{\chi}}\gamma^3[\iota_{\underline{z}}\tilde{v},\tilde{\psi}]\tilde{\check{k}} }{as.6}- \unl{i [\iota_{\tilde{\xi}}\tilde{v},\tilde{\bar{\psi}}]\underline{\tilde{\psi}}^\Finv_n }{as.7} - \unl{i[\iota_{\tilde{\xi}}\tilde{v},\underline{\tilde{\psi}}_n]\tilde{\psi}^\Finv}{as.8} + \unl{i[\iota_{\underline{z}}\tilde{v},\tilde{\bar{\chi}}]\tilde{\psi}^\Finv}{as.9}\\
        &\qquad- \unl{i [\iota_{\tilde{\xi}}\tilde{v},\tilde{\bar{\chi}}]\underline{\tilde{\chi}}^\Finv_n }{as.10} - \unl{\frac{1}{2}\tilde{v}\iota_{\tilde{\varphi}}\underline{\tilde{c}}^\Finv_n }{as.11} + \unl{\frac{1}{2}\iota_{\underline{z}}\tilde{v} \tilde{c}^\Finv_n \tilde{\varphi}^n}{as.12} -\unl{\frac{1}{2}\tilde{\check{k}}\iota_{\tilde{\varphi}}(\tilde{v}\underline{\tilde{\check{k}}}_n  + \iota_{\underline{z}}\tilde{v}\tilde{\check{k}} )}{as.13}\\
        &\qquad + \unl{\frac{1}{2}  \underline{\tilde{\check{k}}}_n \left(  \iota_{\tilde{\varphi}}(\tilde{v}\tilde{\check{k}}) - (\iota_z \tilde{v}\tilde{\check{k}})\tilde{\varphi}^n \right) }{as.14} + \unl{\frac{1}{2}\tilde{v}\underline{\mu}\epsilon_n \tilde{\bar{\psi}}\gamma \tilde{\psi}}{as.15} - \unl{\tilde{v}\underline{\mu}\mathfrak{Q}}{as.16}.
    \end{split}
    \end{equation}
We are then only left to show that equations \eqref{eq: phi_1^*(A_SG)} and \eqref{eq: pullbacl A_L} coincide. 
We start by noticing the following terms match:
\begin{itemize}
\item $\reft{eq: phi_1^*(A_SG)}{as.1}+\reft{eq: phi_1^*(A_SG)}{as.15}\overset{\ref{id: action of omega}}{=}\frac{1}{2}\tilde{v}(\iota_{\underline{z}}\tilde{e} + \mu \epsilon_n  )\bar{\tilde{\psi}}\gamma\tilde{\psi}=\reft{eq: pullbacl A_L}{m.1}$.
\item $\reft{eq: pullbacl A_L}{m.6}=\reft{eq: phi_1^*(A_SG)}{as.7}$.
\item $\reft{eq: pullbacl A_L}{m.8}=\reft{eq: phi_1^*(A_SG)}{as.8}$.
\item $\reft{eq: pullbacl A_L}{m.9}=\reft{eq: phi_1^*(A_SG)}{as.9}$
\end{itemize}
Before we proceed, we need to find an explicit expression of $\mathfrak{Q}$. Indeed, recall that it was defined in \eqref{constr: BV Sugra omega} as the part of $Q_{\mathrm{SG}}(\epsilon_n \tilde{\check{k}} - \tilde{e}\tilde{\check{a}})=0$ depending on $\tilde{\chi}$ and $\tilde{\psi}$. We start by computing $\bm{e}Q_{\mathrm{SG}}\bm{\check{k}}$ in the bulk as in the proof of Proposition \ref{prop: equivalence of BV PC r constraints}. It is a quick computation to see
    \begin{align*}
        \bm{e}Q_{\mathrm{SG}}\bm{\check{k}}=&\bm{e}\left( d_{\bm{\omega}}\bm{e} - \frac{1}{2}\bm{\bar{\psi}}\gamma\bm{\psi} + \mathrm{L}_{\bm{\xi}}^{\bm{\omega}}\bm{\check{k}} - [\bm{c},\bm{\check{k}}] - \frac{1}{2}\iota_{\bm{\varphi}}(\bm{e}\bm{\check{c}}) \right)\\
        &- \frac{1}{4}\bar{\bm{\chi}}\gamma^2[\bm{\check{k}},\gamma]\bm{\psi} - \frac{i}{4}\bar{\bm{\psi}}^\Finv_0\ugam^2\gamma^2 \bm{\chi} - \frac{1}{2}\bm{\check{k}}\iota_{\bm{\varphi}}\bm{\check{k}}.
    \end{align*}
In particular, one has 
    \begin{equation*}
        \mathfrak{Q}=\epsilon_n\left( Q_{\mathrm{SG}}(\tilde{\check{k}})- d_{\tilde{\omega}}\tilde{e} + \frac{1}{2}\tilde{\bar{\psi}}\gamma\tilde{\psi} - \mathrm{L}_{\tilde{\xi}}^{\tilde{\omega}}\tilde{\check{k}} + [\tilde{c},\tilde{\check{k}}]\right).
    \end{equation*}
The only term inside $\mathfrak{Q}$ which can be easily found is given by $-\frac{1}{2}\epsilon_n\left(\bar{\tilde{\chi}}\gamma\tilde{\chi}\tilde{\check{c}}\right)$, as the terms proportional to $\tilde{e}$ can be reabsorbed in the right hand side of the structural constraint \eqref{constr: BV Sugra omega}. In particular, we can define
    \begin{equation*}
        \mathfrak{B}:=\mathfrak{Q}+\frac{1}{2}\epsilon_n\left(\bar{\tilde{\chi}}\gamma\tilde{\chi}\tilde{\check{c}}\right).
    \end{equation*}
Unfortunately, finding $Q_{\mathrm{SG}}\bm{\check{k}}$ and extracting $Q_{\mathrm{SG}}( \tilde{\check{k}})$ as the part not proportional to $dx^n$,  is not convenient in our case, as it involves cumbersome computations.  It is however more convenient to consider the part along $dx^n$ inside $\bm{e}Q_{\mathrm{SG}}(\bm{\check{k}})$, which is given by $\underline{\tilde{e}}_n Q_{\mathrm{SG}}(\tilde{\check{k}}) + \tilde{e}Q_{\mathrm{SG}}(\underline{\tilde{\check{k}}}_n)$. As in the computation of the BV pushforward we are only interested in $\underline{\mu}\tilde{v}\mathfrak{Q}$, we notice that 
    \begin{equation*}
        \tilde{v}\left(\underline{\tilde{e}}_n Q_{\mathrm{SG}}(\tilde{\check{k}}) + \tilde{e}Q_{\mathrm{SG}}(\underline{\tilde{\check{k}}}_n)\right)\tilde{v}\underline{\tilde{e}}_n Q_{\mathrm{SG}}(\tilde{\check{k}}) = - \underline{\mu}\tilde{v}\epsilon_n Q_{\mathrm{SG}}\tilde{\check{k}} - \iota_{\underline{z}}\tilde{v}\tilde{e}Q_{\mathrm{SG}}(\tilde{\check{k}}),
    \end{equation*}
having used $\tilde{e}\tilde{v}=0$. At this point, discarding all the terms that do not depend on $\chi$ and $\psi$, we obtain the term $\reft{eq: phi_1^*(A_SG)}{as.16}$ as
    \begin{equation*}
        -\underline{\mu}\tilde{v}\mathfrak{Q}=\left[\tilde{v}\left(\underline{\tilde{e}}_n Q_{\mathrm{SG}}(\tilde{\check{k}}) + \tilde{e}Q_{\mathrm{SG}}(\underline{\tilde{\check{k}}}_n)\right)+ \iota_{\underline{z}}\tilde{v} \tilde{e}Q_{\mathrm{SG}}(\tilde{\check{k}}) \right]\bigg|_{\psi,\chi}.
    \end{equation*}
In particular, inside $\bm{e}Q_{\mathrm{SG}}\bm{\check{k}}$, what we are interested in are the terms
    \begin{equation*}
         - \bm{e}\frac{1}{2}\iota_{\bm{\varphi}}(\bm{e}\bm{\check{c}})- \frac{1}{4}\bar{\bm{\chi}}\gamma^2[\bm{\check{k}},\gamma]\bm{\psi} - \frac{i}{4}\bar{\bm{\psi}}^\Finv_0\ugam^2\gamma^2 \bm{\chi} - \frac{1}{2}\bm{\check{k}}\iota_{\bm{\varphi}}\bm{\check{k}}.
    \end{equation*}
 We carry out the computation term by term: 
\begin{itemize}
    \item $- \bm{e}\frac{1}{2}\iota_{\bm{\varphi}}(\bm{e}\bm{\check{c}})$. As previously anticipated, this gives $-\frac{1}{2}\epsilon_n\left(\bar{\tilde{\chi}}\gamma\tilde{\chi}\tilde{\check{c}}\right)$. We obtain
        \begin{align*}
            \frac{1}{2}\underline{\mu}\tilde{v}\epsilon_n\left(\bar{\tilde{\chi}}\gamma\tilde{\chi}\tilde{\check{c}}\right)=&-\frac{1}{2}\tilde{v}(\underline{\tilde{e}}_n - \iota_{\underline{z}}\tilde{e})(\iota_{\varphi}\tilde{e} + \tilde{e}_n \tilde{\varphi}^n)\tilde{\check{c}}\\
            =&\frac{1}{2}\tilde{v}\iota_{\varphi}(\underline{\tilde{e}}_n \tilde{e}\tilde{\check{c}})- \frac{1}{2}\iota_{\underline{z}}\tilde{v}\left(\frac{1}{2}\iota_{\tilde{\varphi}}(e^2)\tilde{\check{c}}+\tilde{e}\tilde{e}_n \tilde{\varphi}^n \tilde{\check{c}}\right)\\
            &=\frac{1}{2}\tilde{v}\iota_{\tilde{\varphi}}(\underline{\tilde{c}}^\Finv_n) - \frac{1}{2}\iota_{\underline{z}}\tilde{v}\tilde{c}^\Finv_n\tilde{\varphi}^n,
        \end{align*}    
    having used the fact that $\underline{\tilde{c}}^\Finv_n = \underline{\tilde{e}}_n \tilde{e}\tilde{\check{c}} + \frac{1}{2}\tilde{e}^2\underline{\tilde{\check{c}}}_n$ and that $\tilde{e}\tilde{v}=0$ repeatedly. We see that the terms above exactly cancel out $\reft{eq: phi_1^*(A_SG)}{as.11}+\reft{eq: phi_1^*(A_SG)}{as.12}$. 
    \item $- \frac{1}{4}\bar{\bm{\chi}}\gamma^2[\bm{\check{k}},\gamma]\bm{\psi}$. We have
    \begin{equation}\label{eq: skf1}
    \begin{split}
        &\tilde{v}\left(-\frac{1}{4}\bar{\tilde{\chi}}\gamma^2[\underline{\tilde{\check{k}}}_n\gamma]\tilde{\psi} -\frac{1}{4}\tilde{\bar{\chi}}\gamma^2[\tilde{\check{k}},\gamma]\underline{\tilde{\psi}}_n \right) -\frac{1}{4}\iota_{\underline{z}}\tilde{v}\tilde{\bar{\chi}}\gamma^2[\tilde{\check{k}},\gamma]\psi\\
        &\overset{\eqref{id: action of omega}}{=}\tilde{v}\left(\unl{\tilde{\bar{\chi}}\gamma\tilde{\psi}\underline{\tilde{\check{k}}}_n }{bs.1}+\unl{\tilde{\bar{\chi}}\gamma\underline{\tilde{\psi}}\tilde{\check{k}}}{bs.2}\right)-\frac{1}{3!}\tilde{\bar{\chi}}\gamma^3\left(\unl{[\tilde{v},\tilde{\psi}]\underline{\tilde{\check{k}}}_n}{bs.3} + \unl{[\tilde{v}, \underline{\tilde{\psi}}_n]\tilde{\check{k}}}{bs.4}  \right) - \unl{\iota_{\underline{z}}\tilde{v}\tilde{\bar{\chi}}\gamma\tilde{\psi}\tilde{\check{k}}}{bs.5} +\unl{\frac{1}{3!}\tilde{\bar{\chi}}\gamma^3[\iota_{\underline{z}}\tilde{v},\tilde{\psi}]\tilde{\check{k}}}{bs.6}.
    \end{split}
    \end{equation}
    We see that $\reft{eq: skf1}{bs.1}+\reft{eq: skf1}{bs.2}+\reft{eq: skf1}{bs.5}+\reft{eq: skf1}{bs.6}+\reft{eq: phi_1^*(A_SG)}{as.2}+\reft{eq: phi_1^*(A_SG)}{as.3}+\reft{eq: phi_1^*(A_SG)}{as.5}+\reft{eq: phi_1^*(A_SG)}{as.6}=0$, while $\reft{eq: skf1}{bs.3}=\reft{eq: pullbacl A_L}{m.5}$ and $\reft{eq: skf1}{bs.4}=\reft{eq: pullbacl A_L}{m.4}$
    
    \item $- \frac{i}{4}\bar{\bm{\psi}}^\Finv_0\ugam^2\gamma^2 \bm{\chi} $. Before carrying out the computation, we notice that $- \frac{i}{4}\bar{\bm{\psi}}^\Finv_0\ugam^2\gamma^2 \bm{\chi}=-i[{\bar{\bm{\psi}}}^\Finv,\bm{\chi}]$, where the action on $\bm{\chi}$ is given only by the part in $V$ of $\bm{\psi}^\Finv$. Using Leibniz and the Majorana flip relations, we then have
        \begin{equation}\label{eq: skf2}
        \begin{split}
            &-i\tilde{v}[ \underline{\tilde{\bar{\psi}}}^\Finv_n,\tilde{\chi} ] - i\iota_{\underline{z}}\tilde{v}[\tilde{\bar{\psi}}^\Finv,\tilde{\chi}]\\
            &=\unl{-i[\tilde{v},\tilde{\bar{\chi}}]\underline{\tilde{\psi}}^\Finv_n}{cs.1} +\unl{ i [\iota_{\underline{z}}\tilde{v},\tilde{\bar{\chi}}]\tilde{\psi}^\Finv}{cs.2}.
        \end{split}
        \end{equation}
        We see $\reft{eq: skf2}{cs.2}+\reft{eq: phi_1^*(A_SG)}{as.9}=0$, while $\reft{eq: skf2}{cs.2}=\reft{eq: pullbacl A_L}{m.7}$, which tells us that \eqref{eq: phi_1^*(A_SG)} contains all the terms in \eqref{eq: pullbacl A_L}. We are then left to show that \eqref{eq: phi_1^*(A_SG)} does not contain extra terms. 
    \item $- \frac{1}{2}\bm{\check{k}}\iota_{\bm{\varphi}}\bm{\check{k}}$. We have simply
        \begin{equation}\label{eq: skf3}
        \begin{split}
            &-\frac{1}{2}\tilde{v}\iota_{\tilde{\varphi}}(\underline{\tilde{\check{k}}}_n \tilde{\check{k}}  )+\frac{1}{2}\iota_{\underline{z}}\tilde{v}\tilde{\check{k}}\tilde{\check{k}}_n\tilde{\varphi}^n,
        \end{split}
        \end{equation}
    since $\tilde{\check{k}}\iota_{\tilde{\varphi}}\tilde{\check{k}}=\frac{1}{2}\iota_{\tilde{\varphi}}(\tilde{\check{k}}^2)=0$ because there are no $4$-forms on $\Sigma$ and since $\tilde{\check{k}}_n^2=0$ because of parity. It is easy to see that equation \eqref{eq: skf3} exactly cancels out $\reft{eq: phi_1^*(A_SG)}{as.13}+\reft{eq: phi_1^*(A_SG)}{as.14}$. 
\end{itemize}
The above computation tells us that eq. \eqref{eq: phi_1^*(A_SG)} is equal to eq. \eqref{eq: pullbacl A_L}, hence showing
    \begin{equation*}
        \phi_1^*\left(\mathcal{A}^r_{\mathrm{SG}} +\int_{I\times\Sigma}\frac{1}{2}\tilde{e}_n\tilde{e}[\tilde{v},\tilde{v}]+h(\tilde{v}^\Finv)\right)=\mathcal{A}_{\mathrm{SG}},
    \end{equation*}
\end{proof}
\subsection{AKSZ Symplectomorphism}\label{proof: AKSZ SYMPL}
We now want to show the proof of proposition \ref{prop: AKSZ sympl}. We start by recalling the form of the AKSZ symplectic form:
\begin{equation}\label{eq: AKSZ SG sympform}
    \begin{split}
        \varpi_{\mathrm{SG}}^{AKSZ}=\varpi_{PC}^{AKSZ}+\int_{I\times \Sigma}&\frac{1}{3!}\left(\unl{\underline{\bar{\varsigma}}^\Finv\gamma^3 \delta\psi \delta e }{h1} + \unl{\bar{\psi}\gamma^3 \delta\underline{\varsigma}^\Finv \delta e}{h2} + \unl{\bar{\psi}\gamma^3\delta\psi \delta\underline{f}^\Finv}{h3} + \unl{\underline{f}^\Finv \delta\bar{\psi}\gamma^3 \delta\psi }{h4} \right)\\
        & + \frac{1}{3}\unl{ e \delta\bar{\psi}\gamma^3 \delta\underline{\varsigma}^\Finv}{h5} + i \unl{\delta\underline{\bar{\epsilon}}\delta\theta^\Finv }{h6}+ \unl{i \delta\bar{\chi}\delta\underline{\chi}^\Finv}{h7} + i\delta\underline{\bar{\varsigma}}(\unl{\iota_{\delta\xi} \theta^\Finv}{h8} +\unl{ \iota_\xi \delta\theta^\Finv}{h9} ) \\
        & + i \delta\bar{\psi}\left(\unl{ \iota_{\delta\underline{z}}\theta^\Finv }{h10}+ \unl{\iota_{\underline{z}}\delta\theta^\Finv}{h11} +\unl{\iota_{\delta\xi} \underline{\chi}^\Finv}{h12} +\unl{ \iota_\xi \delta\underline{\chi}^\Finv}{h13}  \right) 
    \end{split}
    \end{equation}
The reduced BV form is given by
    \begin{align*}
        \varpi^r_{\mathrm{SG}}=\varpi^r_{PC}+\int_{I\times\Sigma}i \delta\bar{\un{\psi}}_n\delta\psi^\Finv + i \delta\bar{\psi}\delta\underline{\psi}^\Finv_n + i \delta\bar{\chi}\delta\underline{\chi}^\Finv_n
        \end{align*}
Notice that we have
    \begin{equation}\label{eq: AKSZ PC SG}
    \begin{split}
        \Phi^*_r&(\varpi^r_{PC})=\varpi^{AKSZ}_{PC}+ \\
        &+\int_{I\times\Sigma}\delta e \left(\unl{\frac{1}{3!}\bar{\un{\varsigma}^\Finv}\gamma^3\delta\psi}{k1} + \unl{\frac{1}{3!}\bar{\psi}\gamma^3\un{\varsigma}^\Finv}{k2} - \unl{\frac{1}{3!}\delta(\lambda\mu^{-1})\bar{\varsigma}^\Finv\gamma^3\un{\varsigma}^\Finv }{k3} + \unl{\frac{1}{3} \lambda\mu^{-1}\delta\bar{\varsigma}^\Finv\gamma^3\un{\varsigma}^\Finv}{k4} \right)\\
        &+\delta(\lambda\mu^{-1})f^\Finv\left(\unl{\frac{1}{3!}\bar{\un{\varsigma}^\Finv}\gamma^3\delta\psi}{k5} + \unl{\frac{1}{3!}\bar{\psi}\gamma^3\un{\varsigma}^\Finv}{k6} - \unl{\frac{1}{3!}\delta(\lambda\mu^{-1})\bar{\varsigma}^\Finv\gamma^3\un{\varsigma}^\Finv }{k7} + \unl{\frac{1}{3} \lambda\mu^{-1}\delta\bar{\varsigma}^\Finv\gamma^3\un{\varsigma}^\Finv}{k8} \right)\\
        &-\lambda\mu^{-1}\delta f^\Finv\left(\unl{\frac{1}{3!}\bar{\un{\varsigma}^\Finv}\gamma^3\delta\psi}{k9} + \unl{\frac{1}{3!}\bar{\psi}\gamma^3\un{\varsigma}^\Finv}{k10} - \unl{\frac{1}{3!}\delta(\lambda\mu^{-1})\bar{\varsigma}^\Finv\gamma^3\un{\varsigma}^\Finv }{k11}  \right)\\
        & +  \unl{i \iota_{\delta\xi}\delta\bar{\un{\varsigma}^\Finv}\theta^\Finv}{k12} + \unl{i\iota_{\delta\xi}\bar{\un{\varsigma}^\Finv}\delta\theta^\Finv}{k13} + i \delta(\lambda\mu^{-1})\left( \unl{i \iota_{z}\delta\bar{\un{\varsigma}^\Finv}\theta^\Finv}{k14} + \unl{i\iota_{z}\bar{\un{\varsigma}^\Finv}\delta\theta^\Finv}{k15}  \right)\\
        & + i \lambda\mu^{-1}\left( \unl{i \iota_{\delta z}\delta\bar{\un{\varsigma}^\Finv}\theta^\Finv}{k16} + \unl{i\iota_{\delta z}\bar{\un{\varsigma}^\Finv}\delta\theta^\Finv}{k17}  \right) -\unl{ i \delta(\lambda\mu^{-1})\iota_{\delta\xi}\bar{{\varsigma}}^\Finv\un{\chi}^\Finv }{k18} \\
        &-  i \lambda\mu^{-1}\left(\unl{\iota_{\delta\xi}\delta\bar{{\varsigma}}^\Finv\un{\chi}^\Finv}{k19}+\unl{\iota_{\delta\xi}\bar{{\varsigma}}^\Finv\delta\un{\chi}^\Finv}{k20}  \right) + i\delta(\lambda\mu^{-1})\left(\unl{\delta(\lambda\mu^{-1})\iota_{z}\bar{{\varsigma}}^\Finv\un{\chi}}{k21} +\unl{\lambda\mu^{-1}\iota_{z}\delta\bar{{\varsigma}}^\Finv\un{\chi}}{k22}\right)\\
        &+ i\delta(\lambda\mu^{-1})\left(\unl{\lambda\mu^{-1}\iota_{z}\bar{{\varsigma}}^\Finv\delta\un{\chi}}{k23} + \unl{\lambda\mu^{-1}\iota_{\delta z}\bar{{\varsigma}}^\Finv\un{\chi}}{k24} \right)+\delta(\lambda\mu^{-1})\left(\unl{  \iota_{\delta \un{z}}\bar{{\varsigma}}^\Finv\theta_{\Finv}}{k25} +   \unl{\iota_{\un{z}}\delta \bar{{\varsigma}}^\Finv\theta_{\Finv}}{k26} - \unl{\iota_{\un{z}} \bar{{\varsigma}}^\Finv\delta\theta_{\Finv}}{k27} \right)\\
        &+\frac{1}{3!}\delta(\lambda\mu^{-1})\left( \unl{\delta\un{f}^\Finv\bar{\psi} \gamma^3\varsigma^\Finv}{k28} +\unl{\un{f}^\Finv\delta\bar{\psi} \gamma^3\varsigma^\Finv}{k29}- \unl{\un{f}^\Finv\bar{\psi} \gamma^3\delta\varsigma^\Finv}{k30}+ \unl{\delta e \bar{\varsigma}^\Finv\gamma^3\un{\varsigma}^\Finv}{k31} + \unl{2e \bar{\varsigma}^\Finv\gamma^3\delta\un{\varsigma}^\Finv}{k32} \right)\\
        &- i\delta(\lambda\mu^{-1})\left(\unl{\delta(\lambda\mu^{-1})\iota_{z}\bar{{\varsigma}}^\Finv\un{\chi}}{k33} +\unl{\lambda\mu^{-1}\iota_{z}\delta\bar{{\varsigma}}^\Finv\un{\chi}}{k34} + \unl{\lambda\mu^{-1}\iota_{z}\bar{{\varsigma}}^\Finv\delta\un{\chi}}{k35} + \unl{\lambda\mu^{-1}\iota_{\delta z}\bar{{\varsigma}}^\Finv\un{\chi}}{k36} \right),
    \end{split}
    \end{equation}
and 
    \begin{equation}\label{eq: AKSZ SG pb}
    \begin{split}
        &\Phi_r^*\left( i \delta\bar{\un{\psi}}_n\delta\psi^\Finv + i \delta\bar{\psi}\delta\underline{\psi}^\Finv_n + i \delta\bar{\chi}\delta\underline{\chi}^\Finv_n \right)\\
        &=\frac{1}{3!} \delta \bar{\psi}\left(\unl{ 2\gamma^3 \un{\varsigma}^\Finv \delta e}{l1} + \unl{ e\gamma^3 \delta\un{\varsigma}^\Finv }{l2}-\unl{ \gamma^3\psi \delta \un{f}^\Finv}{l3}  -\unl{\un{f}^\Finv\gamma^3 \delta\psi }{l4}+\unl{ \delta(\lambda\mu^{-1}) f^\Finv\gamma^3\un{\varsigma}^\Finv}{l5} -\unl{\lambda\mu^{-1}\delta f^\Finv\gamma^3\un{\varsigma}^\Finv}{l6}\right)\\
        &-\frac{1}{3!} \unl{\delta \bar{\psi}\lambda\mu^{-1} f^\Finv\gamma^3\delta\un{\varsigma}^\Finv}{l7}+\frac{1}{3!}\delta(\lambda\mu^{-1})\bar{\varsigma}^\Finv\left( \unl{2\gamma^3 \un{\varsigma}^\Finv \delta e}{l8} +  \unl{2e\gamma^3 \delta\un{\varsigma}^\Finv }{l9}-\unl{ \gamma^3\psi \delta \un{f}^\Finv}{l10}  -\unl{\un{f}^\Finv\gamma^3 \delta\psi}{l11} + \unl{\delta(\lambda\mu^{-1}) f^\Finv\gamma^3\un{\varsigma}^\Finv}{l12}\right)\\
        &-\frac{1}{3!}\delta(\lambda\mu^{-1})\bar{\varsigma}^\Finv\left(\unl{ \lambda\mu^{-1}\delta f^\Finv\gamma^3\un{\varsigma}^\Finv}{l13} -\unl{\lambda\mu^{-1} f^\Finv\gamma^3\delta\un{\varsigma}^\Finv}{l14}\right)+\frac{1}{3!}\lambda\mu^{-1}\delta\bar{\varsigma}\left(\unl{ 2\gamma^3 \un{\varsigma}^\Finv \delta e}{l15} +  \unl{e\gamma^3 \delta\un{\varsigma}^\Finv}{l16} - \unl{\gamma^3\psi \delta \un{f}^\Finv }{l17}  \right)\\
        &+\frac{1}{3!}\lambda\mu^{-1}\delta\bar{\varsigma}^\Finv\left( -\unl{\un{f}^\Finv\gamma^3 \delta\psi}{l18}+ \unl{\delta(\lambda\mu^{-1}) f^\Finv\gamma^3\un{\varsigma}^\Finv}{l19}\right)+ i \delta\bar{\psi}\left( \unl{\iota_{\delta\un{z}}\theta^\Finv}{l20} + \unl{\iota_{\un{z}}\delta\theta^\Finv}{l21} \right) \\
        &+ i \delta(\lambda\mu^{-1})\bar{\varsigma}^\Finv\left( \unl{\iota_{\delta\un{z}}\theta^\Finv }{l22}+ \unl{\iota_{\un{z}}\delta\theta^\Finv}{l23} \right) - i\lambda\mu^{-1}\delta\bar{\varsigma}^\Finv\left( \unl{\iota_{\delta\un{z}}\theta^\Finv }{l24}+ \unl{\iota_{\un{z}}\delta\theta^\Finv }{l25}\right) + i \delta\bar{\psi}\left( \unl{\iota_{\delta\xi}\un{\chi}^\Finv}{l26}+\unl{\iota_{\xi}\delta\un{\chi}^\Finv}{l27} \right)\\
        &+i\delta(\lambda\mu^{-1})\bar{\varsigma}^\Finv\left( \unl{\iota_{\delta\xi}\un{\chi}^\Finv}{l28}+\unl{\iota_{\xi}\delta\un{\chi}^\Finv}{l29} \right) - i \lambda\mu^{-1}\delta\bar{\varsigma}^\Finv\left( \unl{\iota_{\delta\xi}\un{\chi}^\Finv}{l30}+\unl{\iota_{\xi}\delta\un{\chi}^\Finv }{l31}\right) + \unl{i \delta\bar{\chi}\delta\un{\chi}^\Finv}{l32} + \unl{i \delta(\lambda\mu^{-1})\iota_\xi \bar{\varsigma}^\Finv \delta\un{\chi}^\Finv }{l33}\\
        &-  \unl{i \lambda\mu^{-1}\iota_{\delta\xi}\bar{\varsigma}^\Finv\delta\un{\chi}^\Finv}{l34}- \unl{i \lambda\mu^{-1}\iota_\xi\delta\bar{\varsigma}^\Finv\delta\un{\chi}^\Finv}{l35} + \unl{i \delta\bar{\un{\epsilon}}\delta\theta^\Finv}{l36} - \unl{i \iota_{\delta\xi}\un{\varsigma}^\Finv\delta\theta^\Finv}{l37} - \unl{i \iota_\xi \delta\bar{\un{\varsigma}}^\Finv\delta\theta^\Finv}{l38} + \unl{i\delta(\lambda\mu^{-1})\iota_{\un{z}}\bar{\varsigma}^\Finv\delta{\theta}^\Finv}{l39} \\
        &-\unl{ i\lambda\mu^{-1}\delta\iota_{\un{z}}\bar{\varsigma}^\Finv\delta{\theta}^\Finv}{l40}- \unl{i\lambda\mu^{-1}\iota_{\delta\un{z}}\bar{\varsigma}^\Finv\delta{\theta}^\Finv}{l41} - \unl{i\lambda\mu^{-1}\delta\iota_{\un{z}}\delta\bar{\varsigma}^\Finv\delta{\theta}^\Finv}{l42}
    \end{split}
    \end{equation}
We can then see that the following terms inside $\Phi_r^*(\varpi^r_{\mathrm{SG}})$ add up to $\varpi^{AKSZ}_{\mathrm{SG}}$:
    \begin{itemize}
        \item $\reft{eq: AKSZ PC SG}{k1} +  \reft{eq: AKSZ SG pb}{l1}=\reft{eq: AKSZ SG sympform}{h1}$
        \item $\reft{eq: AKSZ PC SG}{k2} =\reft{eq: AKSZ SG sympform}{h2}$
        \item $\reft{eq: AKSZ SG pb}{l3} =\reft{eq: AKSZ SG sympform}{h4}$
        \item $\reft{eq: AKSZ SG pb}{l4} =\reft{eq: AKSZ SG sympform}{h4}$
        \item $\reft{eq: AKSZ SG pb}{l2} =\reft{eq: AKSZ SG sympform}{h5}$
        \item $\reft{eq: AKSZ SG pb}{l36} =\reft{eq: AKSZ SG sympform}{h6}$
        \item $\reft{eq: AKSZ SG pb}{l32} =\reft{eq: AKSZ SG sympform}{h7}$
        \item $\reft{eq: AKSZ PC SG}{k12} =\reft{eq: AKSZ SG sympform}{h8}$
        \item $\reft{eq: AKSZ SG pb}{l38} =\reft{eq: AKSZ SG sympform}{h9}$
        \item $\reft{eq: AKSZ SG pb}{l20}+\reft{eq: AKSZ SG pb}{l21} =\reft{eq: AKSZ SG sympform}{h10}+\reft{eq: AKSZ SG sympform}{h11}$
        \item $\reft{eq: AKSZ SG pb}{l26}+\reft{eq: AKSZ SG pb}{l27} =\reft{eq: AKSZ SG sympform}{h12}+\reft{eq: AKSZ SG sympform}{h13}$
    \end{itemize}
while the remaining terms in $\Phi_r^*(\varpi^r_{\mathrm{SG}})$ add up to zero
    \begin{itemize}
        \item $\reft{eq: AKSZ PC SG}{k3}+\reft{eq: AKSZ PC SG}{k31}+\reft{eq: AKSZ SG pb}{l8}=0$
        \item $\reft{eq: AKSZ PC SG}{k4} + \reft{eq: AKSZ SG pb}{l15}=0$
        \item $\reft{eq: AKSZ PC SG}{k5}+\reft{eq: AKSZ PC SG}{k31}+\reft{eq: AKSZ SG pb}{l5}+\reft{eq: AKSZ SG pb}{l29}=0$
        \item $\reft{eq: AKSZ PC SG}{k6} +\reft{eq: AKSZ PC SG}{k30}=0$
        \item $ \reft{eq: AKSZ PC SG}{k7} + \reft{eq: AKSZ SG pb}{l8}=0  $
        \item $ \reft{eq: AKSZ PC SG}{k8}  + \reft{eq: AKSZ SG pb}{l14}+\reft{eq: AKSZ SG pb}{l19}=0  $
        \item $ \reft{eq: AKSZ PC SG}{k9}  + \reft{eq: AKSZ SG pb}{l6}=0  $
        \item $ \reft{eq: AKSZ PC SG}{k10}  + \reft{eq: AKSZ SG pb}{l17}=0  $
        \item $ \reft{eq: AKSZ PC SG}{k11}  + \reft{eq: AKSZ SG pb}{l13}=0  $
        \item $ \reft{eq: AKSZ SG pb}{l7} + \reft{eq: AKSZ SG pb}{l18}=0  $
        \item $ \reft{eq: AKSZ SG pb}{l9} + \reft{eq: AKSZ SG pb}{l18}=0  $
        \item $ \reft{eq: AKSZ SG pb}{l10} + \reft{eq: AKSZ PC SG}{k28}=0  $
        \item $ \reft{eq: AKSZ SG pb}{l16}=0$ because of \eqref{flip:3} and the parity of $\delta\varsigma^\Finv$
        \item $ \reft{eq: AKSZ SG pb}{l22} + \reft{eq: AKSZ PC SG}{k25}=0  $
        \item $ \reft{eq: AKSZ SG pb}{l23} +\reft{eq: AKSZ SG pb}{l39} + \reft{eq: AKSZ PC SG}{k15}+ \reft{eq: AKSZ PC SG}{k27}=0  $
        \item $ \reft{eq: AKSZ SG pb}{l24} + \reft{eq: AKSZ PC SG}{k16}=0  $
        \item $ \reft{eq: AKSZ SG pb}{l25}+\reft{eq: AKSZ SG pb}{l38}=0  $
        \item $\reft{eq: AKSZ SG pb}{l37}+\reft{eq: AKSZ PC SG}{k13}=0  $
        \item $\reft{eq: AKSZ SG pb}{l41}+\reft{eq: AKSZ PC SG}{k17}=0  $
        \item $ \reft{eq: AKSZ PC SG}{k14} + \reft{eq: AKSZ PC SG}{k26}=0 $
        \item $ \reft{eq: AKSZ SG pb}{l28} + \reft{eq: AKSZ PC SG}{k18}=0 $
        \item $ \reft{eq: AKSZ SG pb}{l29} + \reft{eq: AKSZ SG pb}{l33}=0 $
        \item $ \reft{eq: AKSZ SG pb}{l30} + \reft{eq: AKSZ PC SG}{k19}=0 $
        \item $ \reft{eq: AKSZ SG pb}{l31} + \reft{eq: AKSZ SG pb}{l35}=0 $
        \item $ \reft{eq: AKSZ SG pb}{l34}+ \reft{eq: AKSZ PC SG}{k20}=0 $
        \item $ \reft{eq: AKSZ PC SG}{k21}+ \reft{eq: AKSZ PC SG}{k33}=0 $
        \item $ \reft{eq: AKSZ PC SG}{k22}+ \reft{eq: AKSZ PC SG}{k34}=0 $
        \item $ \reft{eq: AKSZ PC SG}{k23}+ \reft{eq: AKSZ PC SG}{k35}=0 $
        \item $ \reft{eq: AKSZ PC SG}{k24}+ \reft{eq: AKSZ PC SG}{k36}=0 $
    \end{itemize}

\nocite{*}
%\emergencystretch=2em
\newrefcontext[sorting=nty]
\sloppy
\printbibliography

@article{DAuria:2020guc,
    author = "D'Auria, Riccardo",
    title = "{Geometric supergravity}",
    eprint = "2005.13593",
    archivePrefix = "arXiv",
    primaryClass = "hep-th",
    month = "5",
    year = "2020"
}

@article{Castellani:2018zey,
    author = "Castellani, Leonardo",
    title = "{Supergravity in the Group-Geometric Framework: A Primer}",
    eprint = "1802.03407",
    archivePrefix = "arXiv",
    primaryClass = "hep-th",
    reportNumber = "ARC-18-02",
    doi = "10.1002/prop.201800014",
    journal = "Fortsch. Phys.",
    volume = "66",
    number = "4",
    pages = "1800014",
    year = "2018"
}

@article{Slavnov:1972fg,
    author = "Slavnov, A. A.",
    title = "{Ward Identities in Gauge Theories}",
    doi = "10.1007/BF01090719",
    journal = "Theor. Math. Phys.",
    volume = "10",
    pages = "99--107",
    year = "1972"
}

@article{Taylor:1971ff,
    author = "Taylor, J. C.",
    title = "{Ward Identities and Charge Renormalization of the Yang-Mills Field}",
    doi = "10.1016/0550-3213(71)90297-5",
    journal = "Nucl. Phys. B",
    volume = "33",
    pages = "436--444",
    year = "1971"
}

@book{anderson,
author={I. M. Anderson},
title={The Variational Bicomplex},
url= {https://ncatlab.org/nlab/files/AndersonVariationalBicomplex.pdf}


}

@inbook{Zuckerman,
author = { Gregg J.   Zuckerman },
title = {Action Principles and Global Geometry},
booktitle = {Mathematical Aspects of String Theory},
chapter = {},
pages = {259-284},
doi = {10.1142/9789812798411_0013},
  }

@article{Fatibene1996,
	doi = {10.1023/a:1018852524599},
	year = 1998,
	month = {09},
	publisher = {Springer Science and Business Media {LLC}},
  	volume = {30},
  	number = {9},
  	pages = {1371--1389},
  	author = {Fatibene, Lorenzo and Ferraris, Marco and Francaviglia, Mauro and Godina, Marco},
  	title = {Gauge Formalism for General Relativity and Fermionic Matter},
  	journal = {General Relativity and Gravitation}
}

@article{FADDEEV196729,
title = {Feynman diagrams for the Yang-Mills field},
journal = {Physics Letters B},
volume = {25},
number = {1},
pages = {29-30},
year = {1967},
issn = {0370-2693},
doi = {https://doi.org/10.1016/0370-2693(67)90067-6},
url = {https://www.sciencedirect.com/science/article/pii/0370269367900676},
author = {L.D. Faddeev and V.N. Popov}
}

@article{FatiNoris22,
author = {Noris, Ruggero and Fatibene, Lorenzo},
title = {Spin frame transformations and Dirac equations},
journal = {International Journal of Geometric Methods in Modern Physics},
volume = {19},
number = {01},
pages = {2250004},
year = {2022},
doi = {10.1142/S0219887822500049},

URL = { 
        https://doi.org/10.1142/S0219887822500049
    
},
eprint = { 
        https://doi.org/10.1142/S0219887822500049
    
}

}

@article{Dir58,
    author ={Dirac, Paul Adrien Maurice},
    doi = {http://doi.org/10.1098/rspa.1958.0141},
    title = {Generalized Hamiltonian Dynamics".},
    journal = {Proc. R. Soc. Lond.},
    year = {1958}
}

@book{spingeom,
author = {H. Blaine Lawson and Marie-Louise Michelsohn},
doi = {doi:10.1515/9781400883912},
title = {Spin Geometry},
year = {1990},
publisher = {Princeton University Press},
ISBN = {9780691085425},
lastchecked = {2022-05-03}
}

@article{Henn,
title = "Elimination of the auxiliary fields in the antifield formalism",
journal = "Physics Letters B",
volume = "238",
number = "2",
pages = "299 - 304",
year = "1990",
author = "Marc Henneaux"
}

@article{BarnichGrigoriev2003,
	Author = {Barnich, Glenn and Grigoriev, Maxim},
	Da = {2005/03/01},
	Doi = {10.1007/s00220-004-1275-4},
	Journal = {Communications in Mathematical Physics},
	Number = {3},
	Pages = {581--601},
	Title = {Hamiltonian BRST and Batalin-Vilkovisky Formalisms for Second Quantization of Gauge Theories},
	Volume = {254},
	Year = {2005}
}

@article{GrigorievDaamgard,
title = "Superfield BRST charge and the master action",
journal = "Physics Letters B",
volume = "474",
number = "3",
pages = "323 - 330",
year = "2000",
author = "M.A. Grigoriev and P.H. Damgaard",
}

@book{HT,
    author = "Henneaux, M. and Teitelboim, C.",
    title = "{Quantization of gauge systems}",
    isbn = "978-0-691-03769-1",
    year = "1992"
}

@article{deWitt,
  title = {Quantum Theory of Gravity. I. The Canonical Theory},
  author = {DeWitt, Bryce S.},
  journal = {Phys. Rev.},
  volume = {160},
  issue = {5},
  pages = {1113--1148},
  numpages = {0},
  year = {1967},
  month = {08},
  publisher = {American Physical Society},
  doi = {10.1103/PhysRev.160.1113},
  url = {https://link.aps.org/doi/10.1103/PhysRev.160.1113}
}

@ARTICLE{AKSZ,
          ids = {AKSZ1997},
       author = {M. Aleksandrov and M. Kontsevich and A. Schwarz and O. Zaboronsky},
        title = {The geometry of the master equation and topological quantum field theory},
 journaltitle = {Int. J. Mod. Phys. A},
         year = {1997},
        issue = {12},
        pages = {1405-1430},
          doi = {10.1142/S0217751X97001031}
}

@ARTICLE{BV3,
          ids = {Batalin:1983},
       author = {{Batalin}, I.~A. and {Fradkin}, E.~S.},
        title = "{A generalized canonical formalism and quantization of reducible gauge theories}",
      journal = {Physics Letters B},
         year = "1983",
        month = "03",
       volume = {122},
       number = {2},
        pages = {157-164},
          doi = {10.1016/0370-2693(83)90784-0}
}

@ARTICLE{BV2,
          ids = {BV1981},
       author = {{Batalin}, I.~A. and {Vilkovisky}, G.~A.},
        title = "{Gauge algebra and quantization}",
      journal = {Physics Letters B},
         year = "1981",
        month = "06",
       volume = {102},
       number = {1},
        pages = {27-31},
          doi = {10.1016/0370-2693(81)90205-7}
}

@article{Baulieu:1990uv,
    author = "Baulieu, Laurent and Bellon, Marc P. and Ouvry, Stephane and Wallet, Jean-Christophe",
    title = "{Balatin-Vilkovisky analysis of supersymmetric systems}",
    reportNumber = "IPNO-TH-90-16, PAR-LPTHE-90-19",
    doi = "10.1016/0370-2693(90)90557-M",
    journal = "Phys. Lett. B",
    volume = "252",
    pages = "387--394",
    year = "1990"
}

@ARTICLE{BV1,
       author = {{Batalin}, I.~A. and {Vilkovisky}, G.~A.},
        title = "{Relativistic S-matrix of dynamical systems with boson and fermion constraints}",
      journal = {Physics Letters B},
         year = "1977",
        month = "08",
       volume = {69},
       number = {3},
        pages = {309-312},
          doi = {10.1016/0370-2693(77)90553-6}
}

@article{Stasheff1997,
author = "Stasheff, Jim",
doi = "10.4310/jdg/1214459757",
fjournal = "Journal of Differential Geometry",
journal = "J. Differential Geom.",
number = "1",
pages = "221--240",
publisher = "Lehigh University",
title = "Homological reduction of constrained Poisson algebras",
volume = "45",
year = "1997"
}

@Article{KS1987,
title = "Symplectic reduction, BRS cohomology, and infinite-dimensional Clifford algebras",
journal = "Annals of Physics",
volume = "176",
number = "1",
pages = "49",
year = "1987",
issn = "0003-4916",
doi = "10.1016/0003-4916(87)90178-3",
author = "Bertram Kostant and Shlomo Sternberg"
}

@article{MW74,
    author = "Marsden, Jerrold and Weinstein, Alan",
    title = "{Reduction of symplectic manifolds with symmetry}",
    doi = "10.1016/0034-4877(74)90021-4",
    journal = "Rept. Math. Phys.",
    volume = "5",
    number = "1",
    pages = "121--130",
    year = "1974"
}

@book{KT79,
    author = "Kijowski, J. and Tulczyjew, W. M.",
    title = "{A SYMPLECTIC FRAMEWORK FOR FIELD THEORIES}",
    year = "1979"
}

@article{Mnev2017,
    author = "Mnev, Pavel",
    archivePrefix = "arXiv",
    eprint = "1707.08096",
    month = "7",
    primaryClass = "math-ph",
    title = "{Lectures on Batalin-Vilkovisky formalism and its applications in topological quantum field theory}",
    year = "2017"
}

@ARTICLE{CMR2,
          ids = {CMR2015},
       author = {{Cattaneo}, Alberto S. and {Mnev}, Pavel and {Reshetikhin}, Nicolai},
        title = "{Perturbative Quantum Gauge Theories on Manifolds with Boundary}",
      journal = {Communications in Mathematical Physics},
         year = "2018",
        month = "01",
       volume = {357},
       number = {2},
        pages = {631-730},
          doi = {10.1007/s00220-017-3031-6}
          }

@ARTICLE{CMR2012,
          ids = { CMR, pippo },
       author = "Cattaneo, Alberto S. and Mnev, Pavel and Reshetikhin, Nicolai",
        title = "Classical BV Theories on Manifolds with Boundary",
      journal = "Communications in Mathematical Physics",
         year = "2014",
       volume = "332",
       number = "2",
        pages = "535--603",
         issn = "1432-0916",
          doi = "10.1007/s00220-014-2145-3",
}

@article{Ati88,
    author = {Atiyah, Michael},
    title = {Topological quantum field theories},
    journal = {Publications Mathématiques de l’Institut des Hautes Scientifiques},
    volume = {68},
    pages = {175--186},
    year = {1988},
    doi = {https://doi.org/10.1007/BF02698547}
}

@article{Graksz21,
    author = "Canepa, Giovanni and Cattaneo, Alberto S. and Schiavina, Michele",
    title = "{General Relativity and the AKSZ Construction}",
    eprint = "2006.13078",
    archivePrefix = "arXiv",
    primaryClass = "math-ph",
    doi = "10.1007/s00220-021-04127-6",
    journal = "Commun. Math. Phys.",
    volume = "385",
    number = "3",
    pages = "1571--1614",
    year = "2021"
}

@misc{canepa20254dpalatinicartangravityhamiltonian,
      title={4D Palatini-Cartan Gravity in Hamiltonian Form}, 
      author={Giovanni Canepa and Alberto S. Cattaneo},
      year={2025},
      eprint={2507.02431},
      archivePrefix={arXiv},
      primaryClass={gr-qc},
      url={https://arxiv.org/abs/2507.02431}, 
}

@misc{C25,
      title={BV Pushforward of Palatini-Cartan gravity}, 
      author={Giovanni Canepa and Alberto S. Cattaneo},
      year={2025},
      eprint={2507.06279},
      archivePrefix={arXiv},
      primaryClass={math-ph},
      url={https://arxiv.org/abs/2507.06279}, 
}

@article{Canepa2019FullyEB,
  title={Fully extended BV-BFV description of General Relativity in three dimensions},
  author={Giovanni Canepa and Michele Schiavina},
  journal={Advances in Theoretical and Mathematical Physics},
  year={2019},
  url={https://api.semanticscholar.org/CorpusID:162184099}
}

@inbook{Cattaneo:2015eip,
    author = "Cattaneo, Alberto S. and Mnev, Pavel and Wernli, Konstantin",
    editor = "Cardona, Alexander and Morales, Pedro and Ocampo, Hernan and Paycha, Sylvie and Reyes Lega, Andres F.",
    title = "{Split Chern-Simons theory in the BV-BFV formalism.}",
    booktitle = "{Proceedings, 9th Summer School on Geometric, Algebraic and Topological Methods for Quantum Field Theory}: {Villa de Leyva, Colombia, July 20-31, 2015}",
    eprint = "1512.00588",
    archivePrefix = "arXiv",
    primaryClass = "math.GT",
    doi = "10.1007/978-3-319-65427-0_9",
    series = "Mathematical Physics Studies",
    pages = "293--324",
    year = "2017"
}

@Inbook{Segal1988,
author="Segal, G. B.",
editor="Bleuler, K.
and Werner, M.",
title="The Definition of Conformal Field Theory",
bookTitle="Differential Geometrical Methods in Theoretical Physics",
year="1988",
publisher="Springer Netherlands",
address="Dordrecht",
pages="165--171",
isbn="978-94-015-7809-7",
doi="10.1007/978-94-015-7809-7_9",
url="https://doi.org/10.1007/978-94-015-7809-7_9"
}

@article{Scht08,
    author = {Schaetz, Florian},
    title = {BFV-Complex and Higher Homotopy Structures},
    journal = {Commun. Math. Phys},
    year = "2009" , 
    pages = {399--443} , 
    doi = {https://doi.org/10.1007/s00220-008-0705-0}
}

@ARTICLE{CMR2012b,
      author         = "Cattaneo, Alberto S. and Mnev, Pavel and Reshetikhin,
                        Nicolai",
      title          = "{Classical and quantum Lagrangian field theories with
                        boundary}",
      booktitle      = "{Proceedings, 11th Hellenic School and Workshops on
                        Elementary Particle Physics and Gravity (CORFU2011):
                        Corfu, Greece, September 4-18, 2011}",
      journal        = "PoS",
      volume         = "CORFU2011",
      year           = "2011",
      pages          = "044",
      doi            = "10.22323/1.155.0044",
      eprint         = "1207.0239",
      archivePrefix  = "arXiv",
      primaryClass   = "math-ph"
}

@article{Cattaneo:2024sfd,
    author = "Cattaneo, A. S. and Moshayedi, N. and Funcasta, A. Smailovic",
    title = "{3D Supergravity In the Batalin--Vilkovisky Formalism}",
    eprint = "2412.14300",
    archivePrefix = "arXiv",
    primaryClass = "hep-th",
    month = "12",
    year = "2024"
}

@incollection{Cattaneo:2023wxd,
title = {Phase Space for Gravity With Boundaries},
editor = {Richard Szabo and Martin Bojowald},
booktitle = {Encyclopedia of Mathematical Physics (Second Edition)},
publisher = {Academic Press},
edition = {Second Edition},
address = {Oxford},
pages = {480-494},
year = {2025},
isbn = {978-0-323-95706-9},
doi = {https://doi.org/10.1016/B978-0-323-95703-8.00052-5},
url = {https://www.sciencedirect.com/science/article/pii/B9780323957038000525},
author = {Alberto S. Cattaneo}
}

@ARTICLE{CS2017,
          ids = {CSPCH},
       author = {{Cattaneo}, Alberto S. and {Schiavina}, Michele},
        title = "{BV-BFV approach to General Relativity: Palatini--Cartan--Holst action}",
         journal = {Advances in Theoretical and Mathematical Physics},
         year = "2019",
        month = "12",
        pages = {2025 - 2059},
       volume = {23},
       issue  = {8},
       doi    = {10.4310/ATMP.2019.v23.n8.a3 }
}

@article{Canepa:2024rib,
    author = "Canepa, Giovanni",
    title = "{On the properties of coframes}",
    eprint = "2410.17682",
    archivePrefix = "arXiv",
    primaryClass = "math-ph",
    month = "10",
    year = "2024"
}

@Article{CS2019,
       author = "Cattaneo, Alberto S. and Schiavina, Michele",
        title = "The Reduced Phase Space of Palatini--Cartan--Holst Theory",
      journal = "Annales Henri Poincar{\'e}",
         year = "2019",
          day = "01",
       volume = "20",
       number = "2",
        pages = "445--480",
          doi = "10.1007/s00023-018-0733-z"
}

@article{CCS2020,
    title={Boundary structure of General Relativity in tetrad variables},
    author={Giovanni Canepa and Alberto S. Cattaneo and Michele Schiavina},
    journal={Adv. Theor. Math. Phys.},
    volume={25},
    pages={327-377},
    year={2021},
    doi={10.4310/ATMP.2021.v25.n2.a3},
    % eprint={2001.11004},
    % archivePrefix={arXiv},
    % primaryClass={math-ph},
      keywords        = {My}
}

@Article{Cartan,
author = {{Cartan}, E.},
title = {Sur une g\'en\'eralisation de la notion de courbure de Riemann et les espaces \`a torsion.},
journal = {C. R. Acad. Sci.},
year = {1922},
volume = {174},
pages = {593-595},
}

@Article{Palatini1919,
author="Palatini, Attilio",
title="Deduzione invariantiva delle equazioni gravitazionali dal principio di Hamilton",
journal="Rendiconti del Circolo Matematico di Palermo (1884-1940)",
year="1919",
volume="43",
number="1",
pages="203--212",
issn="0009-725X",
doi="10.1007/BF03014670"
}

@ARTICLE{Ashtekar1986,
   author = {{Ashtekar}, A.},
    title = "{New variables for classical and quantum gravity}",
  journal = {Physical Review Letters},
     year = 1986,
    month = nov,
   volume = 57,
    pages = {2244-2247},
      doi = {10.1103/PhysRevLett.57.2244}
}

@book{RudolphSchmid,
    author = "Rudolph, Gerd and Schmidt, Matthias",
    title = "{Differential Geometry and Mathematical Physics Part II. Fibre Bundles, Topology and Gauge Fields}",
    doi = "10.1007/978-94-024-0959-8",
    isbn = "978-94-024-0958-1",
    publisher = "Springer Dordrecht",
    month = "3",
    year = "2017"
}

@book{Freedman:2012zz,
    author = "Freedman, Daniel Z. and Van Proeyen, Antoine",
    title = "{Supergravity}",
    isbn = "978-1-139-36806-3, 978-0-521-19401-3",
    publisher = "Cambridge Univ. Press",
    address = "Cambridge, UK",
    month = "5",
    year = "2012"
}

@book{Dabrowski:88,
    author = "Dabrowski, L.",
    title = "{Group actions on spinors: Lectures at the university of Naples}",
    publisher = "Bibliopolis ",
    isbn = "8870882055,9788870882056 ",
    year = "1988"
}

@book{CDF,
    author = "Castellani, L. and D'Auria, R. and Fre, P.",
    title = "{Supergravity and superstrings: A Geometric perspective. Vol. 1: Mathematical foundations}",
    year = "1991"
}

@book{CDF2,
    author = "Castellani, L. and D'Auria, R. and Fre, P.",
    title = "{Supergravity and superstrings: A Geometric perspective. Vol. 1: Mathematical foundations}",
    year = "1991"
}

@article{KT83,
title = {Supersymmetry and the division algebras},
journal = {Nuclear Physics B},
volume = {221},
number = {2},
pages = {357-380},
year = {1983},
issn = {0550-3213},
author = {Taichiro Kugo and Paul Townsend},
}

@article{Scherk:1978fh,
    author = "Scherk, Joel",
    title = "{Extended Supersymmery and Supergravity Theories}",
    reportNumber = "LPTENS-78-21",
    journal = "NATO Sci. Ser. B",
    volume = "44",
    pages = "0479",
    year = "1979"
}

@inproceedings{VanNieuwenhuizen:1985be,
    author = "Van Nieuwenhuizen, P.",
    title = "{An Introduction to Simple Supergravity and the Kaluza-Klein Program}",
    booktitle = "{Les Houches Summer School on Theoretical Physics: Relativity, Groups and Topology}",
    pages = "823--932",
    year = "1985"
}

@article{article,
author = {Wang, Mckenzie Yuen Kong},
year = {1989},
month = {01},
pages = {59-68},
title = {Parallel Spinors and Parallel Forms},
volume = {7},
journal = {Annals of Global Analysis and Geometry},
doi = {10.1007/BF00137402}
}

@book{harvey1990spinors,
  title={Spinors and Calibrations},
  author={Harvey, F.R.},
  isbn={9780123296504},
  lccn={lc89000074},
  url={https://books.google.it/books?id=G1XvAAAAMAAJ},
  year={1990},
  publisher={Elsevier Science}
}

@article{Figueroa,
    author = "Figueroa-O'Farrill, Jose",
    title = "{Majorana Spinors}",
    url=\\ {https://www.maths.ed.ac.uk/~jmf/Teaching/Lectures/Majorana.pdf}
}

@article{BBH,
author = "Barnich, Glenn and Brandt, Friedemann and Henneaux, Marc",
fjournal = "Communications in Mathematical Physics",
journal = "Comm. Math. Phys.",
number = "1",
pages = "57--91",
publisher = "Springer",
title = "Local BRST cohomology in the antifield formalism. I. General theorems",
volume = "174",
year = "1995"
}

@article{Kupka:2024vrd,
    author = "Kupka, Julian and Strickland-Constable, Charles and Valach, Fridrich",
    title = "{Direct derivation of gauged $\mathcal N=1$ supergravity in ten dimensions to all orders in fermions}",
    eprint = "2410.16046",
    archivePrefix = "arXiv",
    primaryClass = "hep-th",
    month = "10",
    year = "2024"
}

@article{Kupka:2024tic,
    author = "Kupka, Julian and Strickland-Constable, Charles and Valach, Fridrich",
    title = "{Supergravity without gravity and its BV formulation}",
    eprint = "2408.14656",
    archivePrefix = "arXiv",
    primaryClass = "hep-th",
    month = "8",
    year = "2024"
}

@article{T2019,
   title={On the Mathematics of Coframe Formalism and Einstein–Cartan Theory—A Brief Review},
   volume={5},
   ISSN={2218-1997},
   DOI={10.3390/universe5100206},
   number={10},
   journal={Universe},
   publisher={MDPI AG},
   author={Tecchiolli, Manuel},
   year={2019},
   month={9},
   pages={206}
}

@misc{delgado2018lagrangianfieldtheoriesindproapproach,
      title={Lagrangian field theories: ind/pro-approach and L-infinity algebra of local observables}, 
      author={Nestor Leon Delgado},
      year={2018},
      eprint={1805.10317},
      archivePrefix={arXiv},
      primaryClass={math-ph},
      url={https://arxiv.org/abs/1805.10317}, 
}

@article{Khudaverdian2000SemidensitiesOO,
  title={Semidensities on Odd Symplectic Supermanifolds},
  author={Hovhannes M. Khudaverdian},
  journal={Communications in Mathematical Physics},
  year={2000},
  volume={247},
  pages={353-390},
  url={https://api.semanticscholar.org/CorpusID:119174961}
}

@article{Severa06,
author = {Severa, Pavol},
year = {2006},
month = {01},
pages = {55-59},
title = {On the Origin of the BV Operator on Odd Symplectic Supermanifolds},
volume = {78},
journal = {Letters in Mathematical Physics},
doi = {10.1007/s11005-006-0097-z}
}

@misc{T1975,
      title={Gauge Invariance in Field Theory and Statistical Physics in Operator Formalism}, 
      author={I. V. Tyutin},
      year={1975},
      eprint={0812.0580},
      archivePrefix={arXiv},
      primaryClass={hep-th}
}

@article{BRS1976,
title = "Renormalization of gauge theories",
journal = "Annals of Physics",
volume = "98",
number = "2",
pages = "287 - 321",
year = "1976",
issn = "0003-4916",
author = "C Becchi and A Rouet and R Stora"
}

@article{Piguet:2000fy,
    author = "Piguet, Olivier",
    title = "{Ghost equations and diffeomorphism invariant theories}",
    eprint = "hep-th/0005011",
    archivePrefix = "arXiv",
    reportNumber = "UFES-DF-OP2000-1",
    doi = "10.1088/0264-9381/17/18/314",
    journal = "Class. Quant. Grav.",
    volume = "17",
    pages = "3799--3806",
    year = "2000"
}

@article{Moritsch:1993eg,
    author = "Moritsch, Otmar and Schweda, Manfred and Sorella, Silvio P.",
    title = "{Algebraic structure of gravity with torsion}",
    eprint = "hep-th/9310179",
    archivePrefix = "arXiv",
    reportNumber = "TUW-93-24",
    doi = "10.1088/0264-9381/11/5/010",
    journal = "Class. Quant. Grav.",
    volume = "11",
    pages = "1225--1242",
    year = "1994"
}

@article{Baulieu:1985md,
    author = "Baulieu, L. and Bellon, Marc P.",
    title = "{$p$ Forms and Supergravity: Gauge Symmetries in Curved Space}",
    reportNumber = "LPTENS 85/7",
    doi = "10.1016/0550-3213(86)90178-1",
    journal = "Nucl. Phys. B",
    volume = "266",
    pages = "75--124",
    year = "1986"
}

@article{CFT24,
author = {Canepa, Giovanni and Cattaneo, Alberto and Fila-Robattino, Filippo and Tecchiolli, Manuel},
year = {2024},
month = {09},
pages = {},
title = {Boundary Structure of the Standard Model Coupled to Gravity},
journal = {Annales Henri Poincaré},
doi = {10.1007/s00023-024-01485-4}
}

@unknown{CFR25,
author = {Cattaneo, Alberto and Fila-Robattino, Filippo},
year = {2025},
month = {03},
pages = {},
title = {BV description of $N = 1$, $D = 4$ Supergravity in the first order formalism},
doi = {10.48550/arXiv.2503.07373}
}

@misc{FR25,
      title={Tools for Supergravity in the spin coframe formalism}, 
      author={Filippo Fila-Robattino},
      year={2025},
      eprint={2503.07355},
      archivePrefix={arXiv},
      primaryClass={math-ph},
      url={https://arxiv.org/abs/2503.07355}, 
}

@misc{grigoriev2025presymplecticbvakszn1d4,
      title={Presymplectic BV-AKSZ for $N=1$ $D=4$ Supergravity}, 
      author={Maxim Grigoriev and Alexander Mamekin},
      year={2025},
      eprint={2503.04559},
      archivePrefix={arXiv},
      primaryClass={hep-th},
      url={https://arxiv.org/abs/2503.04559}, 
}

@misc{blohmann2021hamiltonianliealgebroids,
      title={Hamiltonian Lie algebroids}, 
      author={Christian Blohmann and Alan Weinstein},
      year={2021},
      eprint={1811.11109},
      archivePrefix={arXiv},
      primaryClass={math.SG},
      url={https://arxiv.org/abs/1811.11109}, 
}

\end{document}